\documentclass[
aps,%
12pt,%
oneside,%
onecolumn,%
nobibnotes,%
nofootinbib,%
superscriptaddress,%
noshowpacs,%
centertags]%
{revtex4}

\usepackage{epsfig}
\def\shiftleft#1{#1\llap{#1\hskip 0.04em}}
\def\shiftdown#1{#1\llap{\lower.04ex\hbox{#1}}}
\def\thick#1{\shiftdown{\shiftleft{#1}}}
\def\b#1{\thick{\hbox{$#1$}}}

\begin{document}
\title
{Dibaryon model for nuclear force and the properties
 of the $3N$ system}

\author{V. N. Pomerantsev, V. I. Kukulin,  V. T. Voronchev, 
and A. Faessler\footnotemark[1]\\
 {\em Institute of Nuclear Physics, Moscow State University, Russia. } }
 \noaffiliation 
\footnotetext[1]{Institute for Theoretical Physics, 
 University of T\"ubingen, Germany.}
 
\begin{abstract} \noindent 
The dibaryon model for $NN$ interaction, which implies the formation of
an intermediate six-quark bag dressed by a $\sigma$-field, is applied
to the $3N$ system, where it results in a new three-body force of
scalar nature between the six-quark bag and a third nucleon. A new
multicomponent formalism is developed to describe three-body systems
with nonstatic pairwise interactions and non-nucleonic degrees of
freedom.  Precise variational calculations of $3N$ bound states are
carried out in the dressed-bag model including the new scalar
three-body force. The unified coupling constants and form factors for
$2N$ and $3N$ force operators are used in the present approach, in a
sharp contrast to conventional meson-exchange models. It is shown that
this three-body force gives at least half the $3N$ total binding
energy, while the weight of non-nucleonic components in the $^3$H and
$^3$He wavefunctions can exceed 10\%. The new force model provides a
very good description of $3N$ bound states with a reasonable magnitude
of the $\sigma NN$ coupling constant.  A new Coulomb $3N$ force between
the third nucleon and dibaryon is found to be very important for  a
correct description of the Coulomb energy and r.m.s. charge radius in
$^3$He. In view of the new results for Coulomb displacement energy
obtained here for $A=3$ nuclei, an explanation for the long-term
Nolen--Schiffer paradox in nuclear physics is suggested. The role of
the  charge-symmetry-breaking  effects in the  nuclear force is
discussed.
\end{abstract} 
\maketitle
\setcounter{footnote}{1}

\section{Introduction. current problems in a consistent
description of $NN$ and $3N$ systems with traditional force
models}

 A few historical remarks should be done at first. Current rather high
activity in few-body physics started since the beginning of 1960-s,
after mathematical formulation of the Faddeev equations for three-body
problem. The aim was claimed to establish unambiguously off-shell
properties of the two-body $t$-matrix, which cannot be derived from
two-body scattering data only. It has been hoped in that time that just
accurate solving $3N$ scattering problem is able to put strong
constraints for the off-shell properties of the two-nucleon $t$-matrix.
However, more than forty years passed since that, but still now we are
unable to formulate such a two-nucleon $t$-matrix, which can explain
fully quantitatively the properties of even $3N$ systems.

Moreover, from that time, many puzzles in few-nucleon scattering
experiments have been revealed which could not be explained by the
current force models based on Yukawa concept. Among all such
puzzles, we mention here only the most remarkable ones, such as
the $A_y$ puzzle in $\vec{N}+d$ and $\vec{N}+{}^3{\rm He}$
scattering~\cite{FB,Ay}, disagreements at the minima of
differential cross sections (Sagara puzzle) at $E\sim 150$--200~MeV 
and polarization data for $N+\vec{d}$~\cite{Sagara},
$\vec{N}+d$, $\vec{N}+\vec{d}$~\cite{polnd}, and $\vec{N}+{}^3{\rm
He}$ scattering, and many others.  The strongest discrepancy
between current theories and respective experiments has been found
in studies of the short-range $NN$ correlations in the $^3{\rm
He}(e,e'pp)$~\cite{Nikef}, $^4{\rm He}(\gamma,pp)$~\cite{Grab},
and $^3{\rm He}(e,e'NN)$~\cite{Jef} processes.  In addition to
these particular problems, there are more fundamental problems in
the current theory of nuclear forces, e.g., strong discrepancies
between the $\pi NN$, $\pi N\Delta$, and $\rho NN$ form factors
used both in one-boson-exchange (OBE) models for the description of elastic and
inelastic scattering and in the consistent parametrization of
$2N$ and $3N$ forces~\cite{KuJPG,Gibson,Gibson88,Sauer}. Many of
these difficulties are attributed to a rather poor knowledge of
the short-range behavior of nuclear forces. This behavior was
traditionally associated with the vector $\omega$-meson exchange.
However, the characteristic range of this $\omega$-exchange (for
$m_{\omega}\simeq 780$~MeV) is equal to about $\lambda_{\omega}
\simeq 0.2$--0.3~fm, i.e., is deeply inside the internucleon
overlap region.

In fact, since Yukawa the nucleon-nucleon interaction is explained by a
$t$-channel  exchange of mesons  between two nucleons. The very
successful Bonn, Nijmegen, Argonne, and other modern $NN$ potentials
prove the success of this approach. But the short- and
intermediate-range region  in these potentials is more parametrized than
parameter-free microscopically described.

Besides of the evident difficulties with the description for short-range
nuclear force there are also quite serious problems with consistent
description of basic intermediate-range attraction between nucleons. In
the traditional OBE models  this attraction  is described as a
$t$-channel $\sigma$-exchange with the artificially enhanced $\sigma NN$
vertices. However, the accurate modern calculations of the
intermediate-range $NN$ interaction~\cite{twopi,Oset} within the
$2\pi$-exchange model with the $\pi\pi$ $s$-wave interaction have
revealed that this $t$-channel mechanism cannot give a strong
intermediate-range attraction in the $NN$ sector, which is necessary
for binding of a deuteron and fitting the $NN$ phase shifts. This
conclusion  has  also been corroborated by recent independent
calculations~\cite{Ka04}. Thus, the $t$-channel mechanism of the
$\sigma$-meson exchange should be replaced by some other alternative
mechanism, which should result in the strong intermediate-range
attraction required by even existence of nuclei.

When analyzing the deep reasons for all these failures, we must  look
on a most general element, which is common for all the numerous $NN$
force models tested in few-nucleon calculations for last 40 years. This
common element is just the Yukawa concept for the strong interaction of
nucleons in nuclei. Hence, if, after more than 40 years of development,
we are still unable to explain quantitatively and consistently even the
basic properties of $3N$ and $4N$ systems at low energies and
relatively simple processes like $pp\to pp\gamma$, this concept, which
is a cornerstone of all building of nuclear physics, should be analyzed
critically, especially in the regions where applicability of this
concept looks rather questionable.

Since the quark picture and QCD have been developed, the
"nucleon-nucleon force community" is more and more convinced that at
short ranges the quark degrees of freedom must play  an important role.
One of possible mechanisms for short-range $NN$ interaction  is the
formation of the six-quark bag (dibaryon) in the $s$-channel.
Qualitatively  many would agree with this statement. But to obtain a
quantitative description  of the nucleon-nucleon and the few-nucleon
experimental data with this approach with the same quality as the
commonly used Bonn, Nijmegen, Argonne, and other equivalent potentials is
a quite different problem.

 Within the $6q$ dynamics it has long been
known~\cite{Kus91,Myhrer,Fae83,Progr92,Yam86} that the mixing of the
completely symmetric $s^6[6]$ component with the mixed-symmetry
$s^4p^2[42]$ component can determine the structure of the whole
short-range interaction (in the $S$-wave)\footnote{We will denote the
$NN$ partial waves by capital letters ($S$, $P$ \dots), while the partial waves in all
other cases will be denoted by small letters.}. Assuming a reasonable $qq$
interaction model, many authors (see,
e.g.,~\cite{Oka83,Fujiwara,Stancu,Bartz}) have suggested that this
mixture will result in both strong short-range repulsion (associated
mainly with the $s^6$ component) and intermediate-range attraction
(associated mainly with the above mixed-symmetry $s^4p^2$ component).
However, recent studies~\cite{Stancu,Bartz} for $NN$ scattering on the
basis of the newly developed  Goldstone-boson-exchange (GBE) $qq$
interaction have resulted in  a purely repulsive $NN$
contributions from both $s^6[6]$ and $s^4p^2[42]$ $6q$ components.
There is no need to say that any quark-motivated model for the $NN$
force with $\pi$-exchange between quarks inevitably leads to the
well-established Yukawa $\pi$-exchange interaction between nucleons at
long distances.

Trying to solve the above problems (and to understand more deeply
the mechanism for the intermediate- and short-range $NN$
interaction), the Moscow--T\"ubingen group suggested to replace
the conventional Yukawa meson-exchange ($t$-channel) mechanism (at
intermediate and short ranges) by the contribution of a
$s$-channel mechanism describing the formation of a dressed
$6q$ bag in the intermediate state such as $|s^6
+\sigma\rangle$ or $|s^6+2\pi\rangle$~\cite{KuJPG,KuInt}. It has
been shown that, due to the change in the symmetry of the
$6q$ state in the transition from the $NN$ channel to the
intermediate dressed-bag state, the strong scalar $\sigma$-field
arises around the symmetric $6q$ bag. This intensive
$\sigma$-field squeezes the bag and increases its
density~\cite{Bled}. The high quark density in the symmetric $6q$
bag enhances the meson field fluctuations around the bag and
thereby partially restores the chiral symmetry~\cite{Kuni}.
Therefore, the masses of constituent quarks and $\sigma$ mesons
decrease~\cite{KuJPG}. As a result of this phase transition, the
dressed bag mass decreases considerably (i.e., a large gain in
energy arises), which manifests itself as a strong effective
attraction in the $NN$ channel at intermediate distances. The
contribution of the $s$-channel mechanism would generally be much
larger due to resonance-like
enhancement\footnote{In the theory of nuclear reactions, the
$t$-channel mechanism can be associated with the direct nuclear
reaction, where only a few degrees of freedom are important, while
the  $s$-channel mechanism can be associated with resonance-like
(or compound-nucleus-like) nuclear reactions with much larger
cross sections at low energies.}.

In our previous works~\cite{KuJPG,KuInt} on the basis of the above
arguments we proposed a new dibaryon model for the $NN$
interaction (referred further to as the ``dressed-bag model''
(DBM)),which provided a quite good description of both $NN$ phase
shifts up to 1~GeV and the deuteron structure. The developed model
includes  the conventional $t$-channel contributions (Yukawa
$\pi$- and $2\pi$-exchanges) at long and intermediate distances
and the $s$-channel contributions due to the formation of
intermediate dressed-bag states at short distances. The most
important distinction of such an approach from conventional models
for nuclear forces is the explicit appearance of a non-nucleonic
component in the total wavefunction of the system, which
necessarily implies the presence of new three-body forces (3BF) of
several kinds in the $3N$ system. These new 3BF
differ from conventionally used models for three-body forces. One
important aspect of the novel 3BF should be
emphasized here. In conventional OBE models, the main contribution
to $NN$ attraction is due to the $t$-channel $\sigma$-exchange.
However, the 3BF models suggested until now (such as
Urbana--Illinois or Tucson--Melbourne) are mainly based on the
$2\pi$-exchange with intermediate $\Delta$-isobar production,
and the $\sigma$-exchange is either not taken into account at all,
or is of little importance in these models. In contrast, the
$\sigma$-exchange in our approach dominates in both $NN$ and
$3N$ forces. In fact, in our approach just the unified strong
$\sigma$-field couples both two- and three (and more)-nucleon
systems, i.e., the general pattern of the nuclear interaction
appears to be more consistent.

Our recent considerations have revealed that this dibaryon mode is
extremely useful in the explanation of very numerous facts and
experimental results in nuclear physics, in general. We note here only a
few of them.
\begin{enumerate}
\item The presence of dibaryon degree of freedom (DDF) can result  in
very natural explanation of cumulative effects (e.g., the production of
cumulative particles in high-energy collisions~\cite{65}).
\item DDF leads to automatic enhancement  of near-threshold cross
sections for one- and two-meson production in $pp$, $pd$, etc. collisions,
which is required  by many modern experiments (e.g., the so-called ABC
puzzle~\cite{ABC}).  This is due to an effective enhancement of
meson--dibaryon coupling as compared to meson--nucleon coupling.
\item The incorporation of DDF makes it possible (without the artificial
enhancement of meson--nucleon form factors) to share the large momentum
of an incident probe (e.g., high-energy photon) among other nucleons in
the target  nucleus.
\item The DDF produces in a very natural way a new short-range currents
required by almost all experiments associated with high momentum and
energy transfers.
\item Presence of the dressed $6q$ bag components in nuclear wave
functions leads automatically to a smooth matching between the nucleonic
(at low momentum transferred) and quark currents (at very high
momentum transferred) and, at the same time, results in a correct
counting rules at high momentum transferred.
\end{enumerate}

So, it should be very important to test the above dibaryon concept of
nuclear force in a concise and consistent $3N$ calculations and
to compare  the predictions of  the new model with the results of the
conventional meson-exchange models.

Thus, the aim of this work is to make a comprehensive study of the
properties of the $3N$ system with $NN$ and $3N$ forces  given by
the DBM. However, DBM introduces explicitly
non-nucleonic (quark--meson) channels. Therefore, it is necessary
to introduce a selfconsistent multichannel few-body formalism
for the study of $3N$ system with DBM interaction. We develope in
this work such a general formalism, based on the approach which
was suggested in 1980-s by Merkuriev's group~\cite{1,2,3} for the
boundary-condition-type model for pairwise interactions. This
general formalism leads immediately to a replacement of all
two-body forces related to the dibaryon mechanism to the
respective three (and many)-body forces, leaving two-body
character only for long-range Yukawa $\pi$ and $2\pi$ exchanges, 
which are of little importance for the nuclear binding. Another
straightforward sequence of the formalism developed here is a
strong energy dependence of these many-body forces. In the work we
study all these aspects in detail when applying to the $3N$ system
properties. The preliminary version of this work is published in
\cite{sys3n}.

This paper is organized as follows. In Section 2, we present a new
general multichannel formalism for description of two- and three-body
system of particles having inner degrees of freedom. In Section~3, we
give a brief description of the DBM for the $NN$ system.  In Section~4,
we treat the $3N$ system with DBM interactions, including a new 3BF. In
Section~5, some details of our variational method are discussed,
including calculation of the matrix elements for new Coulomb 3BF.  The
results of our calculations for ground states of $^3$H and $^3$He are
given in Section~6 while in Section~7 we discuss the role of the new
three-body force and present a new  explanation for the Coulomb
displacement energy in $^3$He within our interaction model. A
comprehensive discussion of the most important results found in the
work is given in Section 8. In the Conclusion we summarize the main
results of the work. In the Appendix we give the formulas for the
matrix elements of all DBM interactions taken in the Gaussian
symmetrized variational basis.

\section{The general multicomponent formalism for $2N$ and $3N$
systems with coupled internal and external channels}
 In 1980-s, the former Leningrad group (Merkuriev, Motovilov, Makarov, 
Pavlov, et al.)
 has
constructed and substantiated with mathematical rigor a model of strong
interaction with the coupling of external and internal
channels~\cite{1,2,3,4}. This model was a particular realization of a general
approach to interaction of particles having inner degrees of freedom. The basic
physical hypothesis is  that an energy-dependent interaction appears as
a result of internal structure of interacting particles\footnote{%
From more general point of view, the explicit energy dependence of interaction
reflects its nonlocality in the time, while this time nonlocality, in its
turn, is a result of some excluded degrees of freedom. So, the explicit energy
dependence is signalling about some inner hidden (e.g., quark) degrees of freedom in
$NN$ interaction.}.
A general scheme proposed by Merkuriev et al. has been  based on
assumption on existence of two independent channels: external one, which
describes the motion of particles considered as elementary bodies, i.e.,
neglecting their inner structure, and  internal one, which describes the
dynamics of inner degrees of freedom.   These channels can have quite
different physical and mathematical nature and their dynamics are
governed by independent Hamiltonians. The main issues here were~--  how
to define the coupling between external and internal channels and how to
derive corresponding dynamical equations (of Schr\"odinger or Faddeev
type) for particle motion in the external channel.

In \cite{1,2,3} this coupling has been postulated via boundary conditions on
some hypersurface. Thus, such an approach is well applicable to hybrid models
for $NN$ interaction, which were rather popular in 1980-s, e.g., the quark
compound bag (QCB) model suggested by Simonov~\cite{Sim}.  As for the
$3N$ system, the formalism of incorporation of the internal channels ($6q$ bags) has
been proposed for the first time  also within the QCB model~\cite{Kal}. 
The general scheme
by Merkuriev et al. has allowed to substantiate this formalism. 

In QCB-like models the coupling between the external ($NN$) and the
inner (bag) channels is given just on some hypersurface, similarly
to the well-known $R$-matrix approach in nuclear physics.  Later
on, such a general approach has been applied to the two-channel
Hamiltonian model,  where the internal Hamiltonian had pure
discrete spectrum and the only restriction imposed on the operators
coupling the external and internal channels was their
boundness~\cite{4}. The above general multichannel scheme has
straightforwardly been extended to three-body problem.  In
particular, it has been shown for the two above models that
elimination of the internal channels leads to the following  recipe
for embedding the  energy-dependent pair interactions into
three-body problem: replacement of pair energy by difference
between three-body energy and kinetic energy of third particle:
$\varepsilon_{\alpha} \to E-t_{\alpha}$~\cite{3,4}. It has also
been proved that  the resulted Faddeev equations for external
channel  belong with such energy-dependent potentials to the
Fredholm class  and  are equivalent to four-channel Schr\"odinger
equation.

Our aim here is to extend our new $NN$ force model -- DBM -- by using
the above Merkuriev et al. approach to the $3N$ system. There
are external (nucleon--nucleon) and internal (quark--meson) channels in
our model, and coupling between them is determined within a
microscopical quark--meson approach.  In this section we present a
general multicomponent formalism for description of systems of two and
three particles having internal structure, without  assuming any
specific form for coupling between the external and internal channels.

 \subsection{Two-body system}

We adopt that the total dynamics in two-body system is governed by a
selfconjugated Hamiltonian $h$ acting in the orthogonal sum of spaces:
\[{\cal H}={\cal H}^{\rm ex} \oplus {\cal H}^{\rm in},\]
where  ${\cal H}^{\rm ex}$ is the external Hilbert space of states
describing motion of particles neglecting their internal structure and
${\cal H}^{\rm in}$ is the internal Hilbert space corresponding to internal
degrees of freedom.
 Thus the total state of the system $\Psi \! \in \! {\cal H}$  can be written
  as a two-component column:
\[\Psi= \left (\begin{array}{l} \Psi^{\rm ex}\in \! {\cal H}^{\rm ex}\\
\Psi^{\rm in}\in \! {\cal H}^{\rm in} \end{array} \right )\!.
\]
The two spaces, ${\cal H}^{\rm ex}$ and ${\cal H}^{\rm in}$, can have quite
different nature, e.g., in the case of $NN$ system $\Psi^{\rm ex}$ depends on
the relative coordinate (or momentum) of two nucleons and their spins, while
$\Psi^{\rm in}$ can depend on quark and meson variables. The two independent
Hamiltonians are defined in each of these spaces:
 $h^{\rm ex}$ acts in ${\cal H}^{\rm ex}$  and
 $h^{\rm in}$ acts in ${\cal H}^{\rm in}$.
 Here $h^{\rm ex}$
includes the kinetic energy of relative motion and some part of two-body
interaction $v^{\rm ex}$:
\[ h^{\rm ex}=t+v^{\rm ex}. \]
For $NN$ system $v^{\rm ex}$ includes the peripheral part of meson-exchange potential and
Coulomb interaction between nucleons (if they are protons).
Coupling between external and internal channels is determined formally by some
transition operators:
$ h^{\rm ex,in}= (h^{\rm in,ex})^*$.
 Further, one can write down the  total Hamiltonian $h$  as a matrix operator:
\begin{equation}
h = \left (\begin{array}{ll} h^{\rm ex} & h^{\rm ex,in}\\
 h^{\rm in,ex} & h^{\rm in} \end{array} \right )\!,
\label{h2}
\end{equation}
 not specifying so far the coupling operators (if operators $h^{\rm ex}$ and
 $h^{\rm in}$ are self-adjoint and $h^{\rm ex,in}$ is bounded then the Hamiltonian
 $h$ is the self-adjoint operator in $\cal H$).

Thus one can write down the two-component Schr\"odinger equation
    \[ h\Psi= E\Psi ,\]
and by excluding the internal channel wavefunction one
 obtains an effective Schr\"odinger equation in the external channel
 \begin{equation}  h^{\rm eff}(E)\Psi^{\rm ex} =E\Psi^{\rm ex}
 \label{Sch2}
 \end{equation}
 with an effective ``pseudo-Hamiltonian'':
\begin{equation}
h^{\rm eff}(E) = h^{\rm ex}+h^{\rm ex,in}\,g^{\rm in}(E)\,h^{\rm in,ex} = t +v^{\rm ex} + w(E),
\label{heff2}
\end{equation}
which depends on energy $E$ via the resolvent of internal Hamiltonian
 $g^{\rm in}(E)=(E-h^{\rm in})^{-1}$. (From mathematical point of view, an
 operator depending on the spectral parameter is not operator in all, because its
 domain depends on the spectral parameter. Thus, this object should not be called
 Hamiltonian. However, physicists do not turn their attention to the fact and use
 energy-dependent interactions very widely.)

 Having the solution $\Psi^{\rm ex}$ of effective equation~(\ref{Sch2}), one
 can ``restore''
 the excluded internal state unambiguously:
 \begin{equation}
 \Psi^{\rm in} =g^{\rm in}(E)h^{\rm in,ex}\Psi^{\rm ex}.
 \label{psiin}
 \end{equation}

\subsection{Three-body system}

In three-body system we have three different internal spaces ${\cal H}^{\rm
in}_{i}(i=1,2,3)$ and one common external space ${\cal H}^{\rm ex}_3$. The
three-body internal space ${\cal H}^{\rm in}_{i}$ is a direct product of the
two-body internal space related to the pair ($jk$) and single-particle space
describing motion of third particle ($i$). Here we use the conventional
numbering of particles: $(ijk)=(123),(231),(312)$. The own three-body
Hamiltonian acts in each internal space as:

 \begin{equation}
 H^{\rm in}_{i}= h^{\rm in}_{jk}\otimes
 \mathbb I_{i}+\mathbb I_{jk}\otimes
 t_{i};\;(ijk)=(123),(231),(312),
 \label{hina}
 \end{equation}
where $h^{\rm in}_{jk}$ is the two-body internal Hamiltonian for the  pair $(jk)$,
$\mathbb I$ is unity operator and $t_{i}$ is kinetic energy of third particle
($i$) in respect to the center of mass of the pair ($jk$). (Here and below we
use capital letters for three-body quantities and small letters for two-body
ones.)

The external three-body Hamiltonian acts in the external space ${\cal H}^{\rm ex}_3$
 and includes the total kinetic
energy $T$ and the sum of external two-body interactions, which were incorporated
to the
external two-body Hamiltonians:
\[H^{\rm ex}_3=T+\sum_{i<j}v^{\rm ex}_{ij}.\]

 A state in the full three-body Hilbert space
\[{\cal H}_3={\cal
H}_3^{\rm ex} \oplus\sum_{i}{\cal H}^{\rm in}_{i}\]
 can be written as a four-component column:
\[
\Psi_3=\left ( \begin{array}{l}
  \Psi^{\rm ex} \\
  \Psi^{\rm in}_1 \\
  \Psi^{\rm in}_2 \\
  \Psi^{\rm in}_3
\end{array}\right )\!.
\]

  Thus, the total Hamiltonian, $H_3$,  of the three-body system  acting in
  ${\cal H}_3$ can be written as ($4 \times 4$) matrix:
\begin{equation}
H_3=\left ( \begin{array}{llll}
  H^{\rm ex} & H_1^{\rm ex,in} & H_2^{\rm ex,in} & H_3^{\rm ex,in}  \\
  H_1^{\rm in,ex} & H_1^{\rm in} &    0 & 0 \\
  H_2^{\rm in,ex} & 0   &  H_2^{\rm in} & 0  \\
  H_3^{\rm in,ex} & 0   &    0 &H_3^{\rm in}  \\
\end{array}\right )\!.
\label{Ham3}
\end{equation}
Here we suppose:
\begin{itemize}
\item [(i)] there is no direct coupling between different
internal channels ${\cal H}^{\rm in}_{i}$ and ${\cal H}^{\rm in}_{j}$
for $i \ne j$;
\item [(ii)] the channel coupling operators do
not involve the third (free) particle:
\begin{equation}
H^{\rm ex,in}_{i}=h^{\rm ex,in}_{jk}\otimes
\mathbb I_{i}.
\label{Hexin}
\end{equation}
\end{itemize}

Writing the four-component Schr\"odinger equation with Hamiltonian
(\ref{Ham3}):
\begin{equation}
H_3\Psi_3=E\Psi_3,
\end{equation}
and excluding three internal channels from it (it is simple due to the supposed
absence of direct coupling between different internal channels), one obtains an
effective Schr\"odinger equation for the external three-body wavefunction
$\Psi^{\rm ex}_3$:
 \begin{equation}  H_3^{\rm eff}(E)\Psi_3^{\rm ex} =E\Psi_3^{\rm ex}
 \label{Sch3}
 \end{equation}
 with an effective (pseudo)Hamiltonian:
\begin{equation}
H^{\rm eff}_3(E)=H^{\rm ex}_3+\sum_{i}H^{\rm ex,in}_{i}\,
 G^{\rm in}_{i}(E)\,H^{\rm in,ex}_{i},
 \label{heff3}
\end{equation}
where the resolvent of internal Hamiltonian $G^{\rm in}_{i}$ is a convolution
of the two-body internal resolvent $g^{\rm in}_{jk}$ of pair ($jk$) and the
free motion resolvent for the third particle ($i$):
   \begin{equation}
  G^{\rm in}_{i}=(E-H^{\rm in}_{i})^{-1}=\frac{1}{2\pi\rm i}
 \int\limits_{-\infty}^{\infty} g^{\rm in}_{jk}(z)\,g^0_{i}(E-z)
 {\rm d}z
   =g^{\rm in}_{jk}(E-t_{i})\otimes \mathbb I_{i}.
 \label{Gin}
\end{equation}

Thus the effective Hamiltonian in external three-body channel takes the form:
\begin{equation}
H^{\rm eff}=T+\sum_{jk}\{v^{\rm ex}_{jk}+
w_{jk}(E-t_{i})\},
 \label{heff32}
\end{equation}
i.e., the total effective interaction in external channel of the three-body
system is a sum of the two-body external potentials $v^{\rm ex}_{jk}$ and the
two-body  effective interactions with replacement of pair energy
$\varepsilon_{i}$ with difference between the total three-body energy and
operator for the relative-motion kinetic energy of third particle: 
$\varepsilon_i \to E-t_{i}$.

Just this recipe for inclusion of pair energy-dependent interactions in
three-body problem is widely used in Faddeev calculations. This recipe has been
rigorously proved in  works of Merkuriev et al. for two-channel model without
continuous spectrum in internal channel~\cite{4} and, in particular, for the
boundary condition model~\cite{3}. We see, however, that  this result is a
direct consequence of the above two assumptions and by no means is related to
usage of any specific interaction model\footnote{%
In the literature, however, there were also discussions of the 
alternative variants for
embedding energy-dependent pairwise force into three-body
system~\cite{Motov,Schmid}. These schemes suppose that the effective total
energy $\varepsilon_{12}$ of the two-body subsystem in the three-body system is
obtained from the total three-body energy $E$ in the following way:
$\varepsilon_{12}=E-t_{12}-\langle v_{13}\rangle -\langle v_{23}\rangle $, 
where $v_{ij}$ is the two-body
interaction between particles $i$ and $j$ and an averaging is supposed with the exact
$3N$ wavefunction.}.

The resulted form of the effective three-body Hamiltonian (\ref{heff32}) is suitable for the
Faddeev reduction.  However, it should be emphasized that each term
$W_{\beta\gamma}$ in the effective Hamiltonian (\ref{heff32}) includes a
dependence on the kinetic energy of the third particle, i.e., the each term
$W_{\beta\gamma}$ is, generally speaking,  a three-body force.
 In spite of three-body character of such
effective potentials, the corresponding Faddeev equations have the Fredholm
property and are equivalent to four-channel  Schr\"odinger equation  (it has
been proved for model with discrete internal spectrum~\cite{4}).

\subsection{A new three-body force in the three-body system with
external and internal channels}

In each internal channel one can introduce a new interaction between third
particle and the pair as a whole. This leads to replacement of the operator for
kinetic energy of the third particle $t_{i}$ by some (single-particle) Hamiltonian
$h_i$:
\begin{equation}
t_{i} \Rightarrow h_{i} = t_{i} + v_{i}
\label{halpha}
\end{equation}
in Eq.~(\ref{hina}) for $H^{\rm in}_{i}$, viz.:
 \begin{equation}
 H^{\rm in}_{i}= h^{\rm in}_{jk}\otimes
 \mathbb I_{i}+\mathbb I_{jk}\otimes
 h_{i}.
 \label{hinav}
 \end{equation}
Physical meaning of such interactions will be discussed below and here we treat
only the formal aspects of their introduction. As the internal Hamiltonian
(\ref{hinav}) is still a direct sum of the two-body internal Hamiltonian and
the Hamiltonian corresponding to relative motion of the third nucleon,
then its resolvent can be expressed  as a
convolution of two subresolvents:
   \begin{equation}
   G^{\rm in}_{i}=(E-H^{\rm in}_{i})^{-1}=\frac{1}{2\pi i}
 \int\limits_{-\infty}^{\infty} g^{\rm in}_{jk}(z)\,g_{i}(E-z)
 dz,
 \label{Ginv}
\end{equation}
where $g_{i}(\varepsilon)=(\varepsilon-h_{i})^{-1}.$
Now, of course, the effective interaction in external channel  is not reduced to
sum of pairwise effective interactions with replacement $\varepsilon_i \to
E-t_{i}$. Nevertheless, this interaction includes three terms
$W_{jk}$ and is still suitable for Faddeev reduction. But now  there are
no pure pairwise forces (except $v^{\rm ex}_{jk}$) in the effective
Hamiltonian for the external three-body channel.

Moreover, if even the external interaction $v_{i}$ is disregarded at all, each term $w_{jk}$ in
the effective Hamiltonian (\ref{heff32}) includes a dependence on the kinetic
energy of the third particle, i.e., can be considered, generally speaking, as a
three-body force. This dependence on the third particle momentum reduces the
strength of the effective interaction between two other particles due to a
specific energy dependence of the coupling constants (see below). Therefore,
one can say that there are no pure two-body forces in the three-body
system in such an approach, with the exception of that part of interaction
which is included in $v^{\rm ex}$  (for $NN$ system it is just the peripheral part of
meson exchange).

\section{Dressed-bag model for $NN$ forces}

Here, we give a brief description of the two-component DBM for
the $NN$ interaction. The detailed description has been presented in our
previous papers~\cite{KuJPG,KuInt}. (The effective-field theory description for
the dybarion model of nuclear force has been also developed
recently~\cite{Shikh}.)
The main assumptions of the DBM are following:
\begin{itemize}
 \item[(i)] interacting nucleons form at small and intermediate distances
 ($r_{NN} \sim 1$~fm)
 a compound state --  the
dibaryon or  six-quark bag dressed with $\pi$, $\sigma$, and
$\rho$-fields;
 \item[(ii)] the coupling of the external $NN$ channel with
this state gives the basic attractive force between nucleons at intermediate
and small distances, the $\sigma$-dressed bag giving the main contribution.
\end{itemize}

Thus,  nucleon-nucleon system can be found in two different phase states
(channels): the $NN$ phase and the dressed $6q$ bag phase.
 In the
$NN$ (external) channel the system is described as two nucleons interacting via
OBE; in the internal $6q+\sigma$ channel the system is treated
as a $6q$ bag surrounded by the strong scalar--isoscalar $\sigma$-field (a
``dressed'' bag)\footnote{Full description of the $NN$ interaction at energies
$E\sim 1$~GeV still requires other fields in the bag, such as $2\pi$, $\rho$, and
$\omega$.}. The external two-nucleon Hamiltonian includes the peripheral part
of one-pion- and two-pion-exchange (OPE and TPE, respectively) interaction and Coulomb interaction:
\[h^{\rm ex}=t+\{v^{\rm OPE}+v^{\rm TPE}\}_{(\mbox{\small with soft cutoff})}
 + v^{\rm Coul}.\]

In the simplest version of DBM we used a pole approximation for
the dressed-bag (internal) resolvent $g^{\rm in}$:
 \begin{equation}
g^{\rm in}(E)=\sum_{\alpha}\int\frac{|\alpha,{\bf k}\rangle
\langle \alpha,{\bf k}| d^3k}{E-E_{\alpha }({\bf k})},
\label{resn}
\end{equation}
where $|\alpha\rangle$ is the $6q$ part of the wavefunction for the dressed bag
and $|{\bf k}\rangle$ represents the plane wave of the $\sigma$-meson
propagation.
 Here, $E_{\alpha }({\bf k})$ is the total energy of the dressed bag:
 \begin{equation}
 E_{\alpha }({\bf k}) = m_{\alpha}+\varepsilon_{\sigma}(k),
 \label{eak}
\end{equation}
where
  \begin{equation}
 \varepsilon_{\sigma}(k) = k^2/2m_{\alpha}+\omega_{\sigma}(k) \simeq
 m_{\sigma}+k^2/2\bar{m}_{\sigma},\qquad \bar{m}_{\sigma}=
 \frac{m_{\sigma}m_{\alpha}}{m_{\sigma}+m_{\alpha}},
 \label{epssig}
\end{equation}
 $\omega_{\sigma}(k)=\sqrt{m_{\sigma}^2+k^2}$ is relativistic energy
 of $\sigma$-meson, $m_{\sigma}$ and $m_{\alpha}$ are masses of $\sigma$-meson
 and $6q$ bag, respectively.

 The effective interaction $w(E)$ resulted from the coupling of
the external $NN$ channel to the intermediate dressed-bag state is
illustrated by the graph in Fig.~1.

\begin{figure}[h]
\begin{center}
\epsfig{file=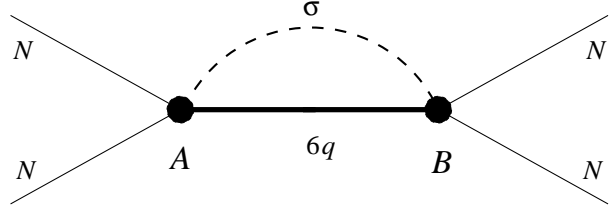,width=0.5\textwidth}
\end{center}
 \setcaptionmargin{0mm} \onelinecaptionsfalse
\captionstyle{flushleft}
\caption
{ Effective $NN$ interaction induced due to the production of an
intermediate dressed bag.}
\label{fig1}
\end{figure}

To derive the effective interaction $w$  for $NN$ channel in such
an approximation, the knowledge of
full internal  Hamiltonian
$h^{\rm in}$ of the dressed bag, as well as the full transition
operator $h^{\rm in,ex}$, is not necessary.  We need only to know how
the transition operator acts
 on those dressed-bag states, which are included into the resolvent (\ref{resn}):
  $h^{\rm in,ex}|\alpha,{\bf k}\rangle$.  The calculation of this quantity
 within a microscopical quark--meson model results in a sum of factorized
 terms~\cite{KuInt}:
  \begin{equation}
  h^{\rm ex,in}|\alpha^{JM},{\bf k}\rangle =
  \sum_{L}|\varphi^{JM}_L\rangle B^J_L({\bf k}),
 \label{vert}
\end{equation}
where $\varphi^{JM}_L \in {\cal H}^{\rm ex}$ is the $NN$ transition
form factor and $ B^J_L({\bf k})$ is the vertex function dependent on the
$\sigma$-meson momentum.

 Here we should elucidate our notation in respect to the quantum numbers of angular
 momenta.
In general, the $6q$-state index $\alpha$ includes all the quantum numbers of the
dressed bag, i.e., $  \alpha\equiv \{J,M,S,T,L_b,L_{\sigma}\}$, where $L_b$,
$S$, $T$, $J$, and $M$ are the orbital angular momentum of the $6q$ bag, its
spin, isospin, total angular momentum, and its projection on the $z$ axis,
respectively, and $L_{\sigma}$ is the orbital angular momentum of the $\sigma$
meson. However, in the present version of the DBM, the $s$-wave state of the
$6q$ bag with the $s^6$ configuration only is taken into account, so that $L_b=0$,
$J=S$, and thus the isospin of the bag is uniquely determined by its spin. The
states of the dressed bag with $L_{\sigma}\ne 0 $ should lie higher than those
with $L_{\sigma} =0$. For this reason, the former states are not included in
the present version of the model. Therefore, the state index $\alpha$ is
specified here by the total angular momentum of the bag $J$ and (if necessary)
by its $z$ projection $M$: $ \alpha \Rightarrow  \{J(M)\}$.

Thus, the effective interaction in the $NN$ channel $w(E)\equiv h^{\rm
ex,in}g^{\rm in}(E) h^{\rm in,ex}$ can be written as a sum of separable terms
in each partial wave:
 \begin{equation}
w(E)=\sum_{J ,L,L'}w^{J}_{LL'}({\bf r},{\bf r}^{\prime},E),
 \label{nqn2}
 \end{equation}
 with
\begin{equation}
w^{J}_{LL^{\prime}}({\bf r},{\bf r'})= \sum_M \varphi^{JM}_{L}({\bf
r})\,\lambda^{J}_{LL^{\prime}}(E)\, {\varphi^{JM}_{L^{\prime}}}^*({\bf r'}).
 \label{zlz}
 \end{equation}
The energy-dependent coupling constants
$\lambda^{J}_{LL^{\prime}}(E)$ appearing in Eq. (\ref{zlz}) are
directly calculated from the loop diagram shown in Fig.~1; i.e.,
they are expressed in terms of the loop integral of the product of
two transition vertices $B$ and the convolution of two propagators
for the meson and quark bag with respect to the momentum $k$:
 \begin{equation}
\lambda^{J}_{LL^{\prime}}(E)=\int\limits^{\infty}_0\, d{\bf k}
\frac{B_L^J({\bf k})\,{B_{L^{\prime}}^J}^*({\bf k})}
{E-E_{\alpha}(k)}.
 \label{lamb}
 \end{equation}

The vertex form factors $B^{J}_{L}({\bf k})$ and the potential form factors
$\varphi^{JM}_{L} \in {\cal H}^{\rm ex}$
 have been calculated in the microscopic quark--meson
model~\cite{KuJPG,KuInt}.

 When the $NN$-channel wavefunction $\Psi^{\rm in}$ is obtained by solving the
Schr\"odinger equation with the effective Hamiltonian $h^{\rm
eff}(E)$, the internal ($6qN$) component of the wavefunction is
found  from Eq.~(\ref{psiin}):
 \begin{equation}
 \Psi^{\rm in}_{JM}(E)=|\alpha^{JM}\rangle
 \underbrace{\sum_{L} \frac{B^J_L({\bf k})}{E-E_{\alpha}({\bf k})}
\langle\varphi^{JM}_L|\Psi^{\rm ex}(E)\rangle},
 \label{psi6q}
 \end{equation}
where the underlined part
 can be interpreted as the mesonic part of the dressed-bag wavefunction.

The weight of the internal dressed-bag component of bound-state wavefunction
(with given value $J$)  is proportional to the norm of $\Psi^{\rm in}_{JM}$:
\begin{equation}
\|\Psi^{\rm in}_{JM}\|^2=\|\alpha^{JM}\|^2
\sum_{LL'}
\langle\varphi^{JM}_L|\Psi^{\rm ex}\rangle
\langle \Psi^{\rm ex}|\varphi^{JM}_{L'}\rangle
\underbrace{\int\frac{B^J_L({\bf k}) {B^J_{L'}}^*({\bf k})}{(E-E_{\alpha}({\bf
  k}))^2} d{\bf k}}_{I_{LL'}^J}
 \label{normin}
 \end{equation}

As one can see from the comparison between Eqs. (\ref{lamb}) and
(\ref{normin}), the integral $I^{J}_{LL'}$ in Eq.(\ref{normin}) is
equal to the energy derivative (with opposite sign) of the
coupling constant $\lambda^J_{LL'}(E)$:
\[ I^{J}_{LL'}=-\frac{{\rm d}\lambda^{J}_{LL'}(E)}{{\rm d}E}, \]
and thus we get an interesting result:
\[ \|\Psi^{\rm in}\|^2 \sim -\frac{{\rm d}\lambda(E)}{{\rm d}E}, \]
i.e., the weight of internal $6qN$ state is proportional to the energy
derivative of the coupling constant of effective $NN$ interaction. In other
words, the stronger the energy dependence of the interaction in $NN$ channel,
the
larger the weight of channel corresponding to non-nucleonic degrees of freedom.
This result is in full agreement with well-known hypothesis:  energy dependence
of interaction is a sequence of underlying inner structure of interacting
particles.

The total wavefunction of the bound state $\Psi$ must be normalized to unity.
Assuming that the external (nucleonic) part of the wavefunction $\Psi^{\rm ex}$
found from the effective Schr\"odinger equation has the standard normalization
 $\|\Psi^{\rm ex}\|=1 $,
 one obtains that the weight of internal, i.e., the dressed-bag component is equal to:
\begin{equation}
P_{\rm in}=\frac{\|\Psi^{\rm in}\|^2}{1+\|\Psi^{\rm in}\|^2}.
\label{P_in2}
\end{equation}

Thus, the $NN$ interaction in DBM approach is a sum of peripheral terms
($v^{\rm OPE}$ and $v^{\rm TPE}$) representing OPE and TPE with soft cutoff
parameter $\Lambda_{\pi NN}$ and an effective interaction $w(E)$ (see Eqs.
({\ref{nqn2}), (\ref{zlz})), which is expressed (in a single-pole
approximation) as an one-term separable potential with the energy-dependent
coupling constants (\ref{lamb}). The potential form factors
$\varphi^{JM}_L(\bf r)$ are taken as the conventional harmonic oscillator 
wavefunctions $|2S\rangle$ and $|2D\rangle$\footnote{It was first
suggested~\cite{Liyaf86} long ago and then confirmed in detailed $6q$
microscopic calculations~\cite{Kus91} that the $6q$ wavefunction in $NN$
channel corresponds just to $2\hbar\Omega$ excited $6q$-bag  components
$|s^4p^2[42]LST\rangle$, while the ground state $|s^6[6]\rangle$  describes
the wavefunction in the bag channel.}. 
Therefore, the total $NN$ potential in DBM model can be represented as:
\begin{equation}
v_{NN}= v^{\rm OPE} + v^{\rm TPE} + v^{\rm Coul}+ w(E) + \lambda \Gamma ,
\end{equation}
where $\Gamma =|\varphi_0\rangle\langle\varphi_0|$ is the projector onto
$|0S\rangle$
harmonic oscillator function and constant $\lambda$ should be taken 
sufficiently large.

\begin{table}[h]
 \setcaptionmargin{0mm} \onelinecaptionsfalse
\captionstyle{flushleft}
\caption {Deuteron properties in DBM and other current $NN$ models}
\medskip

\begin{tabular}{|c|c|c|c|c|c|c|c|}\hline 
Model    & $E_d$, MeV& $P_D$, \%& $r_m$, fm & $Q_d$, fm$^2$ &
$\mu_d$, n.m. & $A_S$,fm$^{-\!1/2}$& $\eta (D/S)$ \\ 
\hline
 RSC      &
2.22461  & 6.47 & 1.957 & 0.2796 & 0.8429 & 0.8776 & 0.0262  \\
 Moscow 99   &2.22452  & 5.52 & 1.966 & 0.2722 & 0.8483 & 0.8844 & 0.0255 \\
 Bonn 2001 &2.224575 & 4.85 & 1.966 & 0.270  & 0.8521 & 0.8846 & 0.0256 \\
 DBM (I) $P_{6q}=3.66$\%  &2.22454 & 5.22&1.9715&0.2754& 0.8548&0.8864& 0.02588\\
 
 DBM (II) $P_{6q}=2.5$\% &2.22459 & 5.31&1.970&0.2768& 0.8538&0.8866& 0.0263\\
Experiment&2.224575 &   --     & 1.971 & 0.2859 & 0.8574 & 0.8846 &
0.0263$^{\rm a)}$ \\
 \hline
\end{tabular}

 {\footnotesize \begin{flushleft} $^{\rm a)}$An average value of the asymptotic mixing
parameter $\eta$ over the results of a few most accurate
experiments is presented here
(see \cite{Hor,Berth,Must,Rodn}).\end{flushleft}}
\label{table1}
\end{table}

The model described above gives a very good description for singlet
$^1S_0$,  the triplet 
$^3S_1- {}^3D_1$ phase shifts, and mixing parameter $\varepsilon_1$ in the energy region from zero up to 1 GeV~\cite{KuInt}. The
deuteron observables obtained in this model without any additional or free parameter
are presented in Table~1 in comparison with some other $NN$ models and
experimental values. The quality of agreement with experimental data for the
$NN$ phase shifts and deuteron static properties found with the presented force
model, in general, is higher than those for the modern $NN$ potential model
such as Bonn, Argonne, etc., especially for the asymptotic mixing parameter
$\eta$ and the deuteron quadrupole moment. The weight of the internal
(dressed-bag) component in the deuteron is varied from 2.5 to 3.6\% in
different versions of the model~\cite{KuJPG,KuInt}.

\section{Three-nucleon system with DBM interaction}

 For description of the three-body
 system with the DBM interaction
 the momentum representation is more appropriate. We will
 employ the same notation for functions both in the coordinate and momentum
 representations. The following notations for coordinates and momenta are
 employed:
 ${\bf r}_i$(${\bf p}_i$) is relative coordinate (momentum) of pair ($jk$), while
  ${\b \rho}_i$
 (${\bf q}_i$) is Jacobi coordinate (momentum) of {\em i}th particle relatively
  to the center of
 mass for the pair ($jk$), and  $\bf k$ is usually a momentum of $\sigma$ meson.

\subsection{Effective interaction due to pairwise $NN$ forces}

One obtains an effective
Hamiltonian for the external $3N$ channel  according to a general recipe for transition from two- to three-particle
system:
 \begin{equation}
H^{\rm eff}=T+\sum_{i}\{v^{\rm ex}_{i}+ W_i(E)\},
\label{heffdbm}
\end{equation}
where each of three effective potentials takes the form:
 \begin{equation}
 W_i(E)=\delta ({\bf q}_i-{\bf q}'_i)w_i(E-q_i^2/2\bar{m}),
\label{weffdbm}
\end{equation}
and $\bar{m}=m_Nm_{\alpha}/(m_N+m_{\alpha})$ is a reduced mass of
 nucleon and $6q$ bag.
 In the pole approximation, this effective interaction reduces to a sum of
 two-body separable potentials with the coupling  constants depending on
 the total three-body energy $E$ and the third-particle momentum $q_i$:
 \begin{equation}
  W_i({\bf p}_i, {\bf p}'_i, {\bf q}_i,{\bf q}'_i;E)=
 \delta ({\bf q}_i-{\bf q}'_i) \sum_{J_iM_i,L_i,L'_i}
 \varphi^{J_iM_i}_{L_i}({\bf p}_i)\,\lambda^{J_i}_{L_iL_i'}
 \left( E-\frac{q_i^2}{2\bar{m}}\right) \,\varphi^{J_iM_i}_{L'_i}({\bf p}'_i).
\label{kernel0}
\end{equation}

When using such an effective interaction, one must also include an additional
3BF due to the meson-exchange interaction between the dressed bag
and the third nucleon (see the next subsection).  The pattern of different
interactions arising in the $3N$ system in such a way is illustrated in Fig.~2.

\begin{figure}[h]
\begin{center}
{\epsfig{file=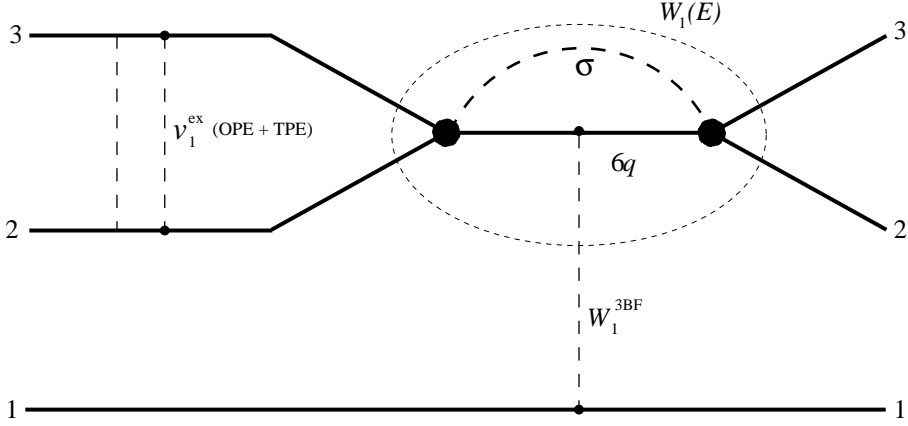,width=0.75\textwidth}}
\end{center}
 \setcaptionmargin{0mm} \onelinecaptionsfalse
\captionstyle{flushleft}
\caption{Different interactions in the $3N$ system for one of
three possible combinations ($1+23$) of three nucleons: the peripheral
two-nucleon interaction $v^{\rm ex}_1$ is
due to OPE $+$ TPE, the effective two-body interaction $W_1(E)$ is induced
by the production of dressed $6q$ bag and meson-exchange 3BF $W^{\rm 3BF}_1$.}
\label{fig2}
\end{figure}

In the single-pole approximation, the internal (dressed-bag) components of the
 total wavefunction
 are expressed in terms of the nucleonic component $\Psi^{\rm
ex}({\bf p}_i, {\bf q}_i)$ as
 \begin{equation}
 \Psi^{\rm in}_i({\bf k,q}_i;E) = \sum_{J_i,M_i,L_i} |\alpha^{J_iM_i}\rangle
\frac{B_{L_i}^{J_i}({\bf k})\,\chi_{L_i}^{J_iM_i}({\bf q}_i)}
{E-E_{\alpha}-\frac{q_i^2}{2m}},
 \label{psi6qNIJ}
 \end{equation}
where $\chi_{L_i}^{J_iM_i}({\bf q}_i)$ are the overlap integrals of the external
$3N$ component and the potential form factors
$\varphi^{J_iM_i}_{L_i}$:
 \begin{equation}
 \chi^{J_iM_i}_{L_i}({\bf q}_i) =
 \int \varphi^{J_iM_i}_{L_i}({\bf p}_i) \,
 \Psi^{\rm ex}({\bf p}_i, {\bf q}_i)\, d{\bf k}_i.
\label{ovJL}
 \end{equation}
  These overlap functions depend on the momentum (or coordinate), spin, and
isospin of the third nucleon. For brevity, the spin--isospin parts
of the overlap functions and corresponding quantum numbers are
omitted unless they are needed. In
Eqs. (\ref{kernel0})--(\ref{ovJL}) and below, we keep the index $i$
in the quantum numbers $L_i$ and $J_i$ in order to distinguish the
orbital and total angular momenta attributed to the $2N$ form
factors from the respective angular momenta $J$ and $L$ of the
whole $3N$ system.

It should be noted that the angular part of the function
$\chi_{L_i}^{J_iM_i}({\bf q}_i)$ in Eq. (\ref{ovJL}) is not equal
to $Y_{L_iM_{L_i}}(\hat{q})$. This part  includes also other
angular orbital momenta due to coupling of the angular momenta and
spins  of the dressed bag and those for the third nucleon.  In the
next section we consider the spin--angular and isospin parts of the
overlaps functions $\chi_{L_i}^{J_iM_i}({\bf q}_i)$ in more
detail.

The norm of each $6qN$ component for the $3N$ bound state
is determined by sum of the integrals:
 \begin{equation}
  \|\Psi^{\rm in}_i\|^2 = \sum_{J_iM_i}\|\alpha^{J_iM_i}\| \sum_{L_iL'_i}
 \int \chi_{L_i}^{J_iM_i}({\bf q}_i)
 \left\{\int \frac{B_{L_i}^{J_i}({\bf k}) B_{L'_i}^{J_i}({\bf k})}
{\left( E-E_{\alpha}-\frac{q_i^2}{2m}\right)^2}
\,d{\bf k}\right \}\,
   \chi_{L'_i}^{J_iM_i}({\bf q}_i)\,d{\bf q}_i.
 \label{norm6qN}
 \end{equation}
The internal loop integral with respect to ${\bf k}$ in
Eq.~(\ref{norm6qN}) (in braces) can be replaced by the energy
derivative of $\lambda^J_L$:
 \begin{equation}
 \int \frac{B_{L_i}^{J_i}({\bf k}) B_{L'_i}^{J_i}({\bf k})\, }
{\left( E-E_{\alpha}-\frac{q_i^2}{2m}\right)^2}\,d{\bf k}
=-\frac{d}{dE}\lambda^{J_i}_{L_iL'_i}\left({\textstyle E-\frac{q_i^2}{2m}}\right).
 \label{repln}
 \end{equation}
 Thus, the weight of the $6qN$ component in the $3N$ system is determined by the
same energy dependence of the coupling constants $\lambda^J_{LL'}(\varepsilon)$ as
the contribution of the $6q$ component in the $NN$ system but at a shifted
energy.

With using Eq. (\ref{repln}), the norm of $6qN$ component can be
rewritten eventually as
 \begin{equation}
 \|\Psi^{\rm in}_i\|^2 = \sum_{J_iM_i}\|\alpha^{J_i}\| \sum_{L_iL'_i}
 \int \chi_{L_i}^{J_iM_i}({\bf q}_i)
 \left (-\frac{d}{dE}\lambda^{J_i}_{L_iL'_i}(E\!-\!q_i^2/2m)\right )
  \chi_{L'_i}^{J_iM_i}({\bf q}_i)\,d{\bf q}_i.
 \label{norm1}
 \end{equation}
Due to explicit presence of the  meson variables in our approach, it is
generally impossible to define the wavefunction describing relative motion  of
the third nucleon $^N\psi (q)$ in the $6qN$ channel. However, by integrating
$\Psi^{\rm in}_{i}({\bf k,q})$ over the meson momentum $\bf k$, one can obtain
an average momentum distribution of the third nucleon in the $6qN$ channel
(i.e., those weighted with the $\sigma$-meson momentum distribution). Based on
Eq. (\ref{repln}), we can attribute the meaning of the third nucleon
wavefunction in the $6qN$ channel to the quantity
 \begin{equation}
 \tilde{\psi}_{L_i}^{J_iM_i}({\bf q}_i) =
  \sqrt{ \left (-\frac{d}{dE}\lambda^{J_i}_{L_iL'_i}(E\!-\!q_i^2/2m) \right )}
  \chi_{L_i}^{J_iM_i}({\bf q}_i).
 \label{npsi}
 \end{equation}
With this ``quasi-wavefunction'', one can calculate the mean value of any operator
depending on the momentum (or coordinate) of the third nucleon. We note that the
derivative $-d\lambda /dE$ is always positive.

\subsection{Three-body forces in the DBM}

\begin{figure}[h]
\begin{center}
\noindent {\epsfig{file=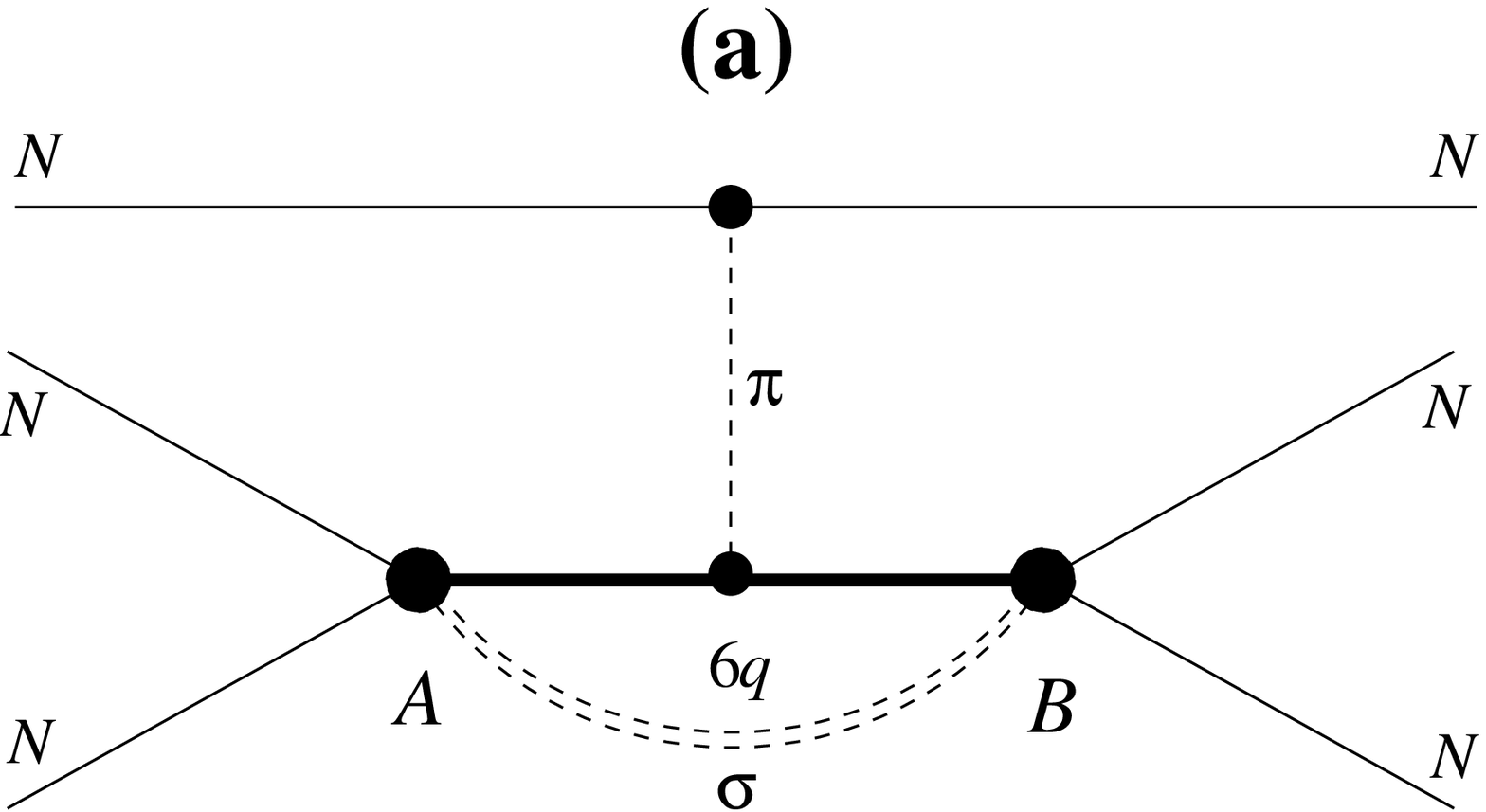,width=0.27\textwidth}}
 \hfill {\epsfig{file=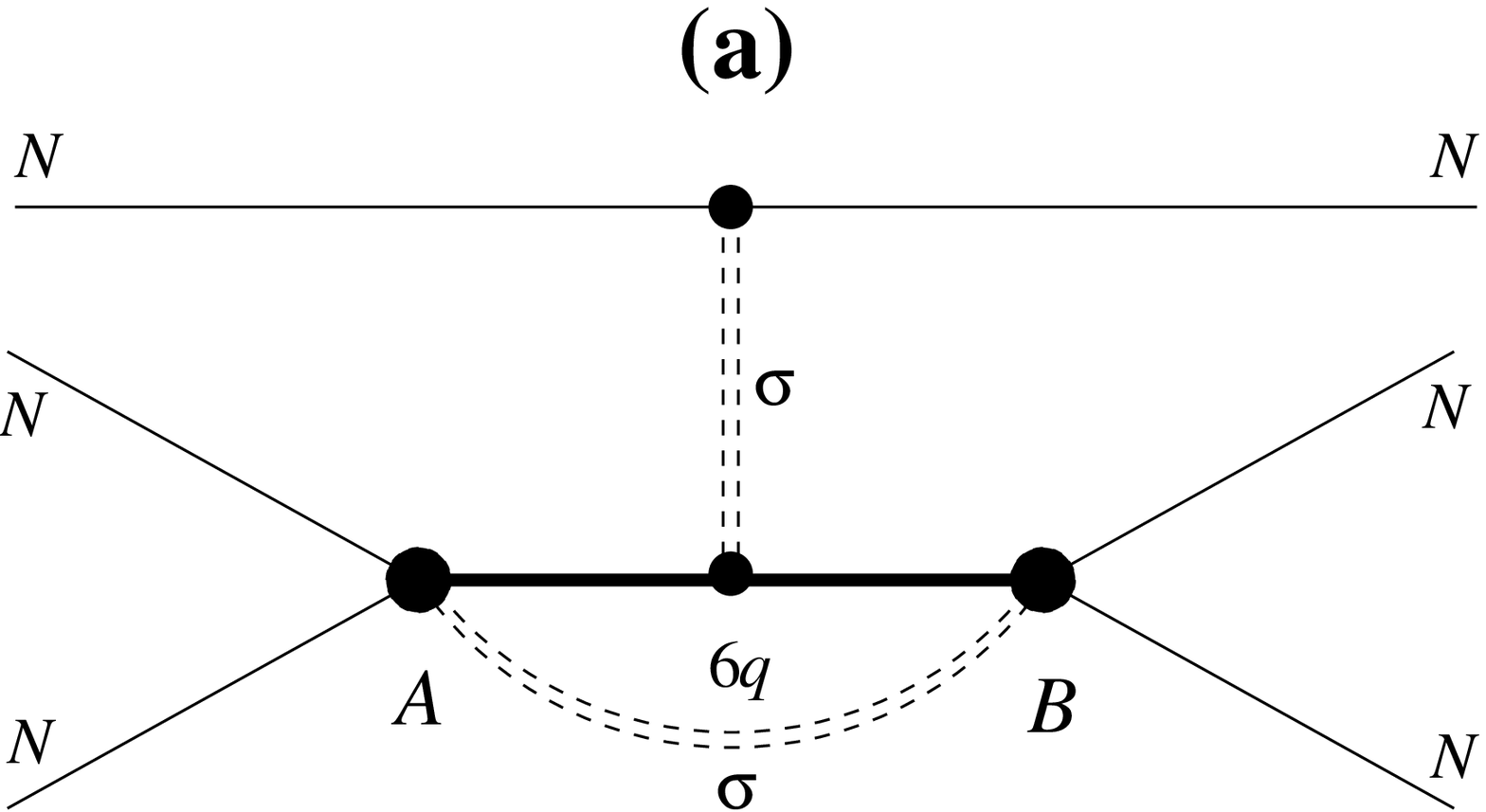,width=0.27\textwidth}}
\hfill {\epsfig{file=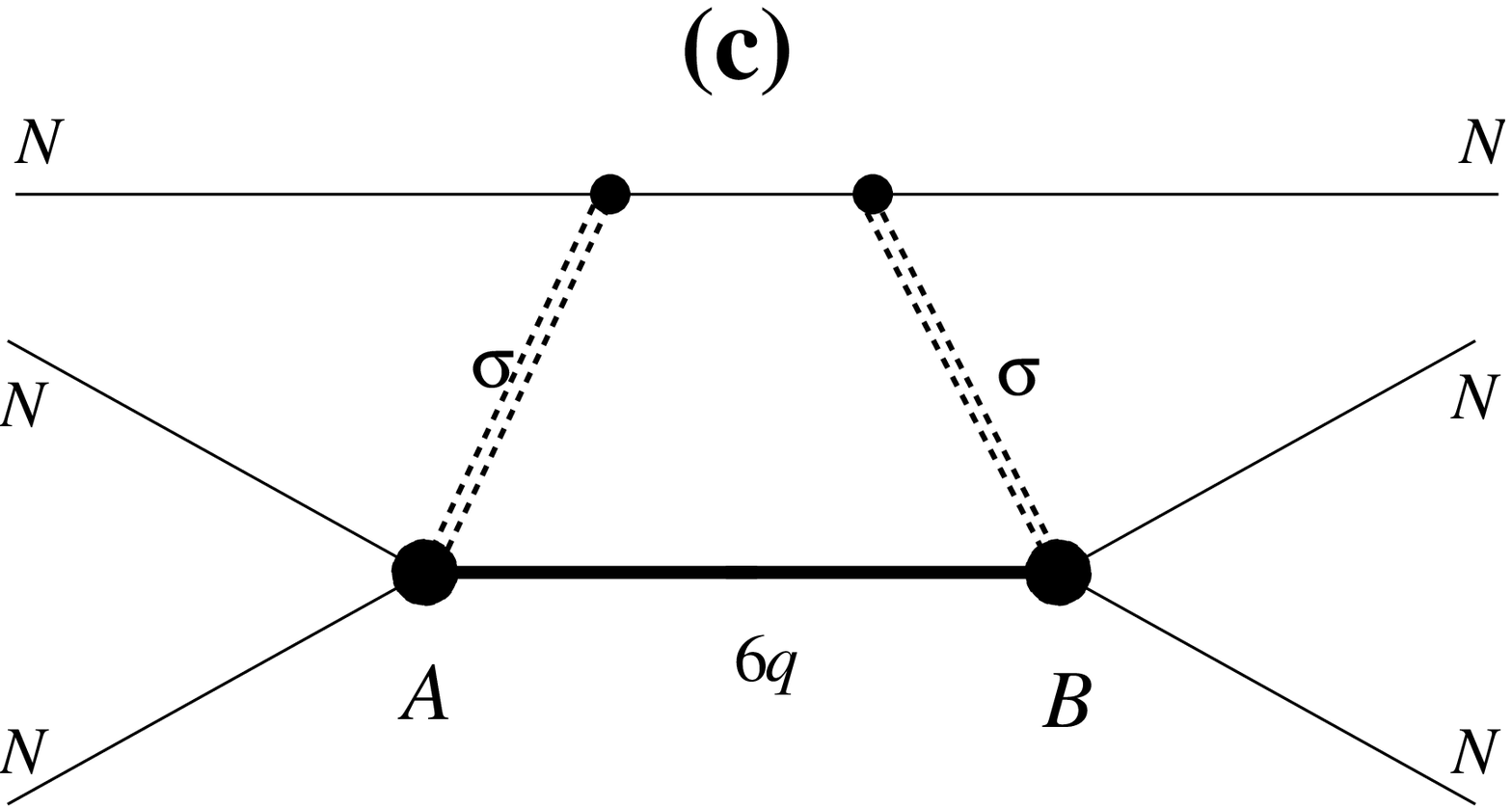,width=0.27\textwidth}}\\
\end{center}
 \setcaptionmargin{0mm} \onelinecaptionsfalse
\captionstyle{flushleft}
\caption{The graphs corresponding to three new types of three-body force.}
\label{fig3}
\end{figure}

 In this study, we employ the effective interaction (\ref{kernel0}) and take
into account the interaction between the dressed bag and the third nucleon
as an additional 3BF. We consider here three types of 3BF: one-meson
exchange ($\pi$ and $\sigma$) between the dressed bag and the third nucleon
(see Figs.~3{\em a}  and 3{\em b}) and the exchange by two $\sigma$-mesons,
where the third-nucleon propagator breaks the $\sigma$-loop of the two-body
force -- $2\sigma$-process (Fig.~3{\em c}).

All these forces can be represented in the effective Hamiltonian for external
$3N$ channel as some integral operators with
factorized kernels:
 \begin{equation}
 W^{\rm 3BF}_{(i)} ({\bf p}_i, {\bf p}'_i, {\bf q}_i, {\bf q}'_i;E)=
 \sum_{JM,J'M',L,L'}\varphi^{JM}_L({\bf p}_i)\, {}^{\rm 3BF}W^{JJ'}_{LL'} 
 ({\bf q}_i,{\bf q}'_i;E)
 \,\varphi^{J'M'}_{L'} ({\bf p}'_i).
 \label{3BF}
\end{equation}
Therefore, matrix elements for 3BF include only the overlap functions, and thus
the contribution of 3BF is proportional to the weight of the internal $6qN$
component in the total $3N$ wavefunction. To our knowledge, the first
calculation of the 3BF contribution  induced by OPE between the $6q$ bag and
the third nucleon was done by Fasano and Lee~\cite{Fasano} in the hybrid QCB
model using perturbation theory. They used the model where the weight of the
$6q$ component in a deuteron was ca. 1.7\%, and thus they obtained a very small
value of --0.041~MeV for the 3BF OPE contribution to the $3N$ binding energy.
Our results for the OPE 3BF agree with the results obtained by Fasano and Lee
(see Table~2 in Section~7), because the OPE contribution to 3BF is proportional
to the weight of the $6q$ component, and  in our case, it should be at least
twice as compared to their calculation. However, we found that a much larger
contribution comes from scalar $\sigma$-meson  exchanges: one-sigma exchange
(OSE) and two-sigma exchange (TSE). We emphasize that, due to (proposed)
restoration of chiral symmetry in our approach, the $\sigma$-meson mass becomes
ca. 400~MeV, and thus the effective radius of the $\sigma$-exchange interaction
is not so small as that in conventional OBE models. Therefore, we cannot use
the perturbation theory anymore to estimate the 3BF contribution and have to do
the full calculation including 3BF in the total three-body Hamiltonian.

\subsubsection{One-meson exchange between the dressed bag and third nucleon}

 For the one-meson exchange (OME) term, the three-body interaction
 $^{\rm 3BF}W^{J_iJ'_i}_{L_iL'_i}$ takes the form:
 \begin{equation}
 {}^{\rm OME}W^{J_iJ_i'}_{L_iL_i'} ({\bf q}_i,{\bf q'}_i;E) =
  \int {d}{\bf k} \frac{B^{J_i}_{L_i} ({\bf k})}
  {E-E_{\alpha}-q_i^2/2m}\,
  V^{\rm OME} ({\bf q}_i,{\bf q}'_i) \,
   \frac{B^{J'_i}_{L'_i} ({\bf k})}
  {E-E_{\alpha}-{q'}_i^2/2m}.
\label{WOME}
\end{equation}
Therefore, the matrix element for OME can be expressed in terms
 of the internal ``bag'' components $\Psi^{\rm in}_{i}$:
 \begin{equation}
 \langle \Psi^{\rm ex} | {\rm OME}| \Psi^{\rm ex} \rangle =
 3\langle\Psi^{\rm in}_i
  |V^{\rm OME}|\Psi^{\rm in}_i\rangle .
 \label{OME}
 \end{equation}
 The integral over $\sigma$-meson momentum $\bf k$ (\ref{WOME}) can be shown to
 be reduced to a difference of the values for constant $\lambda(E-q^2/2m)$, so
 that the vertex functions $B(k)$ can be excluded from formulas for OME 3BF
 matrix elements.
 The details of calculations for such matrix elements are given in the Appendix.

\subsubsection{$2\sigma$-process}

\begin{figure}[h]
\centerline{\epsfig{file=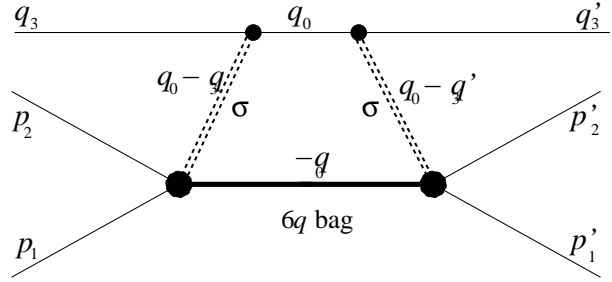,width=0.5\textwidth}}
 \setcaptionmargin{0mm} \onelinecaptionsfalse
\captionstyle{flushleft}
\caption{The graph illustrating three-body scalar force due to two-sigma 
exchange ($2\sigma$-process).}
\label{fig4}
\end{figure}

The $2\sigma$-process (TSE) shown in Fig.~4 also contributes significantly to 3BF.
This $3N$ interaction seems less important than the OSE force, because this
interaction imposes a specific kinematic restriction on the $3N$
configuration\footnote{ It follows from the intuitive picture of this
interaction that this force can be large only if the momentum of the third
nucleon is almost opposite to the momentum of the emitted $\sigma$ meson. Thus,
a specific $3N$ kinematic configuration is required when two nucleons approach
close to each other to form a bag, while the third nucleon has a specific space
localization and momentum.}.

The operator of the TSE interaction includes explicitly the vertex
functions for the transitions ($NN \Longleftrightarrow
6q\!+\!\sigma$)  so that these vertices cannot be excluded
similarly to the case of OME. Therefore, we have to choose some
form for these functions. It is naturally to require that these
vertices should be the same as those assumed in two-body DBM;
i.e., they can be normalized by means of the coupling constants
$\lambda(E)$, which, in turn, are chosen  in the two-nucleon
sector to accurately describe $NN$ phase shifts and deuteron
properties (see below Eq.~(\ref{adjust}) for vertex normalization). We
use the Gaussian form factor for these vertices:

\begin{equation}
B^J_L({\bf k})=B_0^{JL}\frac{{e}^{-b^2k^2}}{\sqrt{2\omega_{\sigma}(k)}},
 \label{Bk}
 \end{equation}
where $\bf k$ is the meson momentum and the parameter $b$ is taken from
the microscopical quark model~\cite{KuInt}:
\begin{equation}
b^2=\frac{5}{24}b_0^2,\qquad b_0=0.5\mbox{ fm}.
 \label{b0}
 \end{equation}
 Then, the vertex constants $B_0^{JL}$ should  be found from the equation:
 \begin{equation}
 \frac{1}{(2\pi)^3}\int d{\bf k}\frac{B_0^{JL}B_0^{JL'}{e}^{-2b^2k^2}}
 {(E\!-\!m_{\alpha}\!-\!\varepsilon_{\sigma}(k))\cdot 2\omega_{\sigma}(k)}
 = \lambda^{J}_{LL'}(E),
 \label{adjust}
 \end{equation}
where $\lambda^{J}_{LL'}(E)$ are the coupling constants employed in the
construction of the  DBM in the  $2N$ sector and are fixed by $NN$ phase
shifts.
For the $\sigma NN$ vertices, we take also the Gaussian form factor:
$g_{\sigma NN}\,{e}^{-\alpha^2k^2}$,  with $ \alpha^2=b_0^2/6$.

Then, the
box diagram in Fig.~4 can be expressed in terms of the integral over
 the momentum ${\bf q}_0$ of the third nucleon in the intermediate
state:
\[
 {}^{\rm TSE}W^{JJ'}_{LL'}({\bf q,q'};E) = \delta_{JJ'}
  g_{\sigma NN}^2\,B_0^{JL}\,B_0^{JL'}\times
\]
 \begin{equation}
  \times\frac{1}{(2\pi)^3}\int d{\bf q}_0
  \frac{{e}^{-(\alpha^2\!+\!b^2)({\bf q}_0\!-\! {\bf q})^2}}
  {m_{\sigma}^2+({\bf q}_0\!-\! {\bf q})^2}\,
  \frac{1}{E\!-\!m_{\alpha}\!-\!q_0^2/2m}\,
  \frac{{e}^{-(\alpha^2\!+\!b^2)({\bf q}_0\!-\!{\bf q'})^2}}
  {m_{\sigma}^2+({\bf q}_0\!-\! {\bf q'})^2}.
\label{WTSE}
\end{equation}
Thus, the matrix element for the total contribution of TSE takes eventually the
form
\begin{equation}
 \langle {\rm TSE}\rangle =3\,\sum_{J_iM_i,L_i,L'_i}
 \int \chi^{J_iM_i}_{L_i}({\bf q})\, {}^{\rm TSE}W^{J_iJ_i}_{L_iL'_i}({\bf q,q'};E)
 \chi^{J_iM_i}_{L_i'}({\bf q'})\,
 d{\bf q}\,d{\bf q'}.
 \label{TSE}
 \end{equation}
After the partial wave decomposition, these six-dimensional integrals can be
reduced to two-fold integrals, which are computed numerically by means
of the appropriate Gaussian quadratures. 

We should emphasize here that both two-nucleon force induced by the DBM and
two parts of 3BF contribution in our approach, i.e., OSE and TSE, are all
taken with unified coupling constants and unified form factors in Eqs.~(37),
(39)--(41), in a sharp contrast to the traditional meson-exchange models (see
also the section~8).

\section{Variational calculations of $3N$ system with DBM interaction}

 The effective Schr\"odinger equation for the external $3N$ part of the total
 wavefunction
$ H^{\rm tot}(E)\Psi^{\rm ex}(E) = E\Psi^{\rm ex}(E) $
 with Hamiltonian
 \begin{equation}
 H^{\rm tot}(E)= T+\sum_{i=1}^3 \{v^{\rm ex}_i+W_i(E)+W^{\rm 3BF}_i(E) \}
 \label{htot}
 \end{equation}
has been solved by variational method using antisymmetrized Gaussian
 basis~\cite{Tur1}. Because of the explicit energy dependence of the three-body total Hamiltonian,
 we used an iterational procedure  in respect to
 the total energy $E$ for solving this equation:
\[ H^{\rm tot}(E^{(n-1)})\Psi^{{\rm ex}(n)}=E^{(n)}\Psi^{{\rm ex}(n)}.
 \]
 Such iterations can be shown to converge, if the energy derivative of
 effective interaction is negative (for our case, this condition is valid
 always). For our calculations,
  5--7 iterations provide usually the accuracy of 5 decimal digits for $3N$
  binding energy.

{\bf Construction of a $3N$ variational basis.} 
Here, we give the form of the basis functions used in this work and the corresponding
notation for the  quantum numbers. The wavefunction of the external
$3N$ channel,
$\Psi^{\rm ex}$, can be written in the  antisymmetrized basis as a sum of the three
terms:
\begin{equation}
 \Psi^{\rm ex}=  \Psi^{(1)}_{\rm ex}+ \Psi^{(2)}_{\rm ex} + \Psi^{(3)}_{\rm ex},
 \label{psi3n}
 \end{equation}
where the label ($i$) enumerates one of three possible set of the
Jacobi coordinates $({\bf r}_i,{\b  \rho}_i)$. Every term in
Eq.~(\ref{psi3n}) takes the form
\begin{equation}
  \Psi^{(i)}_{\rm ex}=\sum_{\gamma}\sum_{n}
  C^{\gamma}_n \Phi^{(i)}_{\gamma n}.
 \label{psi3ni}
 \end{equation}
The basis functions $\Phi^{(i)}_{\gamma n}$ are constructed from Gaussian
functions and corresponding spin-angular and isospin factors:
\begin{equation}
 \Phi^{(i)}_{\gamma n}=N^\gamma_nr_i^{\lambda_i}\rho_i^{l_i}
 \exp\{-\alpha_{\gamma n}r_i^2 -
 \beta_{\gamma n}\rho_i^2\} {\cal
 F}_{\gamma}^{(i)}(\hat{\bf r}_i,\hat{\b \rho}_i){\cal T}_{\gamma}^{(i)},
 \label{Phigamn}
 \end{equation}
 where the spin--angular ${\cal
 F}_{\gamma}^{(i)}(\hat{\bf r}_i,\hat{\b \rho}_i)$ and isospin 
 ${\cal T}_{\gamma}^{(i)}$ components of the basis functions are
given in Appendix and the composite label 
$\gamma\equiv \gamma(i) =\{\lambda_i\,l_i\,L\,S_{jk}\,S\,t_{jk}\}$
represents the respective set of the quantum numbers for the basis functions
(\ref{Phigamn}):
$\lambda_i$ is the orbital angular momentum of the ($jk$) pair; $l_i$
is the orbital angular momentum of  the third nucleon ($i$) relatively to the
center of
mass for the ($jk$) pair; $L$ is the total orbital  angular momentum of the $3N$ system;
$S_{jk}$ and $t_{jk}$ are the spin and isospin of the ($jk$) pair,  respectively; and
$S$ is the total spin of the system. We omit here the total angular momentum $J=1/2$
and its $z$-projection $M$, as well as the total isospin of the system $T=1/2$ and
its projection  $T_z$ (in this work, we neglect the very small contribution of the
$T=3/2$ component). 

The nonlinear parameters of the basis functions $\alpha_{\gamma n}$ and
$\beta_{\gamma n}$ are chosen  on the Chebyshev grid, which provides the
completeness of the basis and fast convergence of variational
calculations~\cite{Cheb}. As was demonstrated earlier~\cite{Gauss}, this
few-body Gaussian basis is  very flexible and can represent quite complicated
few-body correlations. Therefore, it leads to the accurate  eigenvalues and
eigenfunctions. The formulas for the matrix elements of the Hamiltonian (for local
$NN$ interactions) on  antisymmetrized Gaussian basis are given in the
paper~\cite{Tur1}. The matrix elements of DBM interactions on this basis are given in
Appendix.

{\bf Wavefunction in the internal $6qN$ channel.} Having the $3N$ component $\Psi_{3N}$ found in the above
variational calculation, one can construct  the inner
$6qN$-channel wavefunction $\Psi_{\rm in}^{(i)}$, which depends on
the coordinate (or momentum) of the third nucleon and the
$\sigma$-meson momentum and includes the bag wavefunction (see
Eq.~(\ref{psi6qNIJ})). By integrating the modulus squared of this
function with respect to the meson momentum  and inner variables
of the bag, one obtains the density distribution of the third
nucleon relative to  the $6q$ bag in the $6qN$ channel. This
density can be used to calculate further all observables, whose
operators depend on the variables of the nucleons and the bag.
However, it is much more convenient and easier to  deal with the
quasi-wavefunction of the third nucleon in the $6qN$ channel,
which has been introduced  by Eq.~(\ref{npsi}).

To calculate matrix elements of the 3BF Coulomb and OPE forces, one needs the
spin--isospin part of $6qN$ components of the total wavefunction. Here we give
them explicitly. The potential form factors
$\varphi^{J_iM_i}_{L_i}$ now include the spin--isospin part 
${\cal Y}^{J_iM_i}_{L_iS_d}(\hat{\bf p}_i){\cal T}^{(i)}_{t_d}$ 
with quantum numbers corresponding to the dressed bag:
 \begin{equation}
\varphi^{J_iM_it_dt_{d_z}}_{L_iS_d}= \phi^{J_i}_{L_i}({p}_i)
{\cal Y}^{J_iM_i}_{L_iS_d}(\hat{\bf p}_i){\cal T}^{(i)}_{t_d}; \qquad
{\cal T}^{(i)}_{t_d}=|t_jt_k:t_dt_{d_z}\rangle .
 \label {phijl}
 \end{equation}
The full set of the quantum numbers labelling the form factors includes the
total ($J_i$) and orbital ($L_i$) angular momenta, related  to the vertex form
factor, and also the spin and isospin numbers $S_d$, $t_d$, and $t_{d_z}$
related to the dressed bag. However, since the present version of the DBM involves the
bag states with zero orbital
angular momentum, we have $S_d=J_i$,  while the bag spin
and isospin are supplementary to each other: $t_d+S_d=1$. Hence we will omit the
quantum numbers $S_d$ and $t_d$, where they are unnecessary.

The total overlap function
$\chi^{J_iM_i}_{L_i}(i)=\langle\varphi^{J_iM_i}_{L_i}|\Psi_{3N}\rangle$ can be
written (with its spin--isospin part), e.g., as
\begin{equation}
  \chi^{J_iM_i}_{L_i}({\bf q}_i)=
 \sum_{l_i{\cal J}}\Phi^{J_iL_i}_{l_i{\cal J}}(q_i)
 \langle {\cal J}m_{\cal J}J_iM_i|JM\rangle \,
 {\cal Y}^{{\cal J}m_{\cal J}}_{l_i {\frac{1}{2}}}(\hat{\bf q}_i)
 \,\langle t_dt_{d_z}\textstyle {\frac{1}{2}}t_{z_i}|TT_z\rangle
 \,{\cal T}_{{\frac{1}{2}}t_{z_i}}.
 \label {ovfull}
 \end{equation}
 Here, $J$ and $M$ are the total angular momentum of the $3N$ system and its
$z$-projection, $T$ and $T_z$ are the total isospin of the $3N$
system and its $z$-projection, while $l_i$ and $\cal J$ are  the
orbital and total angular momenta of the third ($i$th) nucleon,
respectively, and ${\cal T}_{{\frac{1}{2}}t_{z_i}}$ is isospinor
corresponding to the  third nucleon. In the present calculation
for the ground states of $^3$H and $^3$He (with $J=1/2$), we have
considered two lowest even partial wave components ($S$ and $D$)
in $3N$ wavefunctions only.  Therefore,  $l_i$ can take only two
values: 0 or 2. Moreover, the total angular momentum of the third
nucleon $\cal J$ is uniquely determined by value of $l_i$: ${\cal
J}=1/2$ at $l_i=0$ and ${\cal J}=3/2$ at  $\lambda_i=2$. So,
actually there is no summation over $\cal J$ in Eq.(\ref{ovfull}).

  It is easy to see that the three form factors
$\varphi^{J_i}_{L_i}$ used in the present work  ($\varphi^{0}_0$, $\varphi^1_0$,
and $\varphi^1_2$) determine five radial components of the overlap  function
$\Phi^{J_iL_i}_{l_i{\cal J}}(q_i)$ and five respective components of the
quasi-wavefunction  for the $6qN$ channel. To specify these components it is
sufficient to give three quantum numbers, e.g., $S_d$, $l_i$ and $L_i$,
and we will use notation
 $\Psi^{\rm in}_{S_dl_i,L_i}(q_i)$ for these radial components:
\[
\begin{array}{ll}
 \Psi^{\rm in}_{00,0}\!:\,&(J_i=S_d=0,\,t_d=1,\,L_i=0,\,l_i=0,\,{\cal J}=\frac{1}{2}),\\
 \Psi^{\rm in}_{10,0}\!:\,&(J_i=S_d=1,\,t_d=0,\,L_i=0,\,l_i=0,\,{\cal J}=\frac{1}{2}),\\
 \Psi^{\rm in}_{12,0}\!:\,&(J_i=S_d=1,\,t_d=0,\,L_i=0,\,l_i=2,\,{\cal J}=\frac{3}{2}),\\
 \Psi^{\rm in}_{10,2}\!:\,&(J_i=S_d=1,\,t_d=0,\,L_i=2,\,l_i=0,\,{\cal J}=\frac{1}{2}),\\
 \Psi^{\rm in}_{12,2}\!:\,&(J_i=S_d=1,\,t_d=0,\,L_i=2,\,l_i=2,\,{\cal J}=\frac{3}{2}).
 \end{array}
\]
At last, we give a formula for the total quasi-wavefunction in internal channel
($i$), separating out explicitly its spin--angular and isospin parts, which
include the spin--isospin part of the bag wavefunction:
\begin{equation}
 \Psi^{\rm in}_{i}=\sum_{l_iS_d}\left
 \{\sum_{L_i} \Psi^{\rm in}_{S_dl_i,L_i}(q_i)\right \}
 \textstyle |l_i {\frac{1}{2}}({\cal J})S_d:JM \rangle \,
 |t_d{\frac{1}{2}}:TT_z\rangle.
 \label {psi6qntot}
 \end{equation}
The explicit dependence of this function on the isospin projection $T_z$ is
important for calculation of the Coulomb matrix elements and r.m.s. charge radius.

The interaction matrix elements include the overlap integrals of the potential
form factors with the  basis functions $\Phi_{\gamma ,n}= \Phi^{(1)}_{\gamma
,n} + \Phi^{(2)}_{\gamma ,n} + \Phi^{(3)}_{\gamma ,n}$, where all five above
components of the overlap function enter the matrix elements independently
(certainly, some of  the matrix elements can vanish). The explicit formulas for
the above overlap functions and detailed formulas for the  matrix elements of
all DBM interactions are given in Appendix. When calculating both the
normalization of the internal components and observables, the $6qN$ components
distinguishing only by their radial parts can be summed. Thus, only three
different components of the  $6qN$ quasi-wavefunction remain: the one $S$-wave singlet
 $(S_d=0)$:
 \[\Psi_{00}^{\rm in} \equiv \Psi^{\rm in}_{00,0}, \]
and the two triplet ones  $(S_d=1)$:
\[\Psi_{10}^{\rm in} = \Psi^{\rm in}_{10,0} +  \Psi^{\rm in}_{10,2}, \]
\begin{equation} \Psi_{12}^{\rm in} = \Psi^{\rm in}_{12,0} +
\Psi^{\rm in}_{12,2}.
 \label{3comp}
\end{equation}
The total weight of each of three $6qN$ components is equal to
\begin{equation} P_{\rm in}^{(i)} = \|\Psi^{\rm in}_{00}\|^2 +
\|\Psi^{\rm in}_{10}\|^2 + \|\Psi^{\rm in}_{12}\|^2; \qquad i=1,2,3.
 \label{P6qNi}
\end{equation}
Now, let us introduce the relative weights of the individual $6qN$ components:
\begin{equation}
 P^{\rm in}_{S0} =\frac {\|\Psi^{\rm in}_{00}\|^2}{P^{(i)}_{\rm in}}, \;
 P^{\rm in}_{S1} =\frac {\|\Psi^{\rm in}_{10}\|^2}{P^{(i)}_{\rm in}}, \;
 P^{\rm in}_{D} = \frac {\|\Psi^{\rm in}_{12}\|^2}{P^{(i)}_{\rm in}}.
 \label{P6qNrel}
\end{equation}

After renormalization of the full four-component wavefunction, the total weight
of all internal  components is equal to
\begin{equation}
 P_{\rm in} =\frac {3P^{(i)}_{\rm in}}{1+3P^{(i)}_{\rm in}}
 \label{P6qN}
\end{equation} 
(here, we assume that the $3N$ component of the total wavefunction,
$\Psi^{\rm ex}$, obtained from the variational calculation, is normalized to unity), while
the total weight of the $3N$ component $\Psi_{3N}$ is equal to
\begin{equation}
 P_{\rm ex} =\frac {1}{1+3P^{(i)}_{\rm in}}=1-P_{\rm in}.
 \label{P3N}
\end{equation}

It is also interesting to find the total weight of the $D$ wave with allowance for non-nucleonic
components:
\begin{equation}
 P_{D} = P_D^{\rm ex}(1-P_{\rm in}) + P_D^{\rm in}P_{\rm in}.
 \label{PD}
\end{equation}

Numerical values of all above probabilities for internal and external
components are given below in Table~2.  The total weight of all $6qN$
components $P_{6qN}\equiv P_{\rm in}$ in the $3N$ system turns out to be rather large and
approaches or even exceeds 10\%. Furthermore, taking into account  the
short-range character of these components, the more hard nucleon momentum
distribution (closely associated with the first property) for these components,
and very strong scalar three-body  interaction in the internal $6qN$ channels,
one can conclude that these non-nucleonic components are extremely important
for the properties of nuclear systems.

\section{Coulomb effects in $^3$H\lowercase{e}}

In this section we will demonstrate that the
DBM approach leads to some new features related to the Coulomb effects in
nuclei, and in particular in $^3$He.
First of all, the additional Coulomb force arises because the $6q$ bag and rest
nucleon can have electric charges. We have found that this  new Coulomb
three-body force is responsible for a significant part of the total $^3$He
Coulomb energy (this three-body Coulomb force has been missed fully in previous
$3N$ calculations within hybrid $6qN$ models~\cite{Bakk}).

The second feature of the interaction model used here is the absence of the
local $NN$ short-range  repulsive core. The role of this core is played by the
condition of orthogonality to the confined $6q$  states forbidden in the
external $NN$ channel.  This orthogonality requirement imposed on the
relative-motion $NN$ wavefunction is responsible for the  appearance of some
inner nodes and respective short-range loops in this wavefunction. These
short-range nodes and loops lead to numerous effects and general consequences
for the nuclear structure. One of these consequences is a rather strong
overestimation of the Coulomb contribution when using the interaction between
point-like nucleons. Thus, it is necessary to take into account the finite
radius of the nucleon charge distribution\footnote{%
We remind in this point that the account of the finite radii of nucleons in the
conventional approaches leads to fully negligible corrections to the Coulomb
energy.}.

At last, in order to obtain the accurate Coulomb displacement energy $\Delta
E_{\rm C}=E_B(^3{\rm H})- E_B(^3{\rm He})$, one should take into consideration the
effects associated with the small mass  difference between the proton and
neutron. It is well known~\cite{Gloeckle} that the above mass  difference makes
rather small contribution to the difference between $^3$He and $^3$H binding
energies. Therefore, it was taken usually into account in a perturbation
approach. However, since the  average kinetic energy in our case is twice the
kinetic energy in conventional force models, this correction  is expected also
to be much larger in our case. Hence, we present here the estimation for  such
a correction term without usage of the perturbation theory.

\subsection{``Smeared'' Coulomb interaction}

The Gaussian charge distribution $\rho(r)$, that corresponds to the r.m.s. charge
radius $r_c$ and is normalized to the total charge $z$: $ 4\pi\int\rho r^2dr =
z$,  can be written as
\begin{equation}
 \rho(r)=z\left (\frac{\alpha}{\pi}\right )^{3/2}{e}^{-\alpha r^2},
 \qquad \alpha^{-1}=\frac{2}{3}r_c^2.
 \label{rhogauss}
\end{equation}
The Coulomb potential for the interaction between such a charge distribution $\rho(r)$ and 
a point-like charged particle has the well-known form
\[
 V(R)= \int\frac{d{\bf r}\, \rho(r)}{|{\bf R-r}|} =
 \frac{z}{R}{\rm erf}(R\sqrt{\alpha}).
\]
We have derived here a similar formula for the Coulomb interaction between two 
Gaussian distributions  with different widths $\alpha_1$ and $\alpha_2$ and r.m.s.
radii $r_{c_1}$ and $r_{c_2}$, respectively:
 \begin{equation}
 V(R;\alpha_1,\alpha_2) = \frac{z_1z_2}{R}{\rm erf}(R\sqrt{\tilde{\alpha}}),
 \, \tilde{\alpha} =\frac{\alpha_1\alpha_2}{\alpha_1 + \alpha_2}, \mbox{ or }
 \tilde{\alpha}^{-1}=\frac{2}{3}(r_{c_1}^2 + r_{c_2}^2).
 \label{Vcoul}
\end{equation}
In our calculations, we used the following charge radii for the nucleon and dibaryon:
\[(r_c)_p =0.87\mbox { fm},\]
\[(r_c)_{6q}=0.6\mbox{ fm}\footnotemark{}. \]
\footnotetext{%
This value is simply the r.m.s. charge radius of the $6q$ bag
with the parameters given in \cite{KuInt}. The neutral $\sigma$
field of the bag changes this value only slightly. The evident
difference between the charge radii of the nucleon and dibaryon
can be well understood as follows: the charge radius of the $3q$
core of the nucleon is taken usually as $r_c^{3q}\simeq 0.5$--0.55~fm,  
while remaining 0.3~fm is assumed to come from the
charge distribution of the $\pi^+$ cloud  surrounding the $3q$
core in the proton. In contrast, the meson cloud of the dibaryon
in our approach  is mainly due to the neutral scalar--isoscalar $\sigma$
meson, so that the dibaryon charge distribution is  characterized
by the charge radius of the bare $6q$ core only.} These values
lead to the ``smeared'' Coulomb interactions in the $NN$ and $6qN$
channels:
\[
 V^{\rm Coul}_{NN}(r) = \frac{e^2}{r}{\rm erf}(r\sqrt{\alpha_{NN}}), \qquad
 \alpha_{NN}^{-1/2}=1.005\mbox{ fm},
 \]
  \begin{equation}
  V^{\rm Coul}_{\rm in}(\rho) = \frac{e^2}{\rho}{\rm erf}(\rho\sqrt{\alpha_{\rm in}}),
\qquad \alpha_{\rm in}^{-1/2}=0.863\mbox{ fm}.
 \label{Vcoul1}
\end{equation}

\subsection{Matrix elements of the three-body Coulomb force}

The Coulomb interaction between the charged bag and the third nucleon
in the $6qN$ channel is determined by  the three-particle operator
with the separable kernel (see Eq.~(\ref{3BF})):
\[
   ^{\rm Coul}W^{(i)}({\bf p}_i,{\bf p}'_i;{\bf q}_i,{\bf q}'_i) =\]
\begin{equation}  = \sum_{J_iM_iL_iL'_i}
  \varphi^{J_iM_i}_{L_i}({\bf p}_i)\,
  \frac{1\!+\!\tau_3^{(i)}}{2}\,(1+\hat{t}_{d_z})
  {}^{\rm Coul}W^{J_i}_{L_iL'_i}({\bf q}_i,{\bf q}'_i;E)\,
  \varphi^{J_iM_i}_{L'_i}({\bf p}'_i),
 \label{V3bfCoul}
\end{equation}
where $(1\!+\!\tau_3^{(i)})/2$ is the operator of the $i$th nucleon charge
and $1+\hat{t}_{d_z}$ is operator of the bag charge.
 It is evident that the
matrix element of the operator (\ref{V3bfCoul}) can be expressed in terms of the
integrals of the product of the overlap functions $\chi^{J_iM_i}_{L_i}({\bf q}_i)$ of $NN$ 
form factors and three-body basis functions.
 The method for calculation of such Coulomb integrals is given in Appendix.

\section{Results of calculations}

Here, we present the results of the $3N$ bound-state calculations based on two variants
of the DBM.
\begin{itemize}
\item [(I)] In the first version developed in~\cite{KuInt}, the
dressed-bag propagator includes three loops (two loops are with pions and one
loop is with $\sigma$ meson), two of them are of
the type shown in Fig.~2 of \cite{KuInt}, in which each loop was
calculated within the $^3P_0$ model for quark--meson interaction.
The third loop consists of two such vertices and a convolution of
the $\sigma$-meson and $6q$-bag propagators~\cite{KuInt}. 
\item[(II)] In the second version, we replaced two
above pionic loops with the effective  Gaussian form factor $B(k)$,
which describes the direct $NN \to 6q+\sigma$ transition, i.e.,
the direct transition from the $NN$ channel to the
dressed-dibaryon channel.
\end{itemize}

Both versions have been fitted to the $NN$ phase shifts in low partial waves up
to an energy  1~GeV  with almost the same quality. Therefore, they can be
considered on equal footing. However, version  (II) has one important advantage.
Here, the energy dependence arising from the convolution of the two propagators
involved into the loop, i.e., the propagators of the $\sigma$ meson and bare
dibaryon,  describes (with no further correction) just the energy dependence of
the effective strength of the  $NN$ potential $\lambda^{(II)} (E)$, which is
thereby taken directly from the above loop integral. In contrast, in the
first version of the model, two additional $qq\pi\pi\sigma$ loops give a rather
singular three-dimensional integral for $\lambda^{(I)}(E)$, where the energy
dependence at higher  energies should be corrected by a linear term.

\subsection{Bound-state energies of $^3$H and $^3$He and individual
contributions to them}
\begin{table}[h!]
 \setcaptionmargin{0mm} \onelinecaptionsfalse
\captionstyle{flushleft}
\caption{Results of the $3N$ calculations with two- and three-body
forces for two variants of the DBM}
\begin{tabular}{|*{8}{c|}}  \hline
  Model& $E$, MeV& $P_D$, \% & $P_{S'}$, \%& $P_{6qN}(P_{\rm in})$, \% 
  & \multicolumn{3}{|c|}{Contributions to $H$, MeV} \\
  \cline{6-8}
        &&&&                  &$T$& $T+V^{(2N)}$&$V^{(3N)}$ \\
 \hline
 \multicolumn{8}{|c|}{ $^3$H}\\ \hline
 DBM(I)  $g=9.577^{\rm (a)}$ & --8.482 & 6.87 & 0.67 & 10.99& 112.8 & --1.33 & --7.15  \\
 
 DBM(II) $g=8.673^{\rm (a)}$ & --8.481 & 7.08 & 0.68 & 7.39 & 112.4 & --3.79 & --4.69   \\

 AV18 + UIX$^{\rm (b)}$    & --8.48& 9.3 & 1.05 &  --   & 51.4 & --7.27 & --1.19   \\
 \hline 
\multicolumn{8}{|c|}{ $^3$He}\\ \hline
 DBM(I)  & --7.772  & 6.85 & 0.74 & 10.80 & 110.2 & --0.90 & --6.88   \\
 
 DBM(II) & --7.789  & 7.06 & 0.75 & 7.26  & 109.9 & --3.28 & --4.51   \\
 
 AV18 + UIX$^{\rm (b)}$ & --7.76 & 9.25 & 1.24 & --     & 50.6 & --6.54 & --1.17   \\
 \hline 
\end{tabular}

 {\footnotesize \begin{flushleft} $^{\rm a)}$These values of $\sigma NN$ coupling constant in
$^3$H calculations have been chosen to reproduce the exact binding
energy of $^3$H nucleus. The calculations for $^3$He have been
carried out without any free parameters.

$^{\rm b)}$The values are taken from \cite{Pieper2001}.\end{flushleft}
}
\label{table2}
\end{table}

 The main difference between the results for both
versions is that the energy dependence of  $\lambda(E)$ for the second version
is much weaker than that for the first one. In addition, this  energy
dependence leads to some decrease in the contribution of the $6qN$ component to
all $3N$ observables and thus to the respective increase of the two-body force
contribution as compared to the three-body force one. Table~2 presents the
calculation results for the two above versions for the following
characteristics:
the weights of the internal $6qN$
channels and $D$ wave in the total $3N$ function, as well as the weight of the
mixed-symmetry $S'$ component (only for the $3N$ channel);
the average individual
contributions from the kinetic energy $T$, two-body interactions $V^{(2N)}$
plus the kinetic energy $T$, and three-body force ($V^{(3N)}$) due to OSE 
and TSE to the total Hamiltonian expectation.

For  variant I of the model, we present also the result calculated when 
both 3BF and the $q^2$ dependence of the effective two-body force on the
momentum of the third nucleon are disregarded (the first line). The results in the  second line
of Table 2 are obtained including the $q^2$ dependence of pair forces, but
disregarding 3BF. The percentages of the $D$-wave  and the internal components
given in Table~2 were obtained with incorporation of the three internal
components; i.e., these values correspond to the normalization of the total
(four-component) wavefunction of the system to unity.

To compare the predictions of the new model with the respective
results for the conventional $NN$ potential models, Table~2  also
presents the results of recent calculations with the Argonne
potential AV18 and Urbanna--Illinois 3BF 
UIX~\cite{Pieper2001}. 

\subsection{The densities, r.m.s. radii  and charge distributions in
$^3$H and $^3$He}
 At first, we give definitions of the nucleon and charge distributions
in multichannel system.

{\bf The external $3N$ channel.} The proton ($\rho_p$) and neutron ($\rho_n$)
densities in this channel are defined  by the standard way~\cite{Friar86}:
 \begin{equation}
  \rho^{\rm ex}_{\{{p\atop n}\}}(r) =
 \frac{1}{N_{\{{p\atop n}\}}}\left\langle \Psi^{\rm ex} \sum_{i=1}^3
 \left |\frac{\delta(r-\frac{3}{2}\rho_i)}{r^2} \frac{1\pm \tau_3^{(i)}}{2}
\right |\Psi^{\rm ex}\right\rangle =
 \frac{3}{N_{\{{p\atop n}\}}}\left\langle \Psi^{\rm ex}
 \left |\frac{\delta(r-\frac{3}{2}\rho_1)}{r^2} \frac{1\pm \tau_3^{(1)}}{2}
\right |\Psi^{\rm ex}\right\rangle,
 \label{rhopn}
\end{equation}
 where $\rho_i$ is Jacobi coordinate in the set ($i$)
and  $N_{\{{p\atop n}\}}$ is the number of protons (neutrons).
 Due to property
\[\left\langle \Psi^{\rm ex} \left |\sum_{i=1}^3\frac{1\pm \tau_3^{(i)}}{2}
\right |\Psi^{\rm ex}\right\rangle = {\textstyle \left\langle \Psi^{\rm ex} 
\left |\frac{3}{2}\pm \hat{T}_3
\right |\Psi^{\rm ex}\right\rangle = \frac{3}{2}\pm \hat{T}_3}=N_{\{{p\atop n}\}}, \]
 the above densities are normalized to unity, provided that the external
wavefunction $\Psi^{\rm ex}$ is also normalized to unity:
\[ \int\rho^{\rm ex}_{\{{p\atop n}\}}(r) r^2 {d}r = 1.\]
The matrix element $\langle \Psi^{\rm ex} | \tau_3^{(i)}
|\Psi^{\rm ex}\rangle $ is proportional to $z$-projection of the total isospin $T_3$,
therefore, the nucleon densities can be separated into isoscalar (matter)
density  and isovector parts:
 \begin{equation}
 \rho_s(r) = \rho_m=\left\langle \Psi^{\rm ex}
 \left |\frac{\delta(r-\frac{3}{2}\rho_1)}{r^2} \right |\Psi^{\rm ex}\right\rangle,
 \label{rhosc}
\end{equation}
 \begin{equation}
 \rho_v(r) = \frac{3}{2T_3}\left\langle \Psi^{\rm ex}
\left |\frac{\delta(r-\frac{3}{2}\rho_1)}{r^2} \tau_3^{(1)}
\right |\Psi^{\rm ex}\right\rangle.
 \label{rhovec}
\end{equation}
 Both latter densities are also normalized to unity.
Then the nucleon densities can be  expressed in terms of isoscalar and isovector
 densities as:
 \begin{eqnarray}
 \rho^{\rm ex}_p(^3{\rm He})=\rho^{\rm ex}_n(^3{\rm H})=\frac{1}{4}(3\rho_s+\rho_v), \nonumber \\
 \rho^{\rm ex}_p(^3{\rm H})=\rho^{\rm ex}_n(^3{\rm He})=\frac{1}{2}(3\rho_s-\rho_v).
 \label{ntosv}
\end{eqnarray}
 R.m.s. radii of corresponding distributions are equal to:
 \begin{equation}
  \langle r^2\rangle^{\rm ex}_{\{s,v,p,n\}} =
  \int\rho^{\rm ex}_{\{s,v,p,n\}}(r) r^4 {\rm d}r.
  \label{rms}
  \end{equation}
The r.m.s. charge radius in the $3N$ sector is also defined conventionally:
\begin{equation}
 \langle r_{\rm ch}^2\rangle^{\rm ex} = \langle r^2\rangle^{\rm ex}_p+R_p^2
 +\frac{N_n}{N_p}R_n^2,
 \label{rch3N}
\end{equation}
where $R^2_p=0.7569$~fm$^2$ and $R_n^2=-0.1161$~fm$^2$ are the squared
charge radii of the proton and neutron, respectively.

The various types of one-particle densities (isoscalar, isovector, proton, neutron)
in external $3N$ channel for the $^3$H and $^3$He ground states calculated in 
DBM(I) are shown in Fig.~5. 

\begin{figure}[h]
 \centerline{ \epsfig{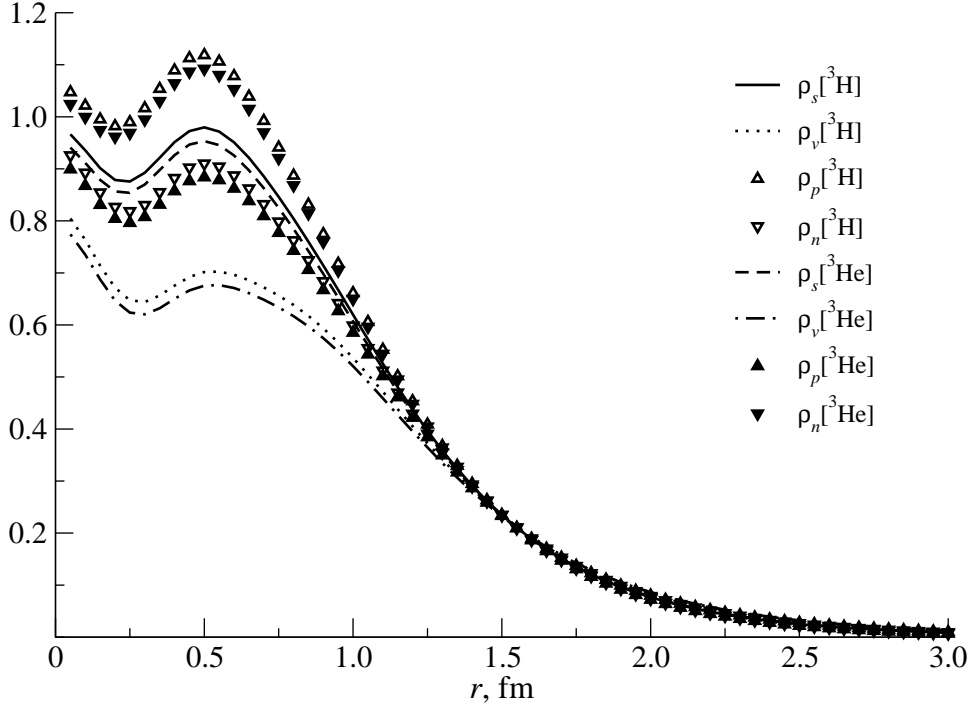}}
 \setcaptionmargin{0mm} \onelinecaptionsfalse
\captionstyle{flushleft}
 \caption {The isoscalar $\rho_s$, isovector $\rho_v$, proton $\rho_p$, 
 and neutron $\rho_n$ densities in external
$3N$ channel for $^3$H and $^3$He systems obtained with DBM (version I).}
\label{fig5}
\end{figure}

Below we present also two-proton density for $^3$He,
which is defined usually as~\cite{Doyle92}:
\begin{equation}
 \rho_{pp}(r)=6\left\langle \Psi^{\rm ex}
 \left |\frac{\delta(r-r_1)}{r^2} \frac{1+ \tau_3^{(2)}}{2}\frac{1+ \tau_3^{(3)}}{2}
\right |\Psi^{\rm ex}\right \rangle.
 \label{rhopp}
\end{equation}
This density is normalized to 2: $\int \rho_{pp}(r)r^2{d}r=2$. (As there is
only a single nucleon in $6qN$ channel, we do not attach the index ``ex'' to
this quantity.) The two-neutron density $\rho_{pp}(r)$ for $^3$H is defined 
similarly (with replacing $1+ \tau_3^{(i)} \to 1- \tau_3^{(i)}$ in
Eq.~(\ref{rhopp}). In Fig.~6 we show both these densities for DBM(I) and 
also the two-proton density for $^3$He found with Bonn $NN$ potential~\cite{Doyle92}.

\begin{figure}[h]
\centerline{ \epsfig{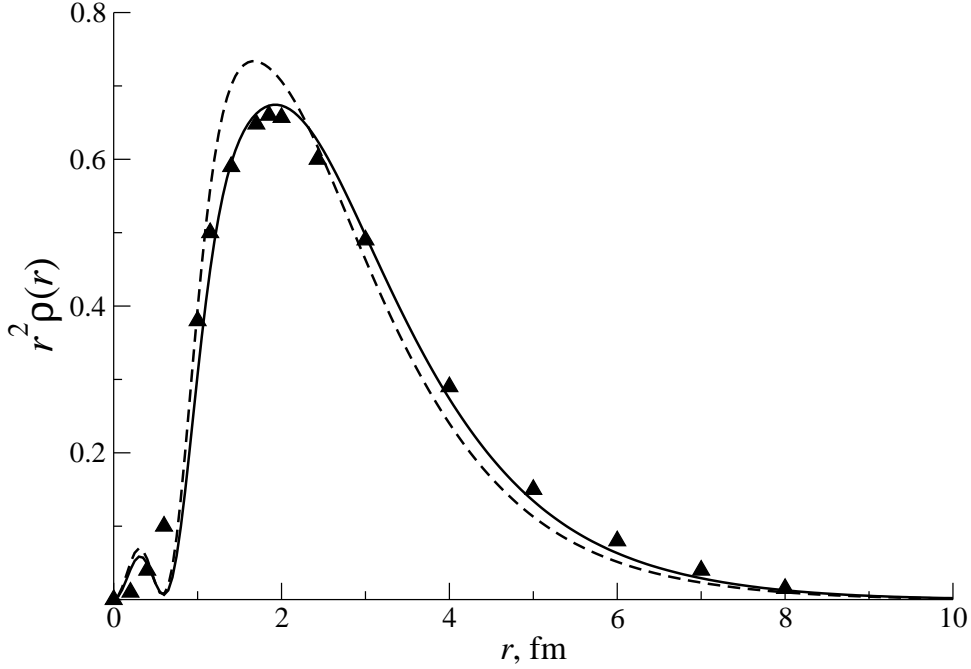}}
 \setcaptionmargin{0mm} \onelinecaptionsfalse
\captionstyle{flushleft}
 \caption {The two-proton density in $^3$He (solid line) and two-neutron 
 density in $^3$H (dashed line) calculated with DBM (version I) in comparison 
 with two-proton density 
 found with Bonn potential~\cite{Doyle92} (triangles).}
\label{fig6} 
\end{figure}

{\bf The internal $6qN$ channels.} Here we define a density (normalized to unity) 
of the pure nucleon distributions  as
 \begin{equation}
 \rho^{\rm in}_{\{{p\atop n}\}}(r) = \frac{1}{N^{\rm in}_{\{{p\atop n}\}}
 P_{\rm in}^{(1)}}
 {\left\langle \Psi^{\rm in}_1
 \left |\frac{\delta(r-\alpha\rho_1)}{r^2} \frac{1\pm \tau_3^{(1)}}{2}
\right |\Psi^{\rm in}_1\right\rangle},
 \label{rhopni}
\end{equation}
 where  $P_{\rm in}^{(1)}=\langle\Psi^{\rm in}_1|\Psi^{\rm in}_1\rangle$
 and the quantity
 \begin{equation}
 N^{\rm in}_{\{{p\atop n}\}}=\frac{1}{P_{\rm in}^{(1}}
 \left\langle \Psi^{\rm in}_{1}
 \left |\frac{1\pm \tau_3^{(1)}}{2}\right |\Psi^{\rm in}_1\right\rangle,
 \label{ninpn}
\end{equation}
 has the meaning of the average number of protons (neutrons)
in the one internal $6qN$ channel (note that $N_p^{\rm in}+ N_n^{\rm in}=1$,
i.e., there is only one nucleon in each internal channel). The number
 $ N^{\rm in}_{\{{p\atop n}\}}$ depends on ratio of norms of $6qN$
 components with different values of isospin of the bag. Therefore, the separation of the
$6qN$-channel density into isoscalar and isovector parts
  has no meaning.

 These average numbers of nucleons  in $6qN$ channel
can be expressed through relative probabilities of the $6qN$ components with
definite value of isospin $t$, which in our case are equal to:
 \begin{eqnarray}
 P^{\rm in}_{0}\equiv P^{\rm in}_{t=0}=P^{\rm in}_{S1}+P^{\rm in}_{D}
 \nonumber,\\
 P^{\rm in}_{1}\equiv P^{\rm in}_{t=1}=P^{\rm in}_{S0},
 \label{pt01}
\end{eqnarray}
 where $P^{\rm in}_{S1}$, $P^{\rm in}_{D}$, and $P^{\rm in}_{S0}$ are determined
 by Eq.~(\ref{P6qNrel}). Hence, $P^{\rm in}_{t=0}+P^{\rm in}_{t=1}=1$.
Then one can write down the average numbers of nucleons as
 \begin{eqnarray}
  N^{\rm in}_p(^3{\rm H})=N^{\rm in}_n(^3{\rm He})=\frac{2}{3}P^{\rm
  in}_{t=1},\nonumber \\
  N^{\rm in}_p(^3{\rm He})=N^{\rm in}_n(^3{\rm H})=
  P^{\rm in}_{t=0}+\frac{1}{3}P^{\rm in}_{t=1}.
 \label{ninpn1}
\end{eqnarray}
 The nucleon densities (\ref{rhopn}) can be expressed by similar formula
 through components of the internal wavefunction with definite value of isospin
 $t$.

{\bf The total densities of nucleon distributions.} The total nucleon densities 
(normalized to unity) for whole $3N$ system with allowance for both the $3N$ and $6qN$
components can be now defined as
\begin{equation}
  \rho_{\{{p\atop n}\}}=\frac{(1-P_{\rm in}) \rho^{\rm ex}_{\{{p\atop n}\}}N_{\{{p\atop n}\}}+
  P_{\rm in} \rho^{\rm in}_{\{{p\atop n}\}}N^{\rm in}_{\{{p\atop n}\}}}{\langle N_{\{{p\atop n}\}}\rangle},
   \label{rhonptot}
 \end{equation}
 where $P_{\rm in}={3P^{\rm in}_1}/(1+3P^{\rm in}_1)$ is the total weight of
 all three internal channels (Eq.~(\ref{P6qN})) and the denominator:
\begin{equation}
 \langle N_{\{{p\atop n}\}}\rangle=(1-P_{\rm in})N_{\{{p\atop n}\}}+
  P_{\rm in} N^{\rm in}_{\{{p\atop n}\}} < N_{\{{p\atop n}\}}
   \label{npntot}
 \end{equation}
  is equal to the average number of protons
 (neutrons) in the whole multicomponent system. 
The densities of the total proton and neutron 
distributions and also external- and internal-channel distributions 
for $^3$He calculated for DBM(i) are presented in Fig.~7. 

\begin{figure}[h]
 \centerline{
 \epsfig{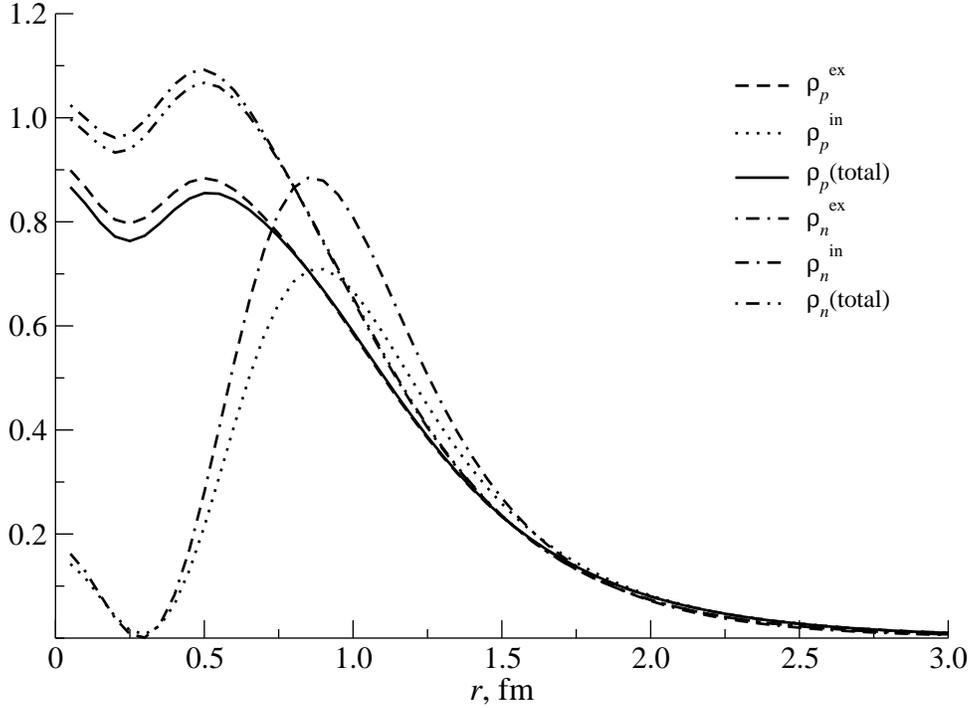}}
 \setcaptionmargin{0mm} \onelinecaptionsfalse
\captionstyle{flushleft}
 \caption {The external $\rho^{\rm ex}$, internal $\rho^{\rm in}$,  and total
 $\rho (\rm total) $ densities of proton- 
 and neutron distributions in $^3$He found with DBM (version I).}
\label{fig7} 
\end{figure}
 
One can define also a (normalized) density  of the matter (or mass)
distribution in the $6qN$ channel as:
 \begin{equation}
 r^2\rho^{\rm in}_m(r) = \frac{1}{(m_N+m_d)\,P_{\rm in}}
 {\langle \Psi^{\rm in}_1
 |{\delta(r-\alpha\rho_1)}m_N + \delta(r-(1-\alpha)\rho_1)m_d
|\Psi^{\rm in}_1\rangle},
 \label{rhomin}
\end{equation}
where $\alpha={m_d}/(m_d+m_N)$, $m_N$ is a nucleon mass, and $m_d$ is mass of
the bag (dibaryon). Then the total matter density (normalized to unity) is equal to
\begin{equation}
  r^2\rho_m(r)=\frac{(1-P_{\rm in}) \rho^{\rm ex}_m 3m_N+
  P_{\rm in} \rho^{\rm in}_m (m_N+m_d)}{\langle m\rangle}.
   \label{rhomtot}
 \end{equation}
 The r.m.s. radius of any distribution normalized to unity is defined
by Eq.~(\ref{rms}). 
The denominator in Eq.~(\ref{rhomtot}) determines the average mass
of the whole system with taking into account non-nucleonic
channels:
\begin{equation}
  \langle m\rangle=(1-P_{\rm in}) 3m_N+
  P_{\rm in} (m_N+m_d)=3m_N+P_{\rm in}(m_d-2m_N)>3m_N.
   \label{mtot}
 \end{equation}

\begin{table}[h]
 \setcaptionmargin{0mm} \onelinecaptionsfalse
\captionstyle{flushleft}
\caption {Isospin structure of $6qN$ channel, average number of
nucleons and average mass calculated with the ground state wavefunctions 
of $^3$H and $^3$He in the DBM approach}

\begin{tabular}{|c|c|c|c|c|}  \hline
 & \multicolumn{2}{|c|}{ $^3$H} &\multicolumn{2}{|c|}{ $^3$He}\\ \hline
                      &  DBM(I) & DBM(II) & DBM(I) & DBM(II)\\ \hline
 $P^{\rm in}_{t=1}$   & 0.6004  & 0.6005  & 0.6044 & 0.6044 \\ 
 $P^{\rm in}_{t=0}$   & 0.3996  & 0.3995  & 0.3956 & 0.3956 \\ 
 $3N^{\rm in}_p$      & 0.799   & 0.799   & 2.209  & 2.209  \\
 $3N^{\rm in}_n$      & 2.201   & 2.201   & 0.791  & 0.791  \\
 $\langle N_p\rangle$ & 0.919   & 0.945   & 1.863  & 1.908  \\  
 $\langle N_n\rangle$ & 1.861   & 1.906   & 0.920  & 0.947  \\ 
 $\langle N\rangle $  & 2.780   & 2.852   & 2.784  & 2.855  \\ 
 $\langle m\rangle /3m_N$
                      & 1.015   & 1.010   & 1.014  & 1.010  \\ \hline
\end{tabular}
\label{table3}
\end{table}

 In the Table~3 we present some characteristics of isospin structure for
 wavefunctions in the $6qN$ channel: the relative probabilities for the 
 components with
 $t=0$ and $t=1$ (i.e., $P^{\rm in}_{t=0}$ and $P^{\rm in}_{t=1}$), average
 numbers of protons and neutrons in all three
 $6qN$ components ($3\langle N_{\{{p\atop n}\}}\rangle$), and also the average
 number of nucleons $\langle N\rangle$ and the average mass $\langle m\rangle$
 (divided by $3m_N$ value) in the whole four-component  $3N$ system. 
 It should be noted that the average number of nucleons in our multicomponent 
 model, $\langle N_{\{{p\atop n}\}}\rangle$, is always less than the numbers of nucleons 
  in $3N$ channel just due to existence of the non-nucleonic components. 
  For example for DBM(I), the average number of protons in $^3$H is approximately equal to 
  the average number of neutrons in $^3$He, viz.  
  $\langle N_{p}\rangle(^3{\rm H})\approx \langle N_{n}\rangle(^3{\rm He})$,
 and is equal 0.92 while the average number of neutrons in $^3$H  
 is approximately equal to 
  the average number of protons in $^3$He, viz.
 $\langle N_{n}\rangle(^3{\rm H})\approx \langle N_{p}\rangle(^3{\rm He})$, 
 and is equal 1.86.
  Hence the average
 number of nucleons found with the total multicomponent $^3$H and $^3$He 
 functions is also always less than 3:
 \[\langle N\rangle=\langle N_p\rangle + \langle N_n\rangle=
 3-2P_{\rm in}<3 .\] 
In our DBM, $\langle N\rangle$ is equal 2.78 and 2.85 for versions I and II,
respectively.

{\bf The charge distributions.} The charge distribution for the point-like 
particles  in the $6qN$
channel can be written as the charge density of a system consisting of a
point-like nucleon and a point-like bag:
  \begin{equation}
 r^2\rho^{\rm in}_{\rm ch-point}(r) = \frac{1}{Z}
 {\left\langle \Psi^{\rm in}_1
\left |{\delta(r-\alpha\rho_1)}\frac{1+ \tau_3^{(1)}}{2} + \delta(r-(1-\alpha)\rho_1)
 (1+\hat{t}_3)\right |\Psi^{\rm in}_1\right\rangle},
 \label{rhochinp}
\end{equation}
where $1+\hat{t}_3$ is operator of the bag charge. The total charge radius in
$6qN$ channel includes the r.m.s. radius of this point-like distribution
$\langle r^2\rangle^{\rm in}_{\rm ch-point}$, the nucleon charge
radius ($R_p$ or $R_n$) and the charge
bag radius $R_d$, which depends on the bag isospin $t$ and its projection $t_3$:
 \begin{equation}
 \langle r^2\rangle^{\rm in}_{\rm ch} =
 \langle r^2\rangle^{\rm in}_{\rm ch-point} +\frac{1}{Z}\left (N^{\rm in}_p R^2_p
 +N^{\rm in}_n R^2_n +
 {\left\langle \Psi^{\rm in}_1
 \left |\sum_{t,t_3}\Gamma_{t,t_3}R^2_d(t,t_3)\right |\Psi^{\rm in}_1\right\rangle}\right ).
 \label{r2chin}
\end{equation}
 The last term in Eq.~(\ref{r2chin}) includes the projectors $\Gamma_{t,t_3}$
 onto the $6q$ bag isospin state with definite values of isospin $t$ and its 
 projection
 $t_3$ and is equal to (for the $^3$H and $^3$He states with  total isospin 
 $T=1/2$):
 \begin{equation}
  \Delta(r^2_{\rm ch})^{\rm in}_{\rm bag} =\frac{1}{Z}\left\{
R^2_d(0,0)P^{\rm in}_{t=0}+\left(\frac{1}{3}R^2_d(1,0)
+\frac{2}{3}R^2_d(1,1)\delta_{T_3,\frac{1}{2}}
+\frac{2}{3}R^2_d(1,-1)\delta_{T_3,-\frac{1}{2}}\right)P^{\rm in}_{t=1}\right\}.
\label{r2chbag}
 \end{equation}
 The own charge radius of the $6q$ bag $R^2_d(t,t_3)$ has, in general, different values in
 different isospin states, (which is related to the different multiquark dynamics
 in the channels with different isospin values) but we suppose in this work 
 that their difference can be ignored, viz.
 \begin{eqnarray}
\nonumber R^2_d(0,0)=R^2_d(1,0)=R^2_d(1,1)=R^2_d=b_0^2,\qquad b_0=0.6\mbox{
fm},\\ \mbox{and } R^2_d(1,-1)=0 \qquad (\mbox{which corresponds to
a $nn$-bag}).
 \label{R2d}
 \end{eqnarray}
With the above assumptions the $6q$ bag contribution to the $^3$H and $^3$He 
charge radius is reduced to:
 \begin{equation}
 \Delta(r^2_{\rm ch})^{\rm in}_{\rm bag} =\frac{R^2_d}{Z}\left (P^{\rm in}_0
 +\frac{2+2T_3}{3}P^{\rm in}_1\right).
\label{r2chbag1}
 \end{equation}
The r.m.s. charge radius of whole multicomponent system is defined as:
\[r^2_{\rm ch}=(1-P_{\rm in})\langle r^2 \rangle_{\rm ch}^{\rm ex}+ 
 P_{\rm in} \langle r^2 \rangle_{\rm in}^{\rm ex}. \]

\begin{table}[h!]
 \setcaptionmargin{0mm} \onelinecaptionsfalse
\captionstyle{flushleft}
\caption {The total r.m.s. radii (in fm) for the proton ($r_p$),
neutron ($r_n$), matter ($r_m$), and charge ($r_{\rm ch}$)
distributions in DBM approach and their separate values for
external and internal channels}

\begin{tabular}{|*{10}{c|}}  \hline
 \multicolumn{2}{|c|}{Model} &\multicolumn{4}{|c|}{$^3$H}&
 \multicolumn{4}{|c|}{$^3$He} \\ \cline{3-10}
 \multicolumn{2}{|c|}{}
       & $r_p$  & $r_n$& $r_m$& $r_{\rm ch}$  & $r_p$ & $r_n$  & $r_m$& $r_{\rm ch}$  \\ \hline
 DBM(I) & $3N$  & 1.625  & 1.770& 1.723& 1.779  & 1.805 & 1.648  & 1.754& 1.989 \\ 
        & $6qN$ & 1.608  & 1.823& 1.142& 1.188  & 1.854 & 1.618  &
1.159& 1.412 \\ 
      & Total & 1.625  & 1.773& 1.663& 1.724  & 1.807 & 1.647  & 1.694& 1.935 \\ \hline
DBM(II) & $3N$  & 1.613  & 1.761& 1.713& 1.769  & 1.795 & 1.636  & 1.744& 1.980 \\ 
        & $6qN$ & 1.573  & 1.797& 1.124& 1.171  & 1.829 & 1.583  &
1.141& 1.396 \\ 
      & Total & 1.613  & 1.762& 1.672& 1.732  & 1.796 & 1.635  & 1.703& 1.944 \\ \hline

 \multicolumn{2}{|c|}{AV18 + UIX$^{\rm (a)}$}
              & 1.59 & 1.73 &  &      &  1.76 & 1.61& &         \\ \hline
 \multicolumn{2}{|c|}{Experiment}
              & 1.60$^{\rm (b)}$  & && 1.755  &  1.77$^{\rm (b)}$ & && 1.95     \\ \hline
\end{tabular}

 {\footnotesize \begin{flushleft}
$^{\rm a)}$Taken from~\cite{Pieper}.

$^{\rm b)}$These ``experimental'' values are taken from
\cite{Pieper}. They have been obtained by substraction of the own
proton and neutron charge radii squared (0.743 and --0.116~fm$^2$,
respectively) from the experimental values of the charge radii
squared.
\end{flushleft}}
\label{table4}
\end{table}

In the Table~4 we give r.m.s. radii for all the above distributions  in $^3$H
and $^3$He found in the impulse approximation, as well as  the respective
experimental values  and results obtained for AV18($2N$) + UIX($3N$) forces. To
demonstrate  the separate contributions of the $3N$ and $6qN$ channels to these
observables, we also present  the values calculated separately with only
nucleonic and $6qN$ parts of the total wavefunction. It is seen from the
Table~4 that both versions of our model  (viz. DBM(I) and DBM(II)) give quite
similar values for all the radii. The most interesting point here is the
importance of $6qN$ component contributions. In fact,  the contribution of the
$6qN$ channel shifts all the radii, i.e., $r_{\rm ch}$ and $r_p$ in $^3$H and
$^3$He,  predicted with pure nucleonic components in our approach, much closer
to the respective experimental values. For example, the value $r_{\rm
ch}=1.822$~fm calculated for $^3$H with only the nucleonic part of the
wavefunction  is essentially larger than the respective experimental value
1.755~fm. However, an admixture of a rather compact $6qN$  component ($r_{\rm
ch}=1.22$~fm) immediately shifts the $^3$H charge radius down to a value of
1.766~fm, which  is very close to its experimental value.

Thus, the dibaryon--nucleon component works in a right way also in this aspect.
It is interesting to  note that, in general, the predictions of our 
two-phase model are quite close to those of the  conventional pure nucleonic
AV18 + UIX model. This means that
our multichannel model is effectively similar to a conventional purely
nucleonic model (at least for many static characteristics). However, this  similarity will surely hold only for the
characteristics that are sensitive mainly to low-momentum transfers, while the
properties and processes involving high-momentum transfers will be treated in
two  alternative approaches in completely different ways.

\subsection{Coulomb displacement energy and charge symmetry
breaking effects}

The problem of accurate description of Coulomb effects in $^3$He in the current
$3N$ approach of the  Faddeev or variational type has attracted much attention
for last three decades (see, e.g., \cite{Friar,Gloeckle}  and the references
therein to the earlier works). It is interesting that the Coulomb puzzle in
$^3$He, being related to the long-range interactions, is treated in a different
manner in our and conventional approaches.

The $\Delta E_{\rm C}$ problem dates back to the first accurate $3N$  calculations
performed on the basis of the Faddeev equations with realistic $NN$ interactions
in the
mid-1970s~\cite{Gignoux}. These pioneer calculations first exhibited a hardly
removable difference of ca. 120 keV between the theoretical prediction  for
$\Delta E_{\rm C}^{\rm th} \simeq 640$~keV and the respective experimental value
$\Delta E_{\rm C}^{\rm exp}  \simeq 760$~keV. In subsequent 30 years, numerous
accurate $3N$ calculations have been performed over the  world using many
approaches, but this puzzle was still generally unsolved. The most plausible
quantitative explanation (but yet not free of serious questions) for the puzzle
has  been recently suggested by Nogga {\em et al.}~\cite{Gloeckle}. They have
observed that the difference in  the singlet $^1S_0$ scattering lengths of $pp$
(nuclear part) and $nn$ systems (originating from the  effects of charge
symmetry breaking (CSB)) can increase the energy difference between $^3$H and
$^3$He  binding energies and thus contribute to $\Delta E_{\rm C}$.

Our results obtained in this work with DBM give an alternative explanation of
the $\Delta E_{\rm C}$
puzzle and other Coulomb  effects in $^3$He without any free parameter. 
The Coulomb displacement energies $\Delta
E_{\rm C}$, together with the individual contributions to the  $\Delta
E_{\rm C}$-value, are presented in Table~5.  

\begin{table}[h!]
 \setcaptionmargin{0mm} \onelinecaptionsfalse
\captionstyle{flushleft}
\caption{Contribution of various terms (in keV) of the Coulomb interaction
to the $^3$H--$^3$He mass difference $\Delta E_{\rm C}$}
\begin{tabular}{|l|c|c|c|}  \hline
Contribution               & DBM(I) & DBM(II) & AV18 + UIX \\ \hline
Point Coulomb $3N$ only    & 598    & 630 & 677 \\ 
Point Coulomb $3N+6qN$     & 840    & 782 & --   \\ 
Smeared Coulomb $3N$  only & 547    & 579 & 648 \\ 
Smeared Coulomb $3N+6qN$   & 710    & 692 & --   \\ 
$np$ mass difference       & 46     & 45  & 14  \\  
Nuclear CSB (see Table~6)  & 0      & 0   &+65  \\  
Magnetic moments and spin-orbit$^{\rm a)}$  & 17 & 17  & 17 \\  
Total                    & 773    & 754 & 754 \\ \hline
\end{tabular}

 {\footnotesize \begin{flushleft} $^{\rm a)}$Here we use the value for this correction from
\cite{Gloeckle}.\end{flushleft}}
\label{table5}
\end{table}

We emphasize three important points, where our results differ from those for
conventional models.

\begin{itemize}
 \parindent=1.5em
\item[(i)] First, we found a serious difference between conventional and our
approaches in the short-range  behavior of wavefunctions even in the nucleonic
channel. Conventional $3N$ wavefunctions are strongly  suppressed along all
three interparticle coordinates $r_{ij}$ due to the short-range local repulsive
core, while our wavefunctions (in the $3N$ channel) have stationary nodes and
short-range loops along  all $r_{ij}$ and the third Jacobi coordinates
$\rho_k$. Such a node along the $\rho$ coordinate is seen also in  the $6qN$
relative-motion wavefunction. This very peculiar short-range behavior of our
wavefunctions leads to a strong enhancement of the high-momentum components of
nuclear wavefunctions, which is required by various modern experiments. On the other hand, these short-range radial loops lead to
significant errors, when using the Coulomb interaction between point-like
particles within our approach. Hence, we must take into account the finite
radii of charge distributions in the proton and $6q$  bag. Otherwise, all
Coulomb energies will be overestimated.

\item[(ii)] Another important effect following from our calculations is a
quite significant contribution of the internal $6qN$ component to $\Delta E_{\rm C}$. In
fact, just this interaction, which is completely missing in  conventional
nuclear force models, makes the main contribution (163 and 113~keV for versions
I and II correspondingly)
to filling the
gap in  $\Delta E_{\rm C}$ between conventional $3N$ calculations and experiment if
the CSB effects are of little significance in $\Delta E_{\rm C}$.

The large magnitude of this three-body Coulomb force contribution 
in our models can be explained by
two factors: first, a rather short average distance
$\langle\rho^2\rangle^{1/2}$ between the $6q$ bag and the third nucleon
(which  enhances the Coulomb interaction in the $6qN$ channels) and,
second, a relatively large weight of the $6qN$ components, where
the $6q$ bag has the charge  +1 (i.e., it is formed from $np$
pair). This  specific Coulomb repulsion in the $6qN$ channel
should appear also in all other nuclei, where the total  weight of
such components is about 10\% and higher. Therefore, it should
strongly contribute to the  Coulomb displacement energies over the
entire periodic table and could somehow explain the long-term
Nolen--Schiffer paradox~\cite{Nollen} in this way.

\item[(iii)]  The third specific effect that has been found in this study and
contributes to the  quantitative explanation of $\Delta E_{\rm C}$ is a strong
increase in the average kinetic energy $\langle  T\rangle$ of the system. This
increase in $\langle T\rangle$ has been already discovered in the first  early
$3N$ calculations with the Moscow $NN$ potential model~\cite{Tur2} and results
in a similar  nodal wavefunction behavior along all interparticle coordinates
but without any non-nucleonic  component.

 The increase in $\langle T\rangle$ leads to the proportional increase
in the $np$ mass difference  correction to $\Delta E_{\rm C}$.
 Since the  average kinetic energy in our case is twice the kinetic energy in conventional force
models, this correction  is expected to be also much larger in our case. Hence, we
 evaluate  such a correction term in the following way (without usage of
the perturbation theory). In the conventional isospin formalism, one can assume
that the $^3$H and $^3$He nuclei consist of the equal-mass nucleons:
\[{m}=\frac{m_p+m_n}{2},\]
so that $m_p= m+\Delta m/2, \; m_n=m-\Delta m/2$, where $\Delta m= m_p-m_n $. The simplest way to
include the correction due to
the mass difference $\Delta m$ is to assume that all particles in $^3$H have the average mass
\[\bar{m}_{\rm H} =\frac{2m_n + m_p}{3}=m-\frac{1}{6}\Delta m,\]
while in $^3$He they have the different average mass
\[\bar{m}_{\rm He} =\frac{2m_p + m_n}{3}=m+\frac{1}{6}\Delta m .\]
In spite of smallness of the parameter $\Delta m/m$, the perturbation
theory in respect of this parameter does not work. So we used the average mass
$\bar{m}_{\rm H}$ in calculation of $^3$H and $\bar{m}_{\rm He}$ in calculation of
$^3$He. The contribution of this $np$ mass difference to the $\Delta E_{\rm C}$ value 
is given in the fifth row of Table~5. As
is seen from the table, this correction is not very small in our case  and
contributes to $\Delta E_{\rm C}$ quite significantly.

 Many other effects attributed to increasing the average kinetic energy
of the system will arise in our approach, e.g., numerous effects associated
with the enhanced Fermi motion of nucleons in nuclei.
\end{itemize}

{\bf Charge symmetry breaking effects in DBM.} 
As was noted above, the best explanation for the $\Delta E_{\rm C}$ value in the framework of
conventional force models  published up to date~\cite{Gloeckle} is based on
the introduction of some CSB effect, i.e., the  difference between $nn$ and
$pp$ strong interactions. At present, two alternative values of the $nn$
scattering length are assumed:
\begin{equation}
a^{(1)}_{nn}= -18.7 \mbox{ fm and } a^{(2)}_{nn}=-16.3 \mbox{ fm}.
\label{ann}
\end{equation}
The first value has been extracted from the previous analysis of
experiments $d(\pi^- ,\gamma)nn$~\cite{dpigam} (see also
\cite{dpig1} and references therein) and is used in all current
$NN$ potential models, while the second value in (\ref{ann}) has
been derived from numerous three-body  breakup experiments $n+d
\to nnp$ done for the last three decades. In recent years, such
breakup  experiments are usually treated in the complete Faddeev
formalism, which includes most accurately both  two-body and
3BF~\cite{a16}. Thus, this $a^{(2)}_{nn}$ value is
considered as a quite reliable one. However, the quantitative
explanation for the $\Delta E_{\rm C}$ value in conventional force
models uses  just the first value of $a_{nn}$ as an essential
point of all the construction. At the same time, the  use of the
second value $a_{nn}(=-16.3\mbox{ fm})$ (which is not less
reliable than the first one) invalidates completely the above
explanation!

Therefore, in order to understand the situation more deeply and to
determine the degree of sensitivity  of our prediction for $\Delta
E_{\rm C}$ to variation in $a_{nn}$,  we made also $3N$ calculations
with  two possible values of $a_{nn}$ from Eq.~(\ref{ann}).  These
 calculations have been carried out with the effective values
of the  singlet-channel coupling constant corresponding to the
$V_{NqN}$ part of the  $NN$ force:

\begin{equation}
 \lambda^{\rm eff}_{^3{\rm He}}(^1S_0)=\frac{1}{3}\lambda_{pp}+
 \frac{2}{3}\lambda_{np},
 \label{effconstHe}
 \end{equation}
\begin{equation}
 \lambda^{\rm eff}_{^3{\rm H}}(^1S_0)=\frac{1}{3}\lambda_{nn}+
 \frac{2}{3}\lambda_{np}.
 \label{effconstH}
 \end{equation}
  In the above calculations, we employ the value $\lambda_{np}=328.9$~MeV that
provides the accurate description of the $^1S_0 $ $np$ phase shifts and the
experimental  value of the $np$ scattering length
$a_{np}=-23.74$~fm~\cite{KuInt}. Here, for $pp$-channel we use the value
$\lambda_{pp}=325.523$~MeV fitted to the well-known experimental magnitude
$a_{pp}=-8.72$~fm and for $nn$-channel two $\lambda_{nn}$ values corresponding to two available
alternative values of the $nn$ scattering length (\ref{ann}) have been tested. 
The calculation results are presented in Table~6.

\begin{table}[h!]
 \setcaptionmargin{0mm} \onelinecaptionsfalse
\captionstyle{flushleft}
 \caption{Contribution of charge symmetry breaking effects to 
 the $^3$H--$^3$He mass difference $\Delta E_{\rm C}$}

 \begin{tabular}{|c|c|c|}\hline
 & \multicolumn{2}{|c|}{$\Delta E_ {\rm C}$, keV }\\ \cline{2-3}
 $a_{nn}$, fm & DBM(I) & DBM(II)\\ \hline
 --16.3 & --18   & --39 \\ 
 --18.9 & +45   & +26 \\ \hline
 \end{tabular}
\label{table6}
\end{table}

As is seen in Tables 5 and 6, the DBM (version I) can precisely reproduce the
Coulomb displacement energy $\Delta E_{\rm C}$ with the lower (in modulus) value
$a_{nn}=-16.3$~fm, while this model overestimates  $\Delta E_{\rm C}$ by 54~keV
(=45 + 9~keV) with
the  larger (in modulus) value $a_{nn}=-18.9$~fm. Thus, the DBM approach, in
contrast  to the conventional force models, prefers the lower (in modulus)
possible value  --16.3~fm of the $nn$ scattering length, which has been
extracted from very numerous $3N$ breakup experiments $n+d\to nnp$~\cite{a16}.

Now, let us discuss shortly the magnitude of CSB effects in our model. The
measure of CSB effects at low energies is used to consider the difference
between $a_{nn}$ and so-called ``pure nuclear'' $pp$ scattering length
$a_{pp}^N$ that is found from $pp$ scattering data, when the Coulomb potential
is disregarded. The model dependence of the latter quantity was actively
discussed in the 1970s--1980s~\cite{SauWall,Rahman,Albev}. However, the majority of
modern  $NN$ potentials fitted to the experimental value $a_{pp}=-8.72$~fm
results in the value $a^N_{pp}=-17.3$~fm, when the Coulomb interaction is
discarded. It is just the value that is adopted now as an ``empirical'' value
of the $pp$ scattering length~\cite{Machl}. Thus, the difference between this
value and $a_{nn}$ is usually considered as the measure of CSB effects.
However, our model (also fitted to the same experimental value 
$a_{pp}=-8.72$~fm) gives a quite surprising result:
 \begin{equation}
 a^N_{pp}({\rm DBM}) = -16.57 \mbox{ fm},
 \label{anpp}
 \end{equation}
 which differs significantly from the above conventional value (by 0.8~fm) due
 to the explicit energy dependence of the $NN$ force in our approach.

 Thus, if the difference $a^N_{pp}-a_{nn}$ is still taken as the measure of
 CSB effects, the smallness of this difference obtained in our model testifies
to a small magnitude of the CSB effects, which is remarkably smaller than the
values derived from conventional OBE models for the $NN$ force.

\section{Discussion}

The $3N$ results presented in the previous section differ significantly from
the results found with any conventional model for $NN$ and $3N$ forces
(based on Yukawa's meson exchange mechanism) and also from the results obtained in
the framework of hybrid models~\cite{hyb}, which include the two-component
representation of the $NN$ wavefunction $\Psi=\Psi_{NN}+\Psi_{6q}$. It is
convenient to discuss these differences in the following order.
 \begin{itemize}
 \parindent=1.5em
 \item[(i)]
 We found that the $q^2$ dependence of pair $NN$ forces on the momentum of the
 third particle in the $3N$ system is more pronounced in our case than in other
 hybrid models~\cite{hyb,Bakk,Weber,Sim}: the $3N$ binding energy decreases by
 ca. 1.7~MeV, from 5.83 to 4.14 MeV when one takes into account the
 $q^2$-dependence (cf. the first and second rows in
 Table~2). From more general point of view, it means that, in our approach,
 pairwise $NN$ interactions (except Yukawa OPE and TPE terms), being ``embedded''
 into a many-body system, loose their two-particle character and become
 substantially many-body forces (i.e., depending on the momenta of other particles of
 the system).

\item [(ii)] Due to such a strong $q^2$ dependence (of
``repulsive'' character), the $3N$ system calculated including
only the pairwise forces turns out to be strongly underbound
($E=-4.14$~MeV). In other words, the ``pairwise'' $NN$ forces
(including their $q^2$ dependence on the momenta of the third
nucleon) give only about half the total $3N$ binding energy,
leaving the second half for the 3BF contribution. Therefore, the
following question is decisively important: can the 3BF 
 (inevitably arising in our approach) give the large missing
contribution to the $3N$ binding energy? Usefulness of the
developed model for the description of nuclear systems depends
directly on the answer to this important question. It is
appropriate here to remind that in the conventional 3BF models
such as Urbana--Illinois or Tucson--Melbourne, the contribution of
3BF to the total $3N$ binding energy does not exceed 1~MeV; i.e.,
this contribution can be considered as some correction ($\sim
15$\%), although it is significant for the precise description of
the $3N$ system.

 \item[(iii)]
 Fortunately, the contribution of 3BF induced by OSE and TSE  enables one to
fill this 4.3~MeV gap between the two-body force contribution and experimental
value. In fact, including both OSE and TSE  contributions to 3BF, taken with
the same coupling constants and  form factors as in the driving $NN$-force
model, together with a quite reasonable value for the $\sigma NN$ coupling
constant, $g_{\sigma NN}=8$--10, one obtains the $3N$ binding energy that is
very close to the experimental value (see rows 3, 4 and 6, 7 in Table~2). Thus,
the presented force model leads to a very reasonable binding energy for the
$3N$ system, however, with the murch larger (as compared to the traditional
$3N$ force model) contribution of 3BF. In fact, the unification of the basic
$2N$ and $3N$ force parameters provides a strong support for the whole force
model suggested here and is   in a sharp contrast with all traditional force
model based on $t$-channel exchange mechanism. We remind to reader that the
$2N$ and $3N$ forces in conventional approaches (where the latter is induced by
an intermediate $\Delta$-isobar production) are taken with different cut-off
parameters values $\Lambda_{\pi NN}$ and $\Lambda_{\pi N\Delta}$ in $2N$ and
$3N$ sectors in order to explain the basic features of $3N$ nuclei and $N+d$
scattering! Thus in the traditional approach, one has some serious
inconsistency in parameter values for $2N$ and $3N$ sectors.

\item[(iv)]
The contributions of the pairwise and different three-body forces to the total $3N$
binding energy for $^3$H are given in the fifth and sixth rows of Table~2. From
the results presented in this table, one can conclude that just the total 3BF
contribution to the $3N$ binding energy dominates and, in fact, determines
the whole structure of the $^3$H and $^3$He ground states\footnote{It should be
noted here that the relative contribution of the pairwise effective force $W(E)$ to
the $3N$ binding energy decreases noticeably when including 3BF (due to 
strengthening of the $q^2$ dependence arising from the pairwise forces).}. Moreover, comparing
the third and fourth rows of the table, one can see a ``nonlinear'' effect of
self-strengthening  for the 3BF contribution. In fact, the comparison of the results
presented in these rows of the table (see the second and the fifth
columns) shows clearly that the binding energy is almost proportional to the
weight $P_{\rm in}$ of the $6qN$ component in the total $3N$ wavefunction.
Thus, when the  weight of the $6qN$ component increases, the 3BF contribution,
which is related directly only to this component of the total wavefunction,
increases accordingly. However, the enhancement of the pure attractive 3BF
contribution squeezes the $3N$ system and thus reduces its r.m.s. radius, i.e.,
the mean distance between nucleons, which, in turn, again increases the weight
of the $6qN$ component. In other words, a some chain process which strengthens
the attraction in the system arises. This process is balanced both by the
weakening of the effective pairwise interaction due to the $q^2$ dependence and by
the repulsive effect of the orthogonalizing pseudopotentials included in each
pair interaction.

There are two another important stabilizing factors weakening the strong
three-body attraction in the $3N$ system. First, the generation of the
short-range repulsive vector $\omega$-field, where all three nucleons are close
to each other~\cite{Bled}. Since the $\omega$-meson is heavy, this field is
located in the deep overlap region of all three nucleons. In the present study, we
omitted the three-body contribution of this repulsive $\omega$-field. This
repulsive contribution will keep the whole system from the further collapse due
to the strong attractive $3N$ force induced by the scalar field.

The second factor weakening slightly the effective $3N$ attraction is
associated with the conservation of the number of scalar mesons generated in
the $2N$ and $3N$ interaction process. The problem is that TSE 
giving the 3BF contribution (see Fig.~4) arises due to the break of the
$\sigma$-meson loop, which induces the main $2N$ force. In other words, the
$\sigma$ meson generated in the transition of pair nucleons from the $NN$ phase
state to the $6q$ state is absorbed either in the $6q$ bag with closing the
loop or by the third nucleon, resulting in the 3BF contribution. Thus, the
appearance of such a 3BF should weaken the attraction between nucleons in the pair.
We carefully estimated the effect of the meson-number conservation for the TSE
contribution on the total $3N$ binding energy. Its magnitude occurred to be
rather moderate in the absolute energy scale (ca. 0.3--0.4~MeV), but quite
noticeable within the whole TSE contribution. However, when the total nucleon
density increases (and the relative TSE contribution also increases), the
effect is enhanced.

\item[(v)] Dependence of the two-body coupling constants $\lambda
(\varepsilon)$ upon the average momentum of other nucleon in $3N$ system (see,
e.g., Eq.~(\ref{kernel0})) can be interpreted generally as a density dependence
of the resulted many-body force in many-nucleon system. It is easy to show the
appearance of the energy-dependent pairwise potentials of the above-mentioned
type leads inavoidably to a repulsive many-body force. In other words, the
effects of the two-body interactions of this type can be reinterpeted in terms
of the conventional static interaction as additional contribution of the
effective repulsive density-dependent many-body force. For example, if to 
remove the $q^2$-dependence from the coupling constant $\lambda(E-q^2/2m)$ 
of our two-body force (this $q^2$-dependence leads to a weakening of the
two-body force in a many-nucleon system, when $q^2$ is rising), then the neglected
$q^2$-dependence must be compensated by an additional repulsive
density-dependent effective three-body force. Thus we can replace this
energy-dependent two-body interaction by an effective static two-body potential
(as it is usually done) plus a repulsive density-dependent 3BF.

On the other hand, it is well known from the Skirme-model calculations of
nuclei that just similar repulsive phenomenological density-dependent
3BF should be added to conventional $2N$- and $3N$-forces to
guarantee the saturation properties of heavy nuclei. Thus, in this respect the
present force model is also in a qualitative agreement with phenomenological
picture of nuclear interactions.
\end{itemize}

\section{Conclusion}

In this paper, we have developed a general formalism for the multicomponent
description of the three-body system with particles having inner degrees of
freedom.  We have applied our new approach to studying $3N$ system with $2N$ and
$3N$ interactions based on the dressed dibaryon intermediate state and
$\sigma$-field generation. It has been shown that the DBM applied to the $3N$
system results inevitably in new three-body scalar and also new (three-body)
Coulomb forces due to the (strong + Coulomb)
interaction  between the dressed dibaryon and the third nucleon. These forces play a
crucial role in the structure of few-nucleon systems and very likely in whole 
nuclear dynamics. Our accurate variational
calculations have demonstrated that new 3BF gives a half of the $3N$ binding energy,
whereas the 3BF contribution in the traditional $NN$-force approaches gives
about 15\% of the total binding energy. Thus, the suggested approach to the
$NN$ and $3N$ interactions can lead to significant revision of relative contributions of
two- and many-body forces in all nuclear systems.

 The developed model gives the precise value for the Coulomb displacement
energy $\Delta E_{\rm C}$ of the  $A=3$ system. Two basic sources of
this contribution, which differ from conventional force models,  should be
indicated here:
\begin{itemize}
\item[ ] the three-body Coulomb force between the dressed bag and
the charged third nucleon;  and
\item[ ] quite significant correction to the kinetic energy of the system due to
the $np$ mass difference and high average kinetic energy.
\end{itemize}

It should be
emphasized that, contrary to other studies based on  conventional force models
(using the $2N$ and $3N$ forces generated via the meson-exchange mechanism),
this explanation  does not require any noticeable CSB effect, although our
model is still  compatible with such effects. However, these CSB effects do not
contribute remarkably to $\Delta E_{\rm C}$  in our approach.

It is crucially important that the DBM leads to a significant non-nucleonic
components in the $3N$ wavefunction (8--11\%), while this component in the
deuteron is ca. 3\%, which results in a reformulation of many basic effects in
few-nucleon systems and other nuclei as well. It is probable that the weight of
such non-nucleonic components in heavy nuclei can be even higher with an
increase in the mass number and nuclear density.

 There is a very specific
new interplay between two- and three-body forces: the stronger the  
three-body force, the smaller the attractive contribution of the two-body
force to the nuclear binding energy! This gives a very important stabilization
in nuclei and nuclear matter. By this way, a very natural density
dependence of nuclear interactions appears from  the beginning. Thus, the
general properties of the $3N$ system, where forces so much differ from any
conventional force model, would appear also to be much differ from the
predictions of any conventional model and, hence, from experiment.

Therefore, it was very surprising for us to find that the static characteristics of the $3N$
system in our case turned out to be very  close to the predictions of the
modern force model (such as AV18 + UIX) and thus to experiment. This  gives us
a good test of the self-consistency and accuracy of the new force model.
However, predictions of the present $NN$- and $3N$-force model in other aspects
will strongly be deviated  from those for conventional models. First, these are the
properties determined by the high-momentum  components of nuclear wavefunctions.
The point is that the system described by our multicomponent wavefunctions
including the dibaryon components explicitly can easily absorb high-momentum
transfers, which can hardly be absorbed by the system consisting of nucleons
only. Therefore, to fit the experimental data
corresponding to large-momentum transfers ($\sim 1\mbox{ GeV}/c$), many  types
of meson-exchange and isobar currents are often introduced into theoretical
frameworks. However,  these currents are often unrelated to the underlying
force model. Hence, it is rather difficult to check the self-consistency of
such calculations, e.g., the validity of gauge invariance, etc.

Numerous modern experiments could corroborate these results. In particular,
according to the recent experiments $^3{\rm He}(e,e'pp)$~\cite{Nikef} and their
theoretical interpretation on the basis of fully realistic $3N$ calculations,
the cross sections for the $^3{\rm He}(e,e'pp)$ process
are underestimated by about five times with
a fully realistic $3N$ model and incorporation of final state interaction 
and meson-exchange
currents.  This important
conclusion has been further confirmed in recent experiments at the Jefferson
Laboratory when the incident electron beam energy has been increased up to
$E_e=2.2$ and 4.4~GeV~\cite{Jef}. The data of the two different experiments
give a clear evidence of very strong short-range $NN$ correlation in the $^3$He
ground state. This correlation still cannot be explained within the traditional
pattern for the $3N$ system.

In addition, our approach has recently been partially supported~\cite{Kask}
 from the other side by considering a model for $2\pi$ production in $pp$
 collisions at $E_p=750$ and 900~MeV. The authors have found that almost
 all particle energy and angular correlations (e.g., $\pi^+ \pi^-,\, pp,
 \, \pi pp$, etc.) can be explained quantitatively by assuming that
 $\pi^+ \pi^-$ production occurs through the generation of an intermediate light
 $\sigma$ meson with the mass $m_{\sigma}\simeq 380$~MeV. These values generally
 agree with the parameters adopted in our $NN$
 model~\cite{KuJPG,KuInt} and drastically disagree with the values
 assumed in OBE and other potential models.

Very interesting general implication of the results presented here is their
evident interrelation to the famous Walecka hadrodynamic model for
nuclei~\cite{Walecka}. It is well known that the Walecka model describes nuclei
and nuclear matter in terms of the scalar $\sigma$- and vector $\omega$-fields,
where the $\sigma$-field gives the attractive contribution, while the vector
$\omega$-field balances this attraction by short-range repulsion. It is very
important that both basic fields appear (in the model) as the explicit
degrees of freedom (together with relativistic nucleons), in contrast to
conventional meson-exchange models for nuclear forces, where mesons appear as
the carriers of forces rather than as the explicit field degrees of
freedom. Our approach does include the $\sigma$-meson (and potentially the
$\omega$-meson) degrees of freedom in an explicit form, similarly to the Walecka
model.  Moreover, since the average kinetic energy of the $3N$ system in our
model is high (it is much higher than that in the conventional OBE approach), nucleon motion is closer to the relativistic case, and
thus the similarity with the Walecka model gets even closer.

There is also an additional strong argument in favor of a tight
interrelation between our and the above Walecka-type nuclear
model. Very recently, we have formulated~\cite{Shikh} the dibaryon
model for $NN$ interaction in terms of relativistic effective
field theory with the intermediate dibaryon being represented as a
color quantum string with color quark clusters at its ends. This
theory includes $\pi$, $\sigma$, $\rho$ and $\omega$-mesons as
a dressing of the dibarion together with the $N\Delta$ and
$\Delta\Delta$ loops. Thus, the $3N$ scalar force introduced in
the present work ``by hands'' can be derived in the field-theory
Lagrange language within the effective field theory. Moreover, in
the mean field approximation this effective field theory approach,
being applied to nuclei, should result in the Walecka--Serot
relativistic model with the dominating collective $\sigma$-field,
which couples the nucleons in a nucleus together.

Thus, the alternative description given here by the new force model looks to be more
self-consistent and  straightforward than the conventional OBE-type models. One aspect of this new picture is evident
-- the present model being applied to any  electromagnetic process on nuclei
leads automatically to a consistent picture of the process as whole:
single-nucleon currents at low-momentum transfers, meson-exchange currents
(including new meson currents)  at intermediate-momentum transfers, and quark
counting rules at very-high-momentum transfers, because  the model wavefunction
includes explicitly multinucleon, meson-exchange, and multiquark components.

From all the above-mentioned arguments one can conclude that the dibaryon concept of
nuclear force advocated in the work results in a new picture for nuclear
structure and dynamics.

We are grateful to Dr. I.~Obukhovsky for discussions and help in
calculation of some matrix elements, and also to many our colleagues in T\"ubingen University  for continuous encouraging in the
course of this work, and also to the staff of Institut f\"ur Theoretische Physik der
Universit\"at T\"ubingen, where the most part of calculations has been
performed.  

This work was supported in part by der Deutsche
Forschungsgemeinschaft (grant no. 436 RUS 113/790) and the Russian Foundation for
Basic Research (grants nos. 02-02-16612, 04-02-04001).

\appendix
\hfill {\bf \em APPENDIX}
\section*{Overlap functions between $3N$ symmetrized basis and $NN$ form factors
 and matrix elements of DBM interactions for a $3N$ system}
\setcounter{section}{1}
\setcounter{equation}{0}

\subsection{The construction of basis functions}
The total
wavefunction in $3N$ channel with the angular momentum $(J,M)$ and the isospin $(T,T_z)$
is written as (below we omit the quantum numbers $JMTT_z$):
\begin{equation}
\Psi^{(JMTT_z)}_{\rm ex}=\Psi^{(1)}_{\rm ex}+\Psi^{(2)}_{\rm ex}+
  \Psi^{(3)}_{\rm ex},
\end{equation}
\begin{equation}
\Psi^{(i)}_{\rm ex}=\sum_{{\gamma},n}C^{\gamma }_n
\Phi_{{\gamma}n}^{(i)} \qquad (i=1,2,3),
\end{equation}
where
\begin{equation}
\Phi_{{\gamma}n}^{(i)}({\bf r}_i,{\b \rho_i})=
N^{\gamma}_nr_i^{\lambda}\rho_i^l\,\exp(-\alpha_nr_i^2-\beta_n\rho_i^2)
{\cal F}^{JMTT_z}_{{\gamma}}(\hat{\bf r}_i,\hat{\b \rho}_i)\, {\cal T}^{(i)}_{t_{jk}}.
\end{equation}
We use the following notations:
 ${\bf r}_i$(${\bf p}_i$) is the relative coordinate (momentum) of the pair ($jk$), while
  ${\b \rho}_i$
 (${\bf q}_i$) is the Jacobi coordinate (momentum) of the  $i$th particle
 relative to the center of mass for the pair ($jk$), $(i,j,k)=(1,2,3)$ or their cyclic
 permutations.
  Here the composite label $\gamma
 =\{\lambda\,l\,L\,S_{jk}\,S\,t_{jk}\}$
represents the set of quantum numbers for the basis functions: the
angular momenta $\lambda$ and $l$ correspond to the Jacobi
coordinates ${\bf r}_i$ and ${\b \rho}_i$, respectively,
$S_{jk}(t_{jk})$ is spin (isospin) of the two-body subsystem
$(jk)$, and $L(S)$ is the total orbital momentum (spin) of the
system. The normalizing coefficient in (A.3) is
\begin{equation}
N^{\gamma}_n=2^{\lambda+l+3}\sqrt{\frac{2\alpha_n^{\lambda +3/2}\beta_n^{l+3/2}}
{\pi (2\lambda +1)!!(2l+1)!!}}.
\end{equation}

The spin-angular ${\cal F}_{{\gamma}}$ and isospin ${\cal T}^{(i)}_{t_{jk}}$ parts of
the basis function
are defined by
 a standard vector-coupling scheme:
\begin{equation}
 {\cal F}^{JMTT_z}_{\gamma}=
 |\{\lambda_i l_i:L\}\{s_js_k(S_{jk})s_i:S\}:JM\rangle ,
  \end{equation}
 \begin{equation}
 {\cal T}^{(i)}_{t_{jk}} = |t_jt_k(t_{jk})t_i:TT_z\rangle ,
 \end{equation}
where $s_i(=1/2)$ and $t_i(=1/2)$ are spin and isospin of the $i$th nucleon.

Now we define the symmetrized basis functions as:
\begin{equation}
\Phi^{\rm sym}_{\gamma n}=\sum_{i=1,2,3}\Phi^{(i)}_{\gamma n},
\end{equation}
so that the total wavefunction in an external ($3N$) channel takes the form:
\begin{equation}
\Psi_{\rm ex}=\sum_{{\gamma},n}C^{\gamma}_n
\Phi_{{\gamma}n}^{\rm sym}.
\end{equation}

\subsection{Nucleon--nucleon form factors}

The $NN$ form factors in the separable DBM interaction and in the
projectors $\varphi^{J_iM_it_dt_{d_z}}_{\lambda_iS_d}({\bf r}_i)$,
corresponding
to the orbital momentum
$\lambda_i$, spin $S_d$, the total angular momentum
$(J_i,M_i)$, and isospin $(t_dt_{d_z})$ of the subsystem $(jk)$ 
($\bf J_i={\b\lambda}_i+S_d$), have the form:
\begin{equation}  \varphi^{J_iM_it_dt_{d_z}}_{\lambda_iS_d}({\bf r}_i)
\equiv
\varphi_{f}({\bf r}_i) =\sum_m D^f_mr^{\lambda_i}_i\,
\textstyle \exp(-\frac{1}{2}\eta^2_mr^2_i) \,
{\cal F}_{f}(\hat{\bf r}_i)\, {\cal T}_f^{(i)},
\end{equation}
where
\begin{equation}
 f\equiv \{\lambda_i, S_d,J_i,M_i,t_d,t_{d_z}\}, \qquad
  {\cal F}_f=|\lambda_iS_d:J_iM_i\rangle, \qquad
  {\cal T}^{(i)}_f=|t_jt_k:t_dt_{d_z}\rangle,
\end{equation}
 and $D^f_m$  and $\eta_m$ are linear and nonlinear parameters, respectively, 
 of the Gaussian expansion. 
 (In this Appendix we have altered the notation
 for the quantum numbers of the $NN$ form factor as compared with the main text of
 the paper: we have replaced  $L_i \to \lambda_i$ for the orbital momentum and
 also included the
 isospin quantum numbers $t_d,t_{d_z}$.) In the
 single-pole approximation the DBM includes only one form factor for each set
 $f$, so that index $f$ determines the form factor uniquely.
In the present version of DBM we use only $0S$, $2S$, and $2D$ oscillator
functions as the form factors. So, we need to expand in Gaussians the $2S$
function only.

\subsection{Overlap integrals}
The total overlap function (\ref{ovfull})
\begin{equation}
\chi^{J_iM_it_dt_{d_z}}_{\lambda_i S_d}(i) \equiv \chi_f(i)
= \langle \varphi_f(i)|\Psi_{\rm ex} \rangle
=\sum_{\gamma n}C^{\gamma}_n \,\langle \varphi_f(i) |
\Phi^{\rm sym}_{\gamma n}\rangle
\end{equation}
and  also matrix elements (m.e.) of any DBM interaction
include the overlap integrals between the  form factors $\varphi_f$
and symmetrized basis
functions $\Phi_{{\gamma} n}^{\rm sym}$:
\begin{equation}
I_{(i)}^{f,{\gamma} n}({\b \rho}_i)=
\langle \varphi_f({\bf r}_i) | \Phi_{{\gamma}n}^{\rm sym}\rangle .
\end{equation}
This overlap integral consists of three terms:
\begin{equation}
I_{(i)}^{f,{\gamma} n}=
I_{(i)i}^{f,{\gamma} n}+
I_{(i)j}^{f,{\gamma} n}+I_{(i),k}^{f{\gamma} n},
\end{equation}
namely, one ``diagonal'' ($I_{(i)i}$) and two non-diagonal ones:
\[
I_{(i)j}^{f,{\gamma} n}({\b \rho}_i)
=\langle \varphi_f({\bf r}_i) |
\Phi_{{\gamma} n}^{(j)}({\bf r}_j,{\b \rho}_j)\rangle =
\langle {\cal T}^{(i)}_f|{\cal T}^{(j)}_{t_{ik}} \rangle
\sum_m D^f_mN^{\gamma}_n \times
\]
\begin{equation}
\times
\int r_i^{\lambda_i}\textstyle \exp(-\frac{1}{2}\eta^2_mr^2_i)
r^{\lambda}_j\rho^{l}_j\,\exp(-\alpha_nr_j^2-\beta_n\rho_j^2)\,
\langle {{\cal F}_f}(\hat{\bf r}_i)|
{\cal F}_{{\gamma}}(\hat{\bf r}_j,\hat{\b \rho}_j)\rangle {\rm d}^3r_i.
\end{equation}

Due to symmetry of the basis functions $\Phi_{{\gamma} n}^{\rm sym}$,
three overlap integras $I_{(i)}^{f,{\gamma} n}({\b \rho}_i)$ ($i=1,2,3$)
are identical, so that further we present formulas for the case of $i=2$.
For example,
\begin{equation}
\chi_f(2)=\sum_{\gamma n}C^{\gamma}_n \left (I^{f,\gamma n}_{(2)2} +
I^{f,\gamma n}_{(2)1}+ I^{f,\gamma n}_{(2)3}\right )
\end{equation}

 {\bf {(i) Diagonal  overlap integrals} $I_{(2)2}$:}
\begin{equation}
I^{f,\gamma n}_{(2)2}({\b \rho}_2)=
\sum_{{\cal J} m}G_{22}^{\varepsilon} \rho_2^l\exp(-\beta_n\rho_2^2)\
{\cal Y}_{l}^{{\cal J} J_i JM}(\hat{\b \rho}_2)\,
\delta_{t_dt_{31}}\,{\cal X}_{2}^{t_dt_{d_z}}; \qquad
\varepsilon\equiv \{{\gamma},f,{\cal J},m\};
\end{equation}
where
\begin{equation}
G_{22}^{\varepsilon}= \delta_{\lambda \lambda_i}
\delta_{S_dS_{31}}(-1)^{\lambda+ l+L} D^f_mN^{\gamma}_n
\frac{(2\lambda+1)!!\sqrt{\pi [L][S][J_i][{\cal J}]}}
{2^{\lambda +2}\,{\alpha}_{nm}^{\lambda +3/2}}
\left \{\begin{array}{ccc}
l&\frac{1}{2} &{\cal J}\\
\lambda &S_{31} &J_i \\
L & S & J   \end {array} \right \},
\end{equation}
\begin{equation}
[X]\equiv 2X+1,
\end{equation}
\begin{equation}
{\alpha}_{nm}=\alpha_n +\frac{1}{2}\eta^2_m,
\end{equation}
\begin{equation}
{\cal Y}_{l}^{{\cal J} J_i JM}(\hat{\b \rho}_2)
= \langle {\cal J} m_{\cal J}J_iM_i|JM\rangle \,
{\cal Y}^{{\cal J}m_{\cal J}}_{l1/2}(\hat{\b \rho}_2),
\end{equation}
\begin{equation}
{\cal X}_{2}^{t_dt_{d_z}} =
 \langle t_dt_{d_z}{\textstyle \frac{1}{2}}t_{2_z}|TT_z\rangle \,
\,|t_2t_{2_z}\rangle.
\end{equation}

{\bf {(ii)} Non-diagonal overlap integrals:}
\begin{equation}
I^{f,\gamma n}_{(2)1}({\b \rho}_2)=(-1)^{S_d+S_{23}}
\sum_{{\cal J} g,t,m}G_{21}^{\tilde{\varepsilon}}
\rho_2^{t}\exp(-{\omega}_{nm}\rho_2^2) \,
{\cal Y}_{g}^{{\cal J} J_i JM}(\hat{\b \rho}_2)
 \tau_{21}(t_d,t_{23}){\cal X}_{2}^{t_dt_{d_z}},
\end{equation}
\begin{equation}
I^{f,\gamma n}_{(2)3}({\b \rho}_2)=(-1)^{\lambda_i+\lambda}
\sum_{{\cal J} g,t,m}G_{21}^{\tilde{\varepsilon}}
\rho_2^{t}\exp(-{\omega}_{nm}\rho_2^2) \,
{\cal Y}_{g}^{{\cal J} J_i JM}(\hat{\b \rho}_2)
 \tau_{23}(t_d,t_{12}){\cal X}_{2}^{t_dt_{d_z}},
\end{equation}

where
\begin{equation}
\tilde{\varepsilon}\equiv \{{\gamma}, n,f,{\cal J},g,t,m\},
\end{equation}
\begin{equation}
G_{21}^{\tilde{\varepsilon}}=\sum_{\xi} G_{21}({\gamma},n,f,{\cal J},g,m,\xi)
\delta_{t,\lambda_i+\lambda +l-L_1-L_3-L_4},\qquad \xi \equiv
\{L_1,L_2,L_3,L_4,j_4,g_1,g_4\}.
\end{equation}
In Eq.~(A.24) the summation is carrying out over all intermediate quantum
numbers incorporated into the composite index $\xi$.
Note that the overlap functions $I_{(2)1}$ and $I_{(2)3}$ are distinguished by 
a phase factor and isospin functions only.

The algebraic coefficients $G_{21}$ in (A.25) are equal to
\[
G_{21}({\gamma},n,f,{\cal J},g,m,\xi)= (-1)^{J_i+g_1+L+1/2-J}D^f_mN^{\gamma}_n
\,A^{\lambda_iL_10}_{\lambda_i0L_1(\lambda_i\!-\!L_1)}({\bf P}_{nm})
\,A^{LL_3L_4}_{\lambda lL_1j_4} ({\bf Q}_{nm}) \times \]
\[\times \frac{[g_1][g_4]\sqrt{[\lambda_i][L][S][S_{23}][S_d][J_i][\lambda_i\!-\!l][j_4][{\cal J}]}}
{(2{\mu}_{nm})^{\frac{L_1+L_3+L_4+3}{2}}} \times \]
\[\times \Gamma(\mbox{\small $\frac{ L_1+L_3+L_4+3}{2}$}) \,
\langle(\lambda_i\!-\!L_1)0j_40|g_0\rangle \,
\left \{\begin{array}{ccc}(\lambda_i-L_1)&L_1&\lambda_i\\S_0&j&g_1\end{array}\right
\}\times\]
\begin{equation}
\times
\left \{\begin{array}{ccc}\frac{1}{2}&\frac{1}{2}&S_{23}\\\frac{1}{2}&S&S_d\end{array}\right \}
\left \{\begin{array}{ccc}{\cal J}&(\lambda_i\!-\!L_1)&g_4\\j_4&\frac{1}{2}&g\end{array}\right \}
\left \{\begin{array}{ccc}J&{\cal J}&J_i\\(\lambda_i\!-\!L_1)&g_1&g_4\end{array}\right \}
\left \{\begin{array}{ccc}
j_4&\frac{1}{2} &g_4\\
L_1 &S_d &g_1 \\
L & S & J   \end {array} \right \}.
\end{equation}
 In the formulas (A.22), (A.23), (A.26) the following notations are used:
 \begin{equation}
{\mu}_{nm}=\mu_n +\frac{1}{2}\eta^2_m; \qquad
{\omega}_{nm}=\nu_n -\frac{\sigma^2_n}{4{\mu}_{nm}},
\end{equation}
 where
\begin{equation}
\mu_n=\frac{1}{4}\alpha_n+\frac{3}{4}\beta_n;\qquad
\nu_n=\frac{3}{4}\alpha_n+\frac{1}{4}\beta_n; \qquad
\sigma_n=\frac{\sqrt{3}}{2}(\alpha_n-\beta_n);
\end{equation}

  The coefficients $A$ in(A.26) are related to rotation of the basis functions from one
 Jacobi set to other one:
\[
A^{LL_1L_2}_{\lambda l j_1j_2}(\hat{\bf R})=
(-1)^{\lambda +l}(R_{11})^{L_1} (R_{12})^{\lambda -L_1}
(R_{21})^{L_2} (R_{22})^{l -L_2}
\left (\begin{array}{ccc}L_1&L_2&J_1\\0&0&0\end{array}\right )
\left (\begin{array}{ccc}\lambda\! -\!L_1&l\!-\!L_2&J_2\\0&0&0\end{array}\right )
\times \]
\begin{equation}
\times \sqrt{\frac{[\lambda ]![l]![\lambda ][l][L_1][L_2][\lambda
-L_1][l-L_2][j_1][j_2]}{[L_1]![L_2]![\lambda-L_1]![l-L_2]!}}
\left \{\begin{array}{ccc}
L_1&\lambda\! -\!L_1&\lambda \\
L_2&l\! -\!L_2& l \\
j_1 & j_2 & L \end {array} \right \}.
\end{equation}

The rotation matrices ${\bf P}_{nm}$ and ${\bf Q}_{nm}$ in (A.26)
have the forms:
\begin{equation}
{\bf P}_{nm}=\left (
\begin{array}{cc}1&-\frac{\sigma_n}{2{\mu}_{nm}}\\0&1\end{array}\right ),
\end{equation}

\begin{equation}
{\bf Q}_{nm}=\left (
\begin{array}{cc}-\frac{1}{2}&-\frac{\sqrt{3}}{2}\\ \frac{\sqrt{3}}{2}&
-\frac{1}{2}\end{array}\right )
{\bf P}_{nm}.
\end{equation}

The overlaps between the basic isospin functions $\tau_{ik}$ are equal to:
\[
\tau_{ij}(t'_{jk},t_{ik})\equiv
\langle {\cal T}^{(i)}_{t'_{jk}} | {\cal T}^{(j)}_{t_{ik}} \rangle =\]
\begin{equation}
=\left\{\begin{array} {l} \delta_{t'_{jk}t_{jk}} \mbox{ for } i=j,\\
\sqrt{(2t'_{jk}+1)(2t_{ik}+1)}
\left \{\begin{array}{ccc}\frac{1}{2}&\frac{1}{2}&t'_{jk}\\
\frac{1}{2}&T&t_{ik}\end{array}\right \}\times \left \{
 \begin{array} {l} (-1)^{t_{ik}} \mbox{ for }(ij)=(13),(21),(32),\\
 (-1)^{t'_{jk}} \mbox{ for }(ij)=(12),(23),(31).
\end{array} \right .
\end{array} \right .
\end{equation}

\subsection{Conversion to momentum representation}
One of the advantages of Gaussian basis is the fact that Gaussian functions have
the same form in both the coordinate and momentum representations. So
expressions for the overlap integrals given below in the coordinate representation can
be directly used for the calculation of m.e. of DBM interaction
operators in the momentum representation. We use a ``symmetrized'' momentum representation:
\begin{equation}
f({\bf p})=\int f({\bf x}){e}^{i{\bf p\cdot x}}\frac{{d}^3x}{(2\pi)^{3/2}}.
\end{equation}
Therefore, due to  properties of the Gaussian functions, the
(normalized) basis functions $\Phi_{{\gamma}n}^{(i)}({\bf
p}_i,{\bf q}_i)$ in momentum representation have the same form
(A.3):
\begin{equation}
\Phi_{{\gamma}n}^{(i)}({\bf p}_i,{\bf q}_i)=
\tilde{N}^{\gamma}_np_i^{\lambda}q_i^l\,
\exp(-\tilde{\alpha}_np_i^2-\tilde{\beta}_nq_i^2)
{\cal F}^{JMTT_z}_{{\gamma}}(\hat{\bf p}_i,\hat{\bf q}_i)\, 
{\cal T}^{(i)}_{t_{jk}},
\end{equation}
where
\begin{equation}
\tilde{\alpha}_n=\frac{1}{4\alpha_n},\;\tilde{\beta}_n=\frac{1}{4\beta_n}.
\end{equation}
Moreover, as all the $NN$ form factors (A.9) are the sums
of Gaussians, the form of the overlap integrals (A.16)-(A.29) keeps
invariable when passing  from coordinate to momentum
representation, if one replaces in these formulas:
\begin{equation}
\alpha_n \to \tilde{\alpha}_n,\;\beta_n \to \tilde{\beta}_n,\;
\eta_m \to \tilde{\eta_m}=\frac{1}{\eta_m}.
\end{equation}
Below we use the symbols with tilde for designation of the corresponding quantities in
momentum representation, e.g., $\tilde{\alpha}_{nm}=\tilde{\alpha}_n
+1/2\tilde{\eta}_m$, etc.

\subsection{Matrix elements for DBM operators}
All quantities related to the non-nucleonic channels in DBM can be
expressed in momentum representation as the sums of integral
operators with factorized kernels (see Eq.~(\ref{3BF})):
\begin{equation}
O^{\rm DBM}_{(i)}= \varphi_{f'}({\bf p}_i)
O^{f'\!f}({\bf q'}_i,{\bf q}_i;E) \varphi_{f}({\bf p}_i).
\end{equation}
Therefore the m.e. of such an operator is equal to the sum of the m.e. for
one-particle operators $O^{f'\!f}({\bf q'}_i,{\bf q}_i;E)$ between the overlap
functions $\chi_f({\bf q}_i)$:
\begin{equation}
M_{2} =\langle \Psi_{\rm ex} | O^{\rm DBM}_{(2)} | \Psi_{\rm ex}\rangle =
\sum_{ff'} \langle \chi^{f'}_{(2)}| O_{(2)} |\chi^{f}_{(2)} \rangle
= \sum_{\gamma n, \gamma' n'} C^{\gamma'}_{n'} C^{\gamma}_n \sum_{i,j=1,2,3}
M^{f'\gamma' n'}_{f\gamma n}(i2j),
\end{equation}
where $M^{f'\gamma' n'}_{f\gamma n}(i2j)$ are the corresponding basis m.e.:
\begin{equation}
M^{f'\gamma' n'}_{f\gamma n}(i2j)=
\langle I^{f',\gamma' n'}_{(2)i} |O_{(2)}|I^{f,\gamma n}_{(2)j}\rangle.
\end{equation}

Any scalar--isoscalar operator $O({\bf q}-{\bf q}')$, which does not depend on
spin and isospin variables (e.g., the DBM two-body force, the projector, the
3BF due to
$\sigma$-exchange, the norm of non-nucleonic component), can be expanded into
spherical harmonics as:
\begin{equation}
O({\bf q}'-{\bf q})=\sum_{LM}O_L({\bf q}',{\bf q})Y^*_{LM}(\hat{\bf q}')
Y_{LM}(\hat{\bf q})
\end{equation}
In this case the spin-angular and isospin parts of the overlaps give:
\begin{equation}
\sum_{M}\langle {\cal Y}_{g'}^{{\cal J}' J_i JM}(\hat{\bf q}')
 | Y^*_{LM}(\hat{\bf q}') Y_{LM}(\hat{\bf q})
 | {\cal Y}_{g}^{{\cal J} J_i JM}(\hat{\bf q}_2)\rangle
 =\delta_{{\cal J}'{\cal J}} \delta_{g'L} \delta_{gL},
\end{equation}
\begin{equation}
\sum_{t_{d_z}}\langle {\cal X}^{t'_dt'_{d_z}} |{\cal X}^{t_dt_{d_z}}
\rangle=\delta_{t'_dt_d}.
\end{equation}

Therefore,  nine  m.e.'s for such an operator $M(i2j)\equiv M_{i2j}
\;(i,j=1,2,3)$ can be
reduced to radial integrals
of four types (here we omit the indices $f\gamma n,\, f'\gamma' n'$ for brevity):
\[M_{222}=\delta_{J'_iJ_i}\delta_{l'l}\delta_{t'_{13}t_{13}}R_{222},\]
\[M_{122}=(-1)^{S_d+S'_{23}}\,\tau_{12}(t'_{23},t_d) R_{122},\]
\[M_{322}=(-1)^{\lambda'_i+\lambda'}\,\tau_{32}(t'_{12},t_d) R_{122},\]
\[M_{221}=(-1)^{S_d+S_{23}}\,\tau_{21}(t_{23},t_d) R_{221},\]
\[M_{223}=(-1)^{\lambda_i+\lambda}\,\tau_{23}(t_{12},t_d) R_{221},\]
\[M_{121}=(-1)^{S_{23}+S'_{23}}\,\tau_{12}(t'_{23},t_d)\tau_{21}(t_{23},t_d)
R_{121},\]
\[M_{323}=(-1)^{\lambda +\lambda' +\lambda_i+\lambda'_i}\,
\tau_{32}(t'_{12},t_d) \tau_{23}(t_{12},t_d) R_{121},\]
\[M_{123}=(-1)^{S_d+S'_{23}+\lambda_i+\lambda}\,
\tau_{12}(t'_{23},t_d) \tau_{23}(t_{12},t_d) R_{121},\]
\[M_{321}=(-1)^{S_d+S_{23}+\lambda'_i+\lambda'}\,
\tau_{32}(t'_{12},t_d) \tau_{21}(t_{23},t_d) R_{121}.\]

Here,
\begin{equation}
R_{121}=\sum_{{\cal J}{\cal J}'gg',mm'tt'} G^{\tilde{\varepsilon}'}_{21}
G^{\tilde{\varepsilon}}_{21}\delta_{{\cal JJ}'}\delta_{gg'}
R^{t'\!t}_g(\tilde{\omega}_{n'm'},\tilde{\omega}_{nm};O),
\end{equation}

\begin{equation}
R_{122}=\sum_{{\cal J}{\cal J}'g',mm't'} G^{\tilde{\varepsilon}'}_{21}
G^{{\varepsilon}}_{22}\delta_{{\cal JJ}'} \delta_{g'l}
R^{t'\!l}_l(\tilde{\omega}_{n'm'},\tilde{\beta}_{n};O),
\end{equation}

\begin{equation}
R_{221}=\sum_{{\cal J}{\cal J}'g,mm't} G^{{\varepsilon}'}_{22}
G^{\tilde{\varepsilon}}_{21}\delta_{{\cal JJ}'} \delta_{gl'}
R^{l'\!t}_{l'}(\tilde{\beta}_{n'},\tilde{\omega}_{nm};O),
\end{equation}

\begin{equation}
R_{222}=\sum_{{\cal J}{\cal J}',mm'} G^{{\varepsilon}'}_{22}
G^{{\varepsilon}}_{22}\delta_{{\cal JJ}'}
R^{ll}_l(\tilde{\beta}_{n'},\tilde{\beta}_{n};O),
\end{equation}

\begin{equation}
R^{t'\!t}_L(\tilde{\omega}',\tilde{\omega};O)=
\int\limits_0^\infty \int\limits_0^\infty (q')^{t'+2}q^{t+2}
{e}^{-\tilde{\omega}'(q')^2} \,{e}^{-\tilde{\omega} q^2 } O_L(q',q){d}q'\,{d}q.
\end{equation}

Below we give the explicit formulas for the radial integrals $R^{t'\!t}_L$
for all
specific terms of the DBM interaction.

{\bf Projector.}
The total projector onto the state $\varphi^{J_i}_{\lambda_i,S_d}$ has the form:
\begin{equation}
\Gamma^{\bar{f}} \equiv \Gamma^{J_i}_{\lambda_i,S_d}=\sum_{M_i,t_{d_z}}
|\varphi^{J_iM_it_dt_{d_z}}_{\lambda_i,S_d}\rangle
\delta({\bf q}-{\bf q'}) \langle\varphi^{J_iM_it_dt_{d_z}}_{\lambda_i,S_d}|.
\end{equation}
After expanding the $\delta$-function into partial waves:
\begin{equation}
\delta({\bf q}-{\bf q'})= \sum_{LM}Y^*_{LM}({\bf q}')Y_{LM}({\bf q})
\delta({q}-{q'})/q^2,
\end{equation}
the corresponding operator $O_L$ in Eq.~(A.47) is reduced to
\begin{equation}
\Gamma^{\bar{f}}_L= \frac{\delta({q}-{q'})}{q^2},
\end{equation}
and the radial integral for projector takes the form:
\begin{equation}
R^{t'\!t}_L(\tilde{\omega}',\tilde{\omega};\Gamma^{\bar{f}})=
\frac{\Gamma(\frac{t'+t+3}{2})}
{2(\tilde{\omega}'+\tilde{\omega})^{\frac{t'+t+3}{2}}},
\end{equation}
where $\Gamma(x)$ is the gamma-function.

{\bf Effective two-body DBM interaction.}
According to Eq.~(\ref{kernel0}), the two-body DBM interaction in
the $3N$ system (between nucleons 1 and 3) has the form:
\begin{equation}
W_2=\sum_{J_i,\lambda'_i \lambda_i} W^{J_i}_{\lambda'_i \lambda_i}
({\bf p}'_2,{\bf p}_2;{\bf q}'_2,{\bf q}_2;E),
\end{equation}
where
\begin{equation}
W^{J_i}_{\lambda'_i \lambda_i}=\sum_{M_i}|\varphi^{J_iM_i}_{\lambda'_i,S_d}\rangle
\delta({\bf q}-{\bf q'}) \lambda^{J_i}_{\lambda'_i
\lambda_i}(E\!-\!q^2/(2\tilde{m}))
\langle\varphi^{J_iM_i}_{\lambda_i,S_d}|,
\end{equation}
therefore
\begin{equation}
\left\{W^{J_i}_{\lambda'_i \lambda_i}\right \}_L= \frac{\delta({q}-{q'})}{q^2}
\lambda^{J_i}_{\lambda'_i\lambda_i}(E\!-\!q^2/(2\tilde{m})).
\end{equation}

In the present version of DBM we employed a rational approximation for the
energy dependence of the coupling constant $\lambda^J_{LL'}(E)$~\cite{KuInt}:
 \begin{equation}
 \lambda^{J}_{LL'}(E) =
  \lambda^{J}_{LL'}(0)\frac{E_0+aE}{E_0-E},
\label{approx}
 \end{equation}
where the parameters $E_0$ and $a$ can be taken to be the same for all
$\lambda$'s. We found that this simple rational form can reproduce quite
accurately the exact energy dependence of the coupling constants
$\lambda^J_{LL'}(E)$ calculated from the loop diagram  in Fig.~1. Therefore,
in the $3N$ system the corresponding coupling constants
$\lambda^{J_i}_{\lambda'_i\lambda_i}$ take the form:
 \begin{equation}
  \lambda^{J_i}_{\lambda'_i\lambda_i}(E\!-\!q^2/(2\tilde{m}))
 = \lambda^{J_i}_{\lambda'_i\lambda_i}(0)\left (-a+(a+1)\frac{E_0}{E_0-E}
 \;\frac{1}{1+\frac{q^2}{2\tilde{m}(E_0-E)}}\right ).
\label{approx3}
 \end{equation}
The first term in Eq.~(A.56) leads to the expression for the radial
matrix element like (A.51). For calculating the second term we
expand the function $1/(1+q^2)$ into a sum of Gaussians:
\begin{equation}
\frac{1}{1+q^2/q_0^2} = \sum_{\cal M} {\cal B}_{\cal M}\exp (-\theta_{\cal
M}q^2/q_0^2),
\end{equation}
where $q_0^2=2\tilde{m}(E_0-E) >0 \; (\mbox{for }E<E_0)$ and the expansion
parameters $\{{\cal B}_{\cal M},\theta_{\cal M}\}$ are universal constants.
Then the total m.e. for two-body DBM interaction $W_2$ takes the form:
\[
R^{t'\!t}_L(\tilde{\omega}',\tilde{\omega};W_2^{\bar{f}'\!\bar{f}})=
\lambda^{J_i}_{\lambda'_i\lambda_i}(0)
\left (-a R^{t'\!t}_L  (\tilde{\omega}',\tilde{\omega};
\Gamma^{\bar{f}'\!\bar{f}})+ \right .\]
\begin{equation}
\left . +(a+1)\frac{E_0}{{E_0\!-\!E}}
\sum_{\cal M} {\cal B}_{\cal M}
\frac{\Gamma(\frac{t'+t+3}{2})}
{2(\tilde{\omega}'+\tilde{\omega}+\theta_{\cal M}/q_0^2)^{\frac{t'+t+3}{2}}}
\right ).
\end{equation}

{\bf Norm of $6qN$ component.}
The norm of $6qN$ component of the total $3N$ wavefunction
determined by Eq.~(\ref{norm1}) can be expressed via a sum of the
m.e. of the operator:
\begin{equation}
{\cal N}({\bf q}',{\bf q})=-\frac{d}{dE}
\lambda^{J_i}_{\lambda'_i\lambda_i}(E\!-\!q^2/(2\tilde{m})) \,
\delta ({\bf q}'-{\bf q}).
\end{equation}
For the energy dependence such as in Eq.~(A.55) the derivative
takes the form:
 \begin{equation}
 -\frac{d}{dE}\lambda(E) = \lambda(0)\,
 \frac{1}{(E_0\!-\!E)^2}.
\label{dlamb}
\end{equation}
Therefore,
\begin{equation}
R^{t'\!t}_L(\tilde{\omega}',\tilde{\omega};{\cal N}^{\bar{f}'\!\bar{f}})=
\lambda^{J_i}_{\lambda'_i\lambda_i}(0) \frac{E_0(1+a)}{(E-E_0)^2}
\sum_{{\cal M}'{\cal M}} {\cal B}'_{{\cal M}'}{\cal B}_{\cal M}
\frac{\Gamma(\frac{t'+t+3}{2})}
{2(\tilde{\omega}'+\tilde{\omega}+(\theta_{\cal M}+\theta_{\cal M}')/q_0^2)
^{\frac{t'+t+3}{2}}}.
\end{equation}

{\bf Three-body force due to OME.}

When calculating the matrix elements for 3BF due to OME
(Eq.~(\ref{WOME})), viz.:
 \begin{equation}
 {}^{\rm OME}W^{J_i'J_i}_{\lambda'_i\lambda_i}({\bf q}_i',{\bf q}_i;E) =
  \int {d}{\bf k} \frac{B^{J_i'}_{\lambda'_i}({\bf k}')}
  {E-E_{\alpha}-\frac{{q'_i}^2}{2m}}\,
  V^{\rm OME}({\bf q}'_i,{\bf q}_i) \,
   \frac{B^{J_i}_{\lambda_i}({\bf k})}
  {E-E_{\alpha}-\frac{{q}_i^2}{2m}},
\label{AWOME}
\end{equation}
we use a similar trick as in
the calculation of the norm for the $6qN$ component. It enables us to exclude
the vertex functions $B^{J_i}_{\lambda_i}({\bf k}) $ from the formulas for the matrix
elements. By replacing the product of propagators in the integral (over
 the meson momentum $\bf k$  in Eq.~(A.62)) with their difference, one
obtains the following expression free of the vertex functions:
 $$
 \int \frac{B^{J_i}_{\lambda'_i}({\bf k})\,B^{J_i}_{\lambda_i}({\bf k})}
 {(E\!-\!\varepsilon(k)\!-\!\frac{q^2}{2m})
 (E\!-\!\varepsilon(k)\!-\!\frac{{q'}^2}{2m})}\,
 {d}{\bf k}= \frac{\lambda^{J_i}_{\lambda'_i\lambda_i}(E\!-\!\frac{{q'}^2}{2m})
 -\lambda^{J_i}_{\lambda'_i\lambda_i}(E\!-\!\frac{{q}^2}{2m})}
 {{q'}^2-{q}^2} =
 $$
 \begin{equation}
 =\Delta\lambda^{J_i}_{\lambda'_i\lambda_i} (q',q)
\label{repl}
 \end{equation}
This quantity is  the finite-difference analogue of the derivative
of $\lambda$ with respect to $q^2$, which, in the case of  the
energy dependence (A.55),  takes the form:
\begin{equation} \Delta\lambda (q',q) =
\lambda(0)\,E_0(1\!+\!a)\frac{1}{E\!-\!q^2/2m}\,\frac{1}{E\!-\!{q'}^2/2m}.
\label{ddlamb}
\end{equation}
Thus, the matrix elements for OME can be found without explicit usage of the
vertex functions $B^{J_i}_{\lambda_i}({\bf k})$.

{\bf Three-body force due to OSE.}
The exchange operator for scalar mesons does not include any
spin--isospin variables. Therefore, Eq.(A.62)  can be simplified
and, in view of the energy dependence given in Eq.~(A.55), reduces
to the form
 \[
 {}^{\rm OSE}W^{J'_iJ_i}_{\lambda'_i\lambda_i}({\bf q}'_i,{\bf q}_i;E) = \delta_{J'_iJ_i}
 \lambda^{J_i}_{\lambda'_i\lambda_i(0)}\,E_0(1+a) \times\]
 \begin{equation}
  \times \frac{1}{E\!-\!E_0\!-\!\frac{q_i^2}{2m}}\,
  \frac{-g^2_{\sigma NN}}{({\bf q}_i-{\bf q}'_i)^2+m_{\sigma}^2}\,
  \frac{1}{E\!-\!E_0\!-\!\frac{{q'_i}^2}{2m}}.
\label{WOSE}
\end{equation}
Using the expansion of the OME interaction into partial waves:
 \begin{equation}
\frac{1}{({\bf q}'_i-{\bf q}_i)^2+m_{\sigma}^2}=
\sum_{LM}Y^*_{LM}({\bf q}')\frac{Q_L(z)}{2qq'}Y_{LM}({\bf q}),
\end{equation}
where $m_\sigma$ is mass of $\sigma$ meson,
$Q_L(z)$ is Legendre function of the second kind, and
\[ z=\frac{{q'}^2+q^2+m_\sigma^2}{2q'q},\]
one gets the following radial integral for OSE m.e.:
\begin{equation}
R^{t'\!t}_L(\tilde{\omega}',\tilde{\omega};{^{\rm OSE}W}^{\bar{f}'\!\bar{f}})=
 g^2_{\sigma NN} \lambda^{J_i}_{\lambda'_i\lambda_i}(0)
\frac{E_0(1+a)}{(E-E_0)^2}
\sum_{{\cal M}'{\cal M}} {\cal B}'_{{\cal M}'}{\cal B}_{\cal M}
{\cal R}(t',\tilde{\omega}'+{\theta}',t,\tilde{\omega}+{\theta};
L,m_\sigma),
\end{equation}
where
\begin{equation}
{\cal R}(t',\omega',t,\omega ;L,m)=\int\int (q')^{t'+2}
{e}^{-\omega' {q'}^2}\frac{ Q_L(z)}{2qq'}{q}^{t+2}{e}^{-\omega q^2}
{d}q' {d}q
\end{equation}

We calculate integrals like those in Eq.~(A.68) in the following way.
In the integral (A.68) $t'\ge L$, $t\ge L$ and it can be shown that
$t'-L$ and $t-L$ are the even numbers.  Introducing the auxiliary
indices $k$ and $k'$, so that
\[ t=L+2k \qquad t'=L+2k',\]
the integral (A.68) can be written as
\begin{equation}
{\cal R}(L+2k',\omega',L+2k,\omega
;L,m)=\left (-\frac{\partial}{\partial\omega'}\right )^{k'}
\left (-\frac{\partial}{\partial\omega}\right )^{k}
{\cal R}(L,\omega',L,\omega;L,m).
\end{equation}
The last integral with $t=t'=L$ is easily calculated in coordinate
representation. Using the well-known formulas
\begin{equation}
\frac{Q_L(\frac{{q'}^2+q^2+m^2}{2q'q})}{2qq'}=\int\limits_0^\infty j_L(q'\rho)\,\frac{{e}^{-m\rho}}{\rho}
j_L(q\rho) \rho^2 {d}\rho
\end{equation}
and
\begin{equation}
\sqrt{\frac{2}{\pi}} \int\limits_0^\infty {\rm e}^{-\beta q^2} q^{L+2}
j_L(q\rho) {\rm d}q = \frac{\rho^L}{(2\beta)^{L+3/2}}{\rm e}^{-\frac{\rho^2}{4\beta}},
 \end{equation}
we get the result:
\begin{equation}
{\cal R} (L,\omega',L,\omega;L,m)= \frac{1}{(4\omega\omega')^{L+3/2}}
J^{\rm Yuk}_L
((\omega^{-1}+{\omega'}^{-1})/4,m),
 \end{equation}
 where
 \begin{equation}
 J^{\rm Yuk}_L(b,m) =
\int\limits_0^\infty \rho^{2L+2}\,\frac{{\rm e}^{-m\rho}}{\rho}\,
{\rm e}^{-b{\rho^2}} {\rm d}\rho .
\end{equation}
This integral of Yukawa potential is calculated by recursions:
\begin{equation}
J^{\rm Yuk}_L(b,m) = \frac{1}{m^{L+1}} Z_L(x), \qquad x=\frac{m}{2\sqrt{b}},
\end{equation}
where
\begin{equation}
Z_0(x)=\sqrt{\pi}x{e}^{x^2}(1-{\rm erf}(x)),
\end{equation}
\begin{equation}
Z_1(x)=2x^2(1-Z_0(x)),
\end{equation}
\begin{equation}
Z_L(x)=2x^2\left((k-2)Z_{L-2}(x)-Z_{L-1}(x)\right )\!.
\end{equation}
These recursions (especially for large $L$) for functions $Z(x)$
and also expressions (A.75)--(A.76) for $Z_0$ and $Z_1$ cannot be
used for large values of $x$. So, at large $x$ we use the following
asymptotic series:
\begin{equation}
J^{\rm Yuk}_{2L}(b,m) = \sum_{i=0}^\infty\left (-\frac{2b}{m^2}\right )^i
\frac{(2L+2i-1)!!(L+i)!}{i!},
\end{equation}
\begin{equation}
J^{\rm Yuk}_{2L+1}(b,m) = \sum_{i=0}^\infty\left (-\frac{2b}{m^2}\right )^i
\frac{(2L+2i+1)!!(L+i)!}{i!}.
\end{equation}

{\bf Three-body force due to OPE.}
For OPE, we take the interaction operator in the standard form
 \begin{equation}
 V^{(i)}_{\rm OPE}= -\frac{g^2_{\pi NN}}{(2m_N)^2}({\b\sigma}^{(i)}\cdot {\bf p})
 \frac{1}{p^2+m_{\pi}^2}({\bf S}_d \cdot {\bf p})({\b\tau}^{(i)}\cdot {\bf T}_d),\,{\bf p}={\bf
 q}-{\bf q'},
\label{OPE}
 \end{equation}
where ${\b\sigma}^{(i)}$ and ${\b\tau}^{(i)}$ are the spin and isospin
operators of the third ($i$th) nucleon, whereas ${\bf S}_d$ and ${\bf T}_d$ are the
operators of the total spin and isospin of the $6q$ bag, respectively. We found
that the contribution of OPE is so small that it is sufficient to include only
$S$ waves in its evaluation. In this case, the central part of the OPE
interaction only is remained:
 \begin{equation}
 V^{\rm OPE}_c= g^2_{\pi NN}\frac{m_{\pi}^2}{(4m_N)^2}
 \frac{1}{3}({\b\sigma}^{(i)}\cdot {\bf S}_d)({\b\tau}^{(i)}\cdot {\bf T}_d)
 \frac{1}{p^2+m_{\pi}^2}.
\label{OPECEN}
 \end{equation}
 The spin--isospin matrix element is nonzero only for a
 singlet--triplet transition:
 \begin{equation}
 \langle S_d\!=\!0,T_d\!=\!1|\frac{1}{3}({\b\sigma}^{(i)}\cdot{\bf S}_d)
 ({\b\tau}^{(i)}\cdot {\bf T}_d)|S_d\!=\!1,T_d\!=\!0\rangle = \frac{4}{9}.
  \label{ST49}
 \end{equation}
 Then, the matrix element of the OPE contribution for $S$-waves takes the form
 (for $S_d\!=\!0,S'_d\!=\!1$  or $S_d\!=\!1,S'_d\!=\!0$):
 \begin{eqnarray}
R^{00}_0(\tilde{\omega}',\tilde{\omega};{^{\rm OPE}W}^{\bar{f}'\!\bar{f}})=
\frac{4}{9}f^2_{\pi NN} \sqrt{\lambda^0_{00}(0)\lambda^1_{00}(0)}
\frac{E_0(1+a)}{(E-E_0)^2}\times \nonumber \\
\times \sum_{{\cal M}'{\cal M}} {\cal B}'_{{\cal M}'}{\cal B}_{\cal M}
{\cal R}(0,\tilde{\omega}'+{\theta}',0,\tilde{\omega}+{\theta};
0,m_\pi).
 \label{OPEfin}
 \end{eqnarray}
 Here, we take the vertex functions $B^0_0$ and $B^1_0$ differing from
 each other by a constant only. Therefore, using Eq.~(A.63), one can
 exclude these functions from the formula for the matrix element.

{\bf Three-body Coulomb force.}
The
m.e. of the operator for three-body Coulomb force (\ref{V3bfCoul}) (for
point-like charges) can be expressed in terms of the
integrals over the overlap functions $\chi_f({\bf q})$:
\begin{equation}
 M_{i2j}({}^{\rm Coul}W) = e^2\sum_{J_iM_iL_iL'_i}
\lambda^{J_i}_{L_iL'_i}(0)
  (1+a)
  \int \frac{\langle I_{(2)i}^{f',\gamma'n'}({\bf q}') |
 \, {\frac{1\!+\!\tau_3^{(2)}}{2}\,(1+\hat{t}_{d_z})}|
 I_{(2)j}^{f,\gamma n}({\bf q})\rangle}{
 (E\!-\!E_0\!-\!\frac{q^2}{2m}) \, ({\bf q}\!-\!{\bf q'})^2\,
 (E\!-\!E_0\!-\!\frac{{q'}^2}{2m})}
 d{\bf q}\,d{\bf q'},
 \label{me3bf}
\end{equation}
where $(1\!+\!\tau_3^{(2)})/(2)$ is the operator of the nucleon charge
(nucleon with number 2).
The isospin part of this m.e. is equal to:
\begin{equation}
\tau^{\rm Coul}(t_d)=\sum_{t_{d_z}}\left\langle {\cal X}^{t_dt_{d_z}}\left |
\frac{1\!+\!\tau_3^{(2)}}{2}\,(1+{t}_{d_z})\right |{\cal X}^{t_d}
\right\rangle=\left \{\begin{array} {l} 1, \mbox{ for } t_d=0, \\
     \frac{1}{3}, \mbox{ for } t_d=1.
     \end{array} \right .
\end{equation}
Thus, for calculation of 3BF Coulomb m.e. one can apply the formulas 
Eqs.~(A.69)--(A.79) for isoscalar operator with this additional isospin 
factor (A.85):
 \begin{equation}
R^{t'\!t}_L(\tilde{\omega}',\tilde{\omega};{^{\rm Coul}W}^{\bar{f}'\!\bar{f}})=
 \delta_{t'_dt_d}\, \tau^{\rm Coul}(t_d) {\rm e}^2 \lambda^{J_i}_{\lambda'_i\lambda_i}(0)
\frac{E_0(1+a)}{(E-E_0)^2}
\sum_{{\cal M}'{\cal M}} {\cal B}'_{{\cal M}'}{\cal B}_{\cal M}
{\cal R}^{\rm Coul}(t',\tilde{\omega}'+{\theta}',t,\tilde{\omega}+{\theta};
L).
\end{equation}
Here the Coulomb integral ${\cal R}^{\rm
Coul}(t,{\omega}',t,{\omega};L)$ for the point-like charges is
obtained from Yukawa integral $\cal R$  (A.68) by putting $m=0$:

\begin{equation}
{\cal R}^{\rm Coul}(t',{\omega}',t,{\omega};L)=
{\cal R}(t',{\omega}',t,{\omega};L,0).
\end{equation}
Hence these Coulomb integrals are reduced by differentiating (see
Eq.~(A.69)) to the integrals:
\[
{\cal R}^{\rm Coul}(L,{\omega}',L,{\omega};L)=
\frac{1}{(4\omega\omega')^{L+3/2}}
\int\limits_0^\infty \rho^{2L+2}\,\frac{1}{\rho}\,
{e}^{-\frac{\rho^2}{4}(\omega^{-1}+{\omega'}^{-1})} {d}\rho \equiv \]
\begin{equation}
\equiv
\frac{1}{(4\omega\omega')^{L+3/2}}J^{\rm Coul}_L
((\omega^{-1}+{\omega'}^{-1})/4),
\end{equation}
\begin{equation}
J^{\rm Coul}_L(b)=\frac{\Gamma(L+1)}{2b^{L+1}}.
\end{equation}

 Now, we
can replace the Coulomb  potential $1/\rho$ between the point-like
charges in the integrand in Eq.~(A.88) with the  corresponding Coulomb potential
between the ``smeared'' charges:
\begin{equation}
J^{\rm Coul}_L(b,a)=\int\limits_0^\infty \rho^{2L+2}\,\frac{{\rm erf}(\rho\sqrt{a})}{\rho}\,
{e}^{-b\rho^2} {d}\rho.
\end{equation}
The latter integral is evaluated  analytically in the form of a
finite sum:
\begin{equation}
  J^{\rm Coul}_L(b,a)
 = \frac{1}{2}\frac{1}{a^{L +1}}\sum_{k=0}^{L}
   C^k_L\,\frac{k!}{(\frac{b}{a})^{k+1}}\,
   \frac{(2L -2k+1)!!}{2^{L-k}(\frac{b}{a}+1)^{L-k+1/2}},
 \label{JCoulba}
\end{equation}
where $C^k_{L}$ are the binomial coefficients.

\end{document}